\documentclass[aip, jcp, reprint, amsmath, amssymb]{revtex4-1}

\usepackage[english]{babel}
\usepackage[latin1]{inputenc}
\usepackage[T1]{fontenc}
\usepackage{graphicx}
\usepackage[caption=false]{subfig}
\usepackage{color}
\usepackage{tabularx}
\usepackage{xfrac}
\usepackage[colorlinks]{hyperref}
\graphicspath{{./figures/}}

\newcolumntype{Y}{>{\centering\arraybackslash}X}

\newcommand{\kT}{k_B T}

\newcommand{\ie}{\textit{i.e.}}
\newcommand{\eg}{\textit{e.g.}}
\newcommand{\etc}{\textit{etc.}}
\newcommand{\etal}{\textit{et al.}}

\newcommand{\bigO}{\ensuremath{{\cal O}}}

\newcommand{\grad}[1]{\ensuremath{\boldsymbol{\nabla}{#1}}}

\newcommand{\dive}[1]{\ensuremath{\boldsymbol{\nabla}\cdot{#1}}}
\newcommand{\lapl}[1]{\ensuremath{\nabla^2{#1}}}

\newcommand{\bF}{\ensuremath{\boldsymbol{F}}}
\newcommand{\bI}{\ensuremath{\boldsymbol{I}}}

\newcommand{\bR}{\ensuremath{\boldsymbol{R}}}
\newcommand{\bS}{\ensuremath{\boldsymbol{S}}}
\newcommand{\bU}{\ensuremath{\boldsymbol{U}}}
\newcommand{\bV}{\ensuremath{\boldsymbol{V}}}

\newcommand{\be}{\ensuremath{\boldsymbol{e}}}

\newcommand{\bn}{\ensuremath{\boldsymbol{n}}}

\newcommand{\bq}{\ensuremath{\boldsymbol{q}}}
\newcommand{\br}{\ensuremath{\boldsymbol{r}}}

\newcommand{\bu}{\ensuremath{\boldsymbol{u}}}
\newcommand{\bv}{\ensuremath{\boldsymbol{v}}}

\newcommand{\bx}{\ensuremath{\boldsymbol{x}}}

\newcommand{\bsig}{\ensuremath{\boldsymbol\sigma}}

\newcommand{\bmu}{\ensuremath{\boldsymbol\mu}}
\newcommand{\bze}{\ensuremath{\boldsymbol\zeta}}

\newcommand{\dd}{\ensuremath{\mathrm{d}}}

\newcommand{\avg}[1]{\ensuremath{\left\langle{#1}\right\rangle}}

\newcommand{\lba}{\ensuremath{\lambda_{\beta\alpha}}}
\newcommand{\gab}{\ensuremath{g_{\alpha\beta}}}
\newcommand{\Sab}{\ensuremath{S_{\alpha\beta}}}
\newcommand{\Hab}{\ensuremath{H_{\alpha\beta}}}

\newcommand{\Dhdta}{\ensuremath{d_{\mathrm{HD,\alpha}}^t}}
\newcommand{\Dhdt}[1]{\ensuremath{d_{\mathrm{HD},#1}^t}}

\newcommand{\Dhdra}{\ensuremath{d_{\mathrm{HD,\alpha}}^r}}
\newcommand{\Dhdr}[1]{\ensuremath{d_{\mathrm{HD},#1}^r}}

\allowdisplaybreaks
\begin{document}
\title{Short-time transport properties of bidisperse suspensions and porous media: a Stokesian Dynamics study}
\author{Mu Wang}
\email[]{mwwang@caltech.edu}
\affiliation{Division of Chemistry and Chemical Engineering, California Institute of Technology, Pasadena, California 91125, USA}

\author{John F. Brady}
\email[]{jfbrady@caltech.edu}
\affiliation{Division of Chemistry and Chemical Engineering, California Institute of Technology, Pasadena, California 91125, USA}

\date{\today}

\begin{abstract}
We present a comprehensive computational study of the short-time transport properties of bidisperse neutral colloidal suspensions and the corresponding porous media.
Our study covers bidisperse particle size ratios up to $4$, and total volume fractions up to and beyond the monodisperse hard-sphere close packing limit.
The many-body hydrodynamic interactions are computed using conventional Stokesian Dynamics (SD) via a Monte-Carlo approach.
We address suspension properties including the short-time translational and rotational self-diffusivities, the instantaneous sedimentation velocity, the wavenumber-dependent partial hydrodynamic functions, and the high-frequency shear and bulk viscosities; and porous media properties including the permeability and the translational and rotational hindered diffusivities.
We carefully compare the SD computations with existing theoretical and numerical results.
For suspensions, we also explore the range of validity of various approximation schemes, notably the Pairwise Additive (PA) approximations with the Percus-Yevick structural input.
We critically assess the strengths and weaknesses of the SD algorithm for various transport properties.
For very dense systems, we discuss in detail the interplay between the hydrodynamic interactions and the structures due to the presence of a second species of a different size.
\end{abstract}


\maketitle

\section{Introduction}
\label{sec:intro}

Understanding the short-time transport properties of colloidal suspensions has been a lasting pursuit of researchers for over a century, dating back to Einstein's inquiry to the effective viscosity of dilute suspensions~\cite{einstein-visc_1906}.
Such understanding has important scientific and technological implications due to colloidal suspensions' rich and complex behaviors---their applications encompass virtually every aspect of our lives.

The principal challenges in investigating colloidal suspensions are ($i$) the long-range and non-pairwise-additive hydrodynamic interactions (HIs) mediated by the solvent, which exhibit sharp transitions when two particles are close, and ($ii$) their sensitive response to the particle configurations, \eg, their shape, size, and physico-chemical environments.
To overcome these difficulties, a wide range of computational techniques have been developed; to name a few: Lattice Boltzmann simulations~\cite{st-motion-lb-simu_ladd_prl_1993, soft-matt-lb_duenweg_adv-pol-sci_2009}, Dissipative Particle Dynamics~\cite{Hoogerbrugge1992, Li2008}, Smoothed Particle Hydrodynamics~\cite{sph_liu_2003, sph_lucy_astroj_1977}, hydrodynamic multipole methods~\cite{hydro-multipole_ladd_jcp_1988, self-diffusion-lubrication-3body_cichocki_jcp1999}, boundary integral methods\cite{Pozrikidis1992, Kumar2012}, the Force Coupling Method~\cite{fcm_maxey_ijmf_2001, fcm-two-phase_maxey_jcompphys_2003, exp-fcm-verification_maxey_ijmf_2002}, and (Accelerated) Stokesian Dynamics~\cite{sd_durlofsky_jfm1987, stokesian-dynamics_brady_anfm1988, asd_sierou_jfm01,asd-brownian_banchio_jcp2003}.
Despite significant advancement, substantial gaps remain in the vast parameter space which leads to the versatility of colloidal suspensions.

In this work we present a comprehensive simulation study of the short-time transport properties of bidisperse colloidal systems, exploring the effects of particle size.
Size polydispersity arises naturally in colloidal systems~\cite{measure-vol-frac_poon_softmatt2012} and is known to affect their phase and packing behavior~\cite{bidisperse-packing_torquato_pre2013} and transport properties~\cite{exp-bimodal-visc_shikata_jor1998,exp-bimodal-latex-visc_wolfe_lang1992, bidisperse-visc-expt_woutersen_jor1993}, particularly at high density.
However, the majority of existing theoretical and simulation work focuses on monodisperse systems.
For polydisperse systems, with a few exceptions~\cite{bidisperse-drag-permeability_kuipers_jfm2005}, earlier studies were restricted to dilute systems~\cite{bimodal-viscosity_wagner_jfm1994,diff-dilute-poly_batchelor_jfm1983,sed-general_batchelor_jfm1982,sed-polydisperse-numerical_batchelor_jfm1982,tracer-diffusivity-bimodal_nagele_jcp2002}, or imposed simplifications on HIs~\cite{sd-bimodal-mc_powell_pof1994, shear-thickening_morris_prl2013}.

To the best of our knowledge, the present work is the first study for polydisperse suspensions with full HIs covering the entire concentration range up to close packing.
Specifically, the following species and mixture properties will be addressed:  (1) the short-time translational self-diffusivity, (2) the short-time rotational self-diffusivity, (3) the instantaneous sedimentation velocity, (4) the hydrodynamic functions, (5) the high-frequency dynamic shear viscosity, and (6) the high-frequency dynamic bulk viscosity.

From a hydrodynamic perspective, flows in porous media are closely related to those in colloidal suspensions.  
In both cases, the fluid motions are governed by the Stokes equation, and, for a given particle configuration, the distinction is that in suspensions the particles are free to move, while in porous media the particles are fixed in space.
Compared to suspensions, the immobile particles give rise to much stronger HIs and qualitatively different behaviors in their transport properties.
Here, we present the following transport properties of bidisperse porous media: (1) the translational drag coefficient, which is related to the permeability, (2) the translational hindered diffusivity, and (3) the rotational hindered diffusivity.

We choose the Stokesian Dynamics (SD)~\cite{sd_durlofsky_jfm1987,brady-sd-ew_jfm_88,stokesian-dynamics_brady_anfm1988} as the computational tool due to the simplicity and effectiveness of its formalism in treating the hydrodynamic interactions.
For monodisperse systems, SD has been used to study the short-time transport properties of hard-sphere suspensions~\cite{sd-ew-transport-coeff-pt1_phillips_pof1988} and porous media~\cite{sd-ew-transport-coeff-pt2_phillips_pof1988}, and its Particle Mesh Ewald (PME) variation, known as Accelerated Stokesian Dynamics (ASD), has been used to study the transport properties of charged colloidal suspensions~\cite{trans-prop-asd_banchio_jcp2008,yukawa-short-time-transport_heinen_jcp2011}.
For polydisperse systems, only partial extensions of SD exist.
Chang \& Powell~\cite{sd-bimodal-monolayer_powell_jfm1993,sd-bimodal-mc_powell_pof1994,sd-bimodal-diffusivity-shearing_powell_jfm1994} extended SD to polydisperse systems without the far-field mobility Ewald summation.
Consequently, their extension is only appropriate for monolayers.
Ando \& Skolnick~\cite{crowding-hydro-cell_skolnick_pnas2010} developed a force-torque level polydisperse SD to investigate the effect of molecular crowding on protein diffusion.
Since stresslet order moments were ignored, their implementation is unsuitable for rheological studies.
In this work we have carefully developed a new simulation program that fully extends SD to polydisperse systems.

The simplicity of the SD framework unfortunately comes at a cost of accuracy for certain transport properties.  However, the error associated with SD cannot be estimated \emph{a priori} and has to be understood by comparing with existing results from other computational techniques.
This leads to the second objective of this work: a careful assessment of the accuracy and effectiveness of SD.

Computing hydrodynamic interactions using conventional SD requires $\bigO(N^3)$ operations, where $N$ is the number of particles in the system.
This makes SD computationally expensive and imposes severe restrictions on the system size accessible to dynamic simulations~\cite{sd-brownian-susp_brady_jfm2000}.
The time limiting step is the explicit inversion of the mobility and resistance tensors.
The scaling can be reduced to $\bigO(N^2)$ by taking advantage of iterative solvers \cite{iterative-book, sd-gpu_hofling_epjst_2012, mpi-sd_dorfler_comp-fluids_2014}, to $\bigO(N\log N)$ in ASD through PME techniques~\cite{asd_sierou_jfm01}, and further down to $\bigO(N)$ using fast multipole methods~\cite{sd-fmm_ichiki_jfm2002}.
However, for computing short-time transport properties in this work, the choice of the $\bigO(N^3)$ algorithm is deliberate.
Here, hydrodynamic computations are performed for independent configurations using a Monte-Carlo approach, and each $\bigO(N^3)$ matrix inversion straightforwardly yields \emph{all} the short-time transport properties associated with the configuration for both the suspension and the porous medium.
In addition, the conventional SD incorporates a mean-field quadrupole contribution in the mobility computation~\cite{brady-sd-ew_jfm_88}, improving its accuracy.

The transport properties of colloidal suspensions can also be approximated via (semi-) analytical expressions.
These approximations are often preferred over full hydrodynamic computations since they are easier to access.
There are two approaches to treat HIs:
One is akin to the diagrammatic methods in liquid state theories~\cite{simpleliquids_hansen}.
For example, the $\delta\gamma$-scheme developed by Beenakker \& Mazur \cite{self-diff-high-dens_beenakker_physa1983, hydro-func-approx_beenakker_physica1984, del-gam-viscosity_beenakker_physicaa_1984} incorporates many-body HIs by resumming an infinite subset of the hydrodynamic scattering series from all particles in the suspension.
In a companion paper~\cite{dg-sd-comp_wang_jcp_2014}, we introduced a semi-empirical extension of the original monodisperse $\delta\gamma$-scheme to approximate the partial hydrodynamic functions of polydisperse suspensions.
The other approach is similar to the virial expansions: Explicit computations of the two-body, three-body, \etc, HIs lead to polynomial expressions of transport properties in powers of concentration.
Its simplest form considers only the two-body HIs and is known as the Pairwise Additive (PA) approximation~\cite{dynamic-charge-susp_nagele_physrep_1996}.  It is asymptotically exact for dilute suspensions, and can conveniently incorporate size polydispersity since the two-body HIs can be computed to arbitrary precision.
At higher concentrations, the many-body HIs become important and the PA approximations break down.
Therefore, the third objective of this work is to assess the validity of the PA approximations for polydisperse suspensions by comparing to the SD simulations.

The remainder of the paper is arranged as follows: in Sec.~\ref{sec:bidisp-susp-poro} we define the bidisperse systems under study and their various transport properties.  Sec.~\ref{sec:polyd-stok-dynam} describes the polydisperse SD algorithm and the simulation procedure.  In Sec.~\ref{sec:pairw-addit-appr} we summarize the equations for the PA approximations, and in Sec~\ref{sec:summ-exist-results} we review the existing analytical results beyond the PA level.
We present and discuss the SD results for bidisperse suspensions and porous media in Sec.~\ref{sec:results-suspensions} and \ref{sec:results-porous-media}, respectively.
We conclude this paper with a few comments in Sec.~\ref{sec:conclusions}.

\section{Bidisperse suspensions and porous media}
\label{sec:bidisp-susp-poro}

\subsection{Static structures}
\label{sec:static-structures}

We consider an unbounded homogeneous isotropic mixture of hard-sphere particles of different radii.
For two particles with radii $a_\alpha$ and $a_\beta$, their interaction potential $u_{\alpha\beta}(r)$ can be written as
\begin{equation}
  u_{\alpha\beta} (r) =\left\{
\begin{array}{l l}
  0      & \text{if $r>a_\alpha+a_\beta$} \\
  \infty & \text{otherwise,}
\end{array}
  \right.
\end{equation}
where $r$ is the center-center distance between the two particles, and  $\alpha,\beta \in \{1, 2\}$ are the species indices for bidisperse systems.
We choose the following dimensionless parameters to describe the configuration:
\begin{align}
  \lambda & = a_2/a_1, \\
  \phi & = \phi_1 + \phi_2,\text{ and} \\
  y_1 & = \phi_1/\phi,
\end{align}
where $\lambda$ is the size ratio, $\phi$ is the total volume fraction, $\phi_\alpha = \tfrac{4}{3}\pi a_\alpha^3 n_\alpha$ is the species volume fraction, and $y_1$ is the volume composition of species $1$.
The species number density $n_\alpha = N_\alpha/V$ with $N_\alpha$ the number of $\alpha$ particles in the system and $V$ the system size.
The total number of particles in the system is $N = N_1 + N_2$, and the total number density is $n = n_1 + n_2$.  The thermodynamic limit corresponds to increasing both $N$ and $V$ to infinity while keeping their ratio constant.  Obviously the volume composition $0 \leq y_1\leq 1$.  For convenience, and without loss of generality, we assume $a_1<a_2$ and thus $\lambda \geq 1$.

The structure of bidisperse systems can be characterized by the partial static strcture factors 
\begin{equation}
  \label{eq:partial-sk}
  S_{\alpha\beta}(q) = \avg{n_{-\bq}^\alpha n_{\bq}^\beta},
\end{equation}
where $q$ is the orientation-averaged wavenumber, and $\avg{\cdot}$ is the average operator in the thermodynamic limit over all orientations.  The species density fluctuation $n_{\bq}^\alpha$ is defined as
\begin{equation}
  \label{eq:species-nq}
   n_{\bq}^\alpha = \frac{1}{\sqrt{N_\alpha}}\sum_{j\in\alpha} e^{-\imath \bq\cdot\br_j},
\end{equation}
with $\imath = \sqrt{-1}$, $\br_j$ the position of particle $j$, and $j\in\alpha$ means summing over all particle $j$ in species $\alpha$.
One way to capture the overall structure of the mixture is the number-number static structure factor 
\begin{equation}
  \label{eq:snn}
  S_{NN} (q) = \sum_{\alpha,\beta} \sqrt{x_\alpha x_\beta} S_{\alpha\beta} (q),
\end{equation}
with $x_\alpha = N_\alpha/N$ the species molar fraction.  However, measurements from scattering experiments are often different from $S_{NN}(q)$, and correspond to a weighted average of $S_{\alpha\beta}(q)$,
\begin{equation}
\label{eq:sk-mixture}
  S_M(q) = \frac{1}{\overline{f^2}(q)}\sum_{\alpha,\beta} \sqrt{x_\alpha x_\beta} f_\alpha (q)f_\beta (q)   S_{\alpha\beta} (q),
\end{equation}
where $f_\alpha(q)$ is the species scattering amplitude, and $\overline{f^2}(q) = \sum_{\alpha} x_{\alpha}  f^2_{\alpha}(q)$ is the square mean scattering amplitude~\cite{hydro-effect-short-time_nagele_pre1993}.
Unless different species have the same constant scattering amplitude $f_\alpha = f_\beta = 1$, we generally have $S_{NN}(q) \neq S_M(q)$, making the interpretation of experiments with polydisperse systems difficult.

The real space characterization of the homogeneous and isotropic mixture structure is described by the partial radial distribution functions $g_{\alpha\beta} (r)$.  It is the probability of finding a particle in species $\beta$ with distance $r$ for a given particle in species $\alpha$.
Accordingly, we have~\cite{dynamic-charge-susp_nagele_physrep_1996}
\begin{equation}
  \label{eq:gab-def}
  g_{\alpha\beta}(r) = \frac{1}{n_\alpha n_\beta}\bigg\langle \sideset{}{'}\sum_{i\in\alpha \atop j\in \beta} \frac{1}{V}\delta(\br - \br_i + \br_j) \bigg\rangle,
\end{equation}
where $\delta(x)$ is the Dirac delta function, and the prime on the summation sign excludes the case of $i=j$.
The radial distribution function is related to the inverse Fourier transform of $S_{\alpha\beta}$ as~\cite{simpleliquids_hansen}
\begin{equation}
 g_{\alpha\beta} (r) = 1+ \frac{1}{2\pi^2 r \sqrt{n_\alpha n_\beta} }\int_0^\infty [S_{\alpha\beta}(q) -\delta_{\alpha\beta}] q \sin(qr) \dd q,
\end{equation}
where  $\delta_{\alpha\beta}$ is the Kronecker delta.  
Accordingly, the mixture total radial distribution function is 
\begin{equation}
  \label{eq:gr}
 g(r) = \sum_{\alpha,\beta} x_\alpha x_\beta g_{\alpha\beta} (r).
\end{equation}

\subsection{The short-time hydrodynamics}
\label{sec:short-time-hydr}

Colloidal suspensions and porous media exhibit different behaviors depending on the time scale~\cite{colloidal_dispersions_1989}, and in this work we are interested in the short-time properties.
For a Newtonian solvent with shear viscosity $\eta_0$ and density $\rho_0$, by ``short-time'' we mean a coarse grained time scale $t$ satisfying
\begin{equation}
  \tau_H \sim \tau_I \ll t \ll \tau_D,
\end{equation}
where $\tau_H$ is the hydrodynamic time, $\tau_I$ is the inertia time, and $\tau_D$ is the diffusion time.

The hydrodynamic time $\tau_H = \rho_0 a_2^2 / \eta_0$ characterizes the time required for the fluid momentum to diffuse a length scale of the (larger) particle.  
With $\tau_H \ll t$, the Reynolds number $\mathrm{Re} = \tau_H/t \ll 1$, and therefore the HIs are dominated by the viscous stresses.
Consequently, the fluid motion is governed by the Stokes equation and the incompressibility constraint,
\begin{equation}
\label{eq:stokes}
  \grad p(\bx)  = \eta_0 \lapl \bv(\bx)\; \text{ and } \dive{\bv(\bx)} = 0,
\end{equation}
where $p(\bx)$ and $\bv(\bx)$ are the fluid pressure and velocity field, respectively.  We further supplement the above equations with the no-slip boundary conditions on the particle surfaces.

The particle inertia time, $\tau_I = \tfrac{2}{9}\rho_2 a_2^2/\eta_0 $, where $\rho_2$ is the density of the (larger) particle, describes the time required for the \emph{particle} momentum to dissipate by interacting with the solvent.
The consequence of $\tau_I\ll t$ is that the particle momentum almost dissipates instantaneously and the particle dynamics are completely overdamped in the time scale we are interested in.
As a result, the HIs in the suspension are solely determined by the instantaneous particle configurations $\br^N = \{\br_1, \br_2, \dots, \br_N\}$.
This allows the use of Monte-Carlo type approaches to study the short-time transport properties, as each independent configuration is equivalent.

The diffusion time $\tau_D = 6\pi \eta_0 a_1^3 / \kT$, where $\kT$ is the thermal energy, sets the upper limit of the short-time regime. It characterizes the time for a smaller particle to move a distance of its own size when driven by thermal fluctuations, \ie, $\tau_D = a_1^2/d_{0,1}^t$, where $d_{0,1}^t=\kT/(6\pi\eta_0 a_1)$ is the Stokes-Einstein-Sutherland (SES) translational diffusivity of a single particle with radius $a_1$.
In dense suspensions, $\tau_D$ also characterizes the time required for a particle to move a distance of the interparticle gap spacing  $\xi a_1 \ll a_1$.
This is because for nearly touching particles, the relative mobility scales as $\xi$, and the characteristic velocity from thermal fluctuations is $\sim d_{0,1}^t(\xi/a_1)$.
Therefore, $\tau_D$ is a valid diffusion time scale at any suspension volume fraction.
At the time scale $t\sim \tau_D$, the (smaller) particles wander far from their original positions and interact directly with the neighboring particles.  Such interactions significantly affect the suspension dynamics.
For the moderate size ratios considered in this work, the time scale $\tau_D$ is always several orders of magnitude larger than $\tau_I$ and $\tau_H$, leaving a well-defined short-time regime.

The Stokes equation in Eq.~(\ref{eq:stokes}) governs the HIs in the suspension.  Its linearity gives rise to the linear dependence between the forces $\bF$, torques $\boldsymbol{T}$, and stresslets $\bS$ and the linear and angular velocities $\bU$ and $\boldsymbol{\Omega}$, respectively.
For all particles in the suspension, we have~\cite{Kim2005}
\begin{equation}
\label{eq:resis-form}
\begin{pmatrix}
\boldsymbol{\cal F} \\ 
\bS \end{pmatrix} = 
- \boldsymbol{\cal R}\cdot
\begin{pmatrix} \boldsymbol{\cal U} - \boldsymbol{\cal U}^\infty \\ 
-\be^\infty
\end{pmatrix},  
\end{equation}
where $\boldsymbol{\cal R}$ is the grand resistance tensor, $\boldsymbol{\cal F}=\{ \bF, \boldsymbol{T}  \}$ is the generalized force, $\boldsymbol{\cal U} - \boldsymbol{\cal U}^\infty =\{ \bU- \bu^\infty, \boldsymbol{\Omega} - \boldsymbol{\omega}^\infty \}$ is the generalized velocity disturbance, and 
$\bu^\infty$, $\boldsymbol{\omega}^\infty$, and $\be^\infty$ are the imposed linear velocity, angular velocity, and strain rate, respectively.
The unsubscripted symbols suggest all particles are involved, \eg, $\bF = \{ \bF_1, \bF_2, \dots, \bF_N\}$.
Each element of the grand resistance tensor depends on the configuration of the entire system, \ie, $\boldsymbol{\cal R} = \boldsymbol{\cal R}(\br^N)$, and the minimum dissipation theorem of the Stokes flow requires $\boldsymbol{\cal R}$ to be symmetric and positive definite~\cite{Kim2005}.
We can partition the grand resistance tensor $\boldsymbol{\cal R}$ as
\begin{equation}
\label{eq:grand-res-tensor}
\boldsymbol{\cal R} (\br^N) = 
\begin{pmatrix}\bR_{FU} & \bR_{FE} \\
\bR_{SU} &  \bR_{SE} \end{pmatrix},  
\end{equation}
where, for example, $\bR_{FU}$ describes the coupling between the generalized force and the generalized velocity.
The resistance tensor $\bR_{FU}$ and its inverse $\bR_{FU}^{-1}$ play a particularly important role in the short-time transport properties of suspensions and porous media, and can be further partitioned as
\begin{align}
   \bR_{FU} & = 
\begin{pmatrix}
    \bze^{tt} & \bze^{tr} \\
    \bze^{rt} & \bze^{rr} 
\end{pmatrix}, \label{eq:res-tensor}\\
   \bR_{FU}^{-1} & = 
\begin{pmatrix}
    \bmu^{tt} & \bmu^{tr} \\
    \bmu^{rt} & \bmu^{rr} 
\end{pmatrix},\label{eq:mob-tensor}
\end{align}
where each sub-matrix contains coupling between the translational ($t$) and rotational ($r$) velocities and forces.

\subsection{Suspension transport properties}
\label{sec:transp-prop-susp}

The dynamic structural evolution of a colloidal mixture can be described by the dynamic partial structure factors
\begin{equation}
\label{eq:sab-qt}
  S_{\alpha\beta}(q,t) = \avg{n_{-\bq}^\alpha(0) n_{\bq}^\beta(t)},
\end{equation}
where $n_{\bq}^\beta(t)$ is the density fluctuations measured at time $t$ from Eq.~(\ref{eq:partial-sk}), and thus $S_{\alpha\beta}(q) = S_{\alpha\beta}(q,0)$.
The dynamics of $S_{\alpha\beta}(q,t)$ are governed by the Smoluchowski equation~\cite{Dhont1996,dynamic-charge-susp_nagele_physrep_1996}, and one can show that in short times
\begin{equation}
 \lim_{t\rightarrow 0} \bS(q,t) = \bS(q) \exp[-t q^2 \boldsymbol{D}(q)],
\end{equation}
where $\bS(q,t)$ for bidisperse suspensions is a $2\times 2$ matrix with elements of $S_{\alpha\beta}(q,t)$, and $\boldsymbol{D}(q)$ is the $q$-dependent diffusivity matrix depending on the suspension structure and HIs.
The hydrodynamic contribution to diffusivity matrix $\boldsymbol{D}(q)$ is extracted as 
\begin{equation}
\boldsymbol{H}(q) = \boldsymbol{D}(q) \cdot \bS(q) /(\kT),
\end{equation}
and $\boldsymbol{H}(q)$ is known as the hydrodynamic matrix with elements $H_{\alpha\beta}(q)$, the partial hydrodynamic functions.
The microscopic definition of $H_{\alpha\beta}(q)$ is
\begin{equation}
  \label{eq:hq-partial}
  H_{\alpha\beta}(q) = \frac{1}{\sqrt{N_{\alpha} N_{\beta}}}  \bigg\langle\sum_{i\in\alpha \atop j\in\beta} \hat{\bq} \cdot \bmu^{tt}_{ij} (\br^N) \cdot \hat{\bq} e^{\imath \bq\cdot(\br_i-\br_j)}\bigg\rangle,
\end{equation}
where $\hat{\bq} = \bq/|\bq|$ is the unit vector of $\bq$. The mobility tensors $\bmu_{ij}^{tt}$ are elements of the tensor $\bmu^{tt}$ in Eq.~(\ref{eq:mob-tensor}), and describe the coupling between the velocity disturbance of the particle $i$ due to an imposed force on the particle $j$.  

It is convenient to split $H_{\alpha\beta}(q)$ as 
\begin{equation}
H_{\alpha\beta}(q) = \delta_{\alpha\beta} d_{s,\alpha}^t/(\kT) + H^d_{\alpha\beta}(q),
\end{equation}
where $H^d_{\alpha\beta}(q)$ is the $q$-dependent distinct part of the partial hydrodynamic function, and $d_{s,\alpha}^t$ is the short-time translational self-diffusivity of species $\alpha$.  Note that we use the lowercase symbol to signify its $q$-independence.
The microscopic definition of $d_{s,\alpha}^t$ is
\begin{equation}\label{eq:dt-def}
 d_{s,\alpha}^t = \frac{\kT}{ N_\alpha} \bigg\langle\sum_{i\in\alpha}\hat{\bq}\cdot\bmu^{tt}_{ii}\cdot\hat{\bq}\bigg\rangle,
\end{equation}
and it describes the short-time mean-square displacement of species $\alpha$ in a Brownian suspension
\begin{equation}
  d_{s,\alpha}^t = \lim_{t\rightarrow 0}\frac{\dd}{\dd t}\avg{\tfrac{1}{6}[\br_i(t) - \br_i(0)]^2}, i\in\alpha.
\end{equation}
Comparing Eq.~(\ref{eq:hq-partial}) and (\ref{eq:dt-def}), we see that in the short wave-length limit,
\begin{equation}
  d_{s,\alpha}^t = \lim_{q\rightarrow\infty} \kT H_{\alpha\alpha}(q).
\end{equation}

The suspension hydrodynamic functions can be obtained by dynamic scattering experiments, but $H_{\alpha\beta}(q)$ is often difficult to access directly unless special techniques such as selective index of refraction matching are employed~\cite{Williams2001}.
Otherwise, the measured hydrodynamic function $H_{M}(q)$ is related to $H_{\alpha\beta}(q)$ as~\cite{hydro-effect-short-time_nagele_pre1993}
\begin{equation}
\label{eq:hm-def}
  H_M(q) = \frac{1}{\overline{f^2}(q)}\sum_{\alpha,\beta} \sqrt{x_\alpha x_\beta} f_\alpha (q)f_\beta (q)   H_{\alpha\beta} (q),
\end{equation}
where $f_\alpha(q)$, $f_\beta(q)$ and $\overline{f^2}(q)$ are defined in Eq.~(\ref{eq:sk-mixture}).
In the hypothetical case of $f_\alpha = f_\beta = 1$, the number-number mixture hydrodynamic function is
\begin{equation}
\label{eq:hnn-def}
H_{NN}(q) =\sum_{\alpha,\beta} \sqrt{x_\alpha x_\beta} H_{\alpha\beta} (q).
\end{equation}


The rotational Brownian motion of colloidal suspensions can be observed by introducing optical anisotropy to the otherwise spherical particles using depolarized dynamic light scattering techniques~\cite{tracer-diffusivity-bimodal_nagele_jcp2002,validity-se-rot-diff_koenderink_faradiss2003}.
The optical anisotropy is characterized by the orientation unit vector $\hat{\bn}_i(t)$ for particle $i$ at time $t$.
The short-time decay of the rotational correlation function of particles of species $\alpha$, 
\begin{equation}
  S_r^\alpha(t) =\avg{P_2[\hat{\bn}_i(t) \cdot  \hat{\bn}_i(0) ]}, i\in \alpha
\end{equation}
where $P_2(x)$ is the Legendre polynomial of the second order, defines the short-time rotational self-diffusivity
\begin{equation}
  d_{s,\alpha}^r = -\tfrac{1}{6}\lim_{t\rightarrow 0} \frac{\dd}{\dd t} \ln[S^\alpha_r(t)].
\end{equation}
Microscopically, $d_{s,\alpha}^r$ is defined as
\begin{equation}
d_{s,\alpha}^r = \frac{\kT}{ N_\alpha} \bigg\langle{\sum_{i\in\alpha}\hat{\bq}\cdot\bmu^{rr}_{ii}\cdot\hat{\bq}}\bigg\rangle,
\end{equation}
where $\bmu_{ij}^{rr}$ are elements of $\bmu^{rr}$ in Eq.~(\ref{eq:mob-tensor}), and describe the angular velocity disturbance on particle $i$ due to an imposed torque on particle $j$.


Sedimentation occurs when the particle density $\rho_\alpha$ is different from the solvent density $\rho_0$.
The net body force exerted on species $\alpha$ depends on the species radius $a_\alpha$ and the density difference $\Delta \rho_\alpha= \rho_\alpha - \rho_0$.
In bidisperse suspensions, the instantaneous sedimentation velocities depend on the size ratio $\lambda$ and the density ratio~\cite{sed-general_batchelor_jfm1982,bidisperse-sed_batchelor_jfm1986}
\begin{equation}
  \gamma = \Delta \rho_2 / \Delta \rho_1.
\end{equation}
The ratio of the mean forces between the two species is $F_2/F_1 = \lambda^3\gamma$.
Examination of Eq.~(\ref{eq:hq-partial}) reveals that the species instantaneous sedimentation velocities, $U_{s,1}$ and $U_{s,2}$, can be expressed in terms of $H_{\alpha\beta}(0)=\lim_{q \rightarrow 0}H_{\alpha\beta}(q)$ as
\begin{align}
\frac{ U_{s,1}}{U_{0,1}} = & \frac{1}{\mu_{0,1}}\bigg[H_{11}(0) + \lambda^3\gamma\sqrt{\frac{x_2}{x_1}}  H_{12}(0)\bigg],\label{eq:us-1}\\
  \frac{U_{s,2}}{U_{0,2}} = & \frac{1}{\mu_{0,2}}\bigg[\frac{1}{\lambda^3\gamma}\sqrt{\frac{x_1}{x_2}}H_{21}(0) +   H_{22}(0)\bigg],\label{eq:us-2}
\end{align}
where, for species $\alpha$, $\mu_{0,\alpha} = (6\pi\eta_0 a_\alpha)^{-1}$ is the single particle mobility and $U_{0,\alpha} = \mu_{0,\alpha} F_\alpha$ is the single particle sedimentation velocity.  For simplicity, we only consider the case $\gamma = 1$.

A distinguishing feature of sedimentation in polydisperse suspensions is that $U_{s,\alpha}$ can be negative.
The motion of one species can give rise to a strong back flow that reverses the sedimentation velocity of the second species, \ie, the particles move in a direction opposite to the imposed body force, especially when the body force is weak~\cite{sed-general_batchelor_jfm1982}.  
For monodisperse suspensions, on the other hand,  the positive definiteness of the mobility tensor $\bmu$ requires the sedimentation velocity to be positive.

Eq. (\ref{eq:us-1}) and (\ref{eq:us-2}) also reveal the close connection between $U_{s,\alpha}$ and $H_{\alpha\beta} (q)$.
At different wavenumber $q$, $H_{\alpha\beta}(q)$ probes the suspension HIs at different length scales: single particle behaviors as $q\rightarrow \infty$, and collective dynamics as $q\rightarrow 0$.
The wavenumber corresponding to the maximum of $H_{\alpha\beta}(q)$ is closely related to the size of the structures that dominate the suspension short-time dynamics~\cite{trans-prop-asd_banchio_jcp2008}.

The suspension rheological properties are obtained from the volume average of the Cauchy stress~\cite{Brady1993,bulk-viscosity_brady_jfm2006},
\begin{equation}
  \label{eq:bulk-stress-st}
\avg{\bsig} =  -\langle p\rangle_f \bI + 2\eta_0 \avg{\be^\infty} + (\kappa_0 - \tfrac{2}{3}\eta_0)\avg{\dive{\bu^\infty}}\bI + n \langle\bS^H \rangle,  
\end{equation}
where $p$ is the solvent pressure, $\langle \cdot \rangle_f$ is the fluid phase averaging operator, $\bI$ is the idem tensor, $\kappa_0$ is the solvent bulk viscosity, and $\langle \bS^H \rangle$ is the stresslet due to the presence of particles.  
In the short-time limit and without the interparticle forces,
\begin{equation}
\label{eq:part-se-def}
  \langle\bS^H \rangle = - \langle \bR_{SU}\cdot\bR_{FU}^{-1}\cdot \bR_{FE} - \bR_{SE} \rangle : \avg{\be^\infty}.
\end{equation}
Eq.~(\ref{eq:bulk-stress-st}) ignores the stress contributions from the Brownian motion, and therefore is only valid in the short-time limit.
To measure the transport properties associated with $\avg{\bsig}$ defined in Eq.~(\ref{eq:bulk-stress-st}) and (\ref{eq:part-se-def}), the rheological experiments have to be performed with high-frequency, low-amplitude deformations, such that the suspension microstructures are only slightly perturbed from the equilibrium hard-sphere structures, and the Brownian stress contribution is out of phase with the applied oscillating deformation~\cite{sd-ew-transport-coeff-pt1_phillips_pof1988}.
In a high-frequency shear experiment with an imposed strain rate amplitude $\dot{\gamma}$, the suspension high-frequency dynamic shear viscosity is
\begin{equation}
  \eta_s = \eta_0 + n\langle S^H \rangle_{12}/\dot{\gamma},
\end{equation}
where the subscript $12$ denotes the velocity-velocity gradient component of the stresslet.
In a high-frequency expansion experiment with an imposed expansion rate amplitude $\dot{e}$, the high-frequency dynamic bulk viscosity is
\begin{equation}
\kappa_s = \kappa_0 + \tfrac{1}{3}n\langle \bS^H \rangle:\bI/\dot{e}.  
\end{equation}
Note that for solvent with a finite bulk viscosity $\kappa_0$, the incompressibility condition of the Stokes equation is violated.
However, as is shown in Ref.~\onlinecite{bulk-viscosity_brady_jfm2006}, the fluid velocity disturbance remains incompressible and satisfies the Stokes equation.
The rigid colloidal particles, unable to expand with the fluid, therefore contribute to the suspension bulk viscosity.

\subsection{Porous medium transport properties}
\label{sec:transp-prop-poro}

When a fluid passes through a porous medium, the fixed particles resist the flow, creating a pressure drop across the material.
The resistance behavior is often characterized by the dimensionless drag coefficient $F_{\alpha}$~\cite{bidisperse-drag-permeability_kuipers_jfm2005}, defined through
\begin{equation}
6\pi\eta_0 a_\alpha F_\alpha \bV  = \bF_{d,\alpha},
\end{equation}
where $\bV$ is the superficial fluid velocity and $\bF_{d,\alpha}$ is the mean drag force for particle species $\alpha$ including the back pressure gradient contribution from the fluid.
A force balance considering both the fluid and the particles shows that the average force for each particle is 
\begin{equation}
  \bF_{d,\alpha} = \frac{1-\phi}{N_\alpha} \bigg\langle\sum_{i\in\alpha}\sum_{j=1}^N \bze^{tt}_{ij} \bigg\rangle \cdot \bV,
\end{equation}
where $\bze^{tt}_{ij}$ are from the resistance tensor $\bze^{tt}$ in Eq.~(\ref{eq:res-tensor}), and describe the coupling between the force on particle $i$ due to a velocity on particle $j$.

For a porous medium containing particles of different sizes, the average drag coefficient is defined as
\begin{equation}
\label{eq:avgf-def}
  \avg{F} = \sum_\alpha \frac{y_\alpha}{z_\alpha^2} F_\alpha,
\end{equation}
where the diameter fraction $z_\alpha = a_\alpha / \avg{a}$ and $\avg{a} = (\sum_\alpha y_\alpha/a_\alpha)^{-1}$.
As is shown in Ref.~\onlinecite{bidisperse-drag-permeability_kuipers_jfm2005}, Eq.~(\ref{eq:avgf-def}) allows convenient extension of the Darcy's equation to polydisperse systems.
The porous medium permeability $\avg{K}$ is closely related to $\avg{F}$ in Eq.~(\ref{eq:avgf-def}) as
\begin{equation}
  \avg{K} = \frac{\avg{F}}{1-\phi}.
\end{equation}

The diffusive behaviors of particles in porous media are characterized by the translational and rotational hindered diffusivities, denoted as $\Dhdta$ and $\Dhdra$, respectively.  They describe the short-time Brownian motions of a \emph{single} mobile particle in a matrix of fixed particles.  In terms of the resistance tensors, we have~\cite{sd-ew-transport-coeff-pt2_phillips_pof1988}
\begin{align}
  \Dhdta = & \frac{\kT}{N_\alpha} \bigg\langle\sum_{i\in\alpha} \hat{\bq}\cdot (\bze^{tt}_{ii})^{-1} \cdot\hat{\bq}\bigg\rangle,\\
\Dhdra  = &\frac{\kT}{ N_\alpha} \bigg\langle\sum_{i\in\alpha} \hat{\bq}\cdot (\bze^{rr}_{ii})^{-1}\cdot\hat{\bq} \bigg\rangle,
\end{align}
where $\bze^{rr}_{ij}$ are elements of $\bze^{rr}$ in Eq.~(\ref{eq:res-tensor}), and describe the torque-angular velocity coupling between particles $i$ and $j$.

\section{The polydisperse Stokesian Dynamics}
\label{sec:polyd-stok-dynam}

The framework of Stokesian Dynamics (SD) has been extensively discussed elsewhere~\cite{sd_durlofsky_jfm1987,brady-sd-ew_jfm_88,stokesian-dynamics_brady_anfm1988,asd-brownian_banchio_jcp2003,swaroopthesis2010} and here we only present the aspects pertinent to the extension to polydisperse systems.
The grand resistance tensor $\boldsymbol{\cal R}$ in Eq. (\ref{eq:grand-res-tensor}) is computed in SD as 
\begin{equation}
\boldsymbol{\cal R} = (\boldsymbol{\cal M}^{\infty})^{-1} + \boldsymbol{\cal R}_{2B} - \boldsymbol{\cal R}_{2B}^\infty,
\end{equation}
where the far-field mobility tensor $\boldsymbol{\cal M}^{\infty}$ is constructed pairwisely from the multipole expansions and Fax\'{e}n's laws of Stokes equation up to the stresslet level, and its inversion captures the long-range many-body HIs.
The near-field lubrication correction $(\boldsymbol{\cal R}_{2B} -\boldsymbol{\cal R}_{2B}^\infty)$ is based on the exact two-body solutions with the far-field contributions removed, and it accounts for the singular HIs when particles are in close contact.
The SD recovers the exact solutions of two-particle problems and was shown to agree well with the exact solution of three-particle problems~\cite{three-spheres_wilson_jcp2013}.

Extending SD to polydisperse systems retains the computational framework above.
The far-field polydisperse mobility tensor $\boldsymbol{\cal M}^{\infty}$ is computed using multipole expansions as Ref.~\onlinecite{sd-bimodal-monolayer_powell_jfm1993} and the results are extended to infinite periodic systems using Beenakker's technique~\cite{ewald-sum-rotne-prager_beenaker_jcp86,ewald-poly_hase_pof01}.
The lubrication corrections $(\boldsymbol{\cal R}_{2B}-\boldsymbol{\cal   R}_{2B}^\infty)$ for a particle pair with radii $a_\alpha$ and $a_\beta$ are based on the exact solutions of two-body problems in series form~\cite{resist-func_jeffrey_jfm1984, resist-func_jeffrey_pof1992, pres-moment_jeffrey_pof1993, compres-res_khair_pof2006} up to $s^{-300}$, where $s = 2r/(a_\alpha+a_\beta)$ is the scaled center-center particle distance.
In the simulations, the lubrication corrections are invoked when $r<2(a_\alpha + a_\beta)$, and the analytic lubrication expressions are used when $r<1.05(a_\alpha + a_\beta)$.
To avoid singularities in the grand resistance tensor due to particle contact, we enforced a minimum separation of $10^{-6}(a_i+a_j)$ between particles.
Note that in Ref.~\onlinecite{compres-res_khair_pof2006} there is an extra $(n+1)$ that should be removed in the denominator of the fraction in front of $P_{s(q-s)(p-n-1)}$ for the expression of $P_{npq}$.

Our polydisperse SD program treats the solvent as a compressible fluid and computes the fluid velocity disturbance due to the presence of rigid particles.  As a result, the trace of the particle stresslet is no longer zero and has to be computed. 
The solvent compressibility allows the quantities related the pressure moment to be directly incorporated to the grand resistance tensor $\boldsymbol{\cal R}$, augmenting its size from $11N\times 11N$ to $12N \times 12 N$.
This is more convenient compared to the earlier approaches, where the the pressure related quantities are treated as a separate problem and sometimes require iterations~\cite{pressure-sd_morris_jor2008,rheo-non-colloidal-suspension_brady_jor02,pres-moment_jeffrey_pof1993}.

A subtlety in incorporating the fluid compressibility is that in the mobility problem, a compressible flow disturbance can only be generated by the trace of the stresslet.
As a result, the pairwisely constructed far-field grand mobility tensor $\boldsymbol{\cal M}^\infty$ is not symmetric.
This asymmetry is necessary to eliminate the spurious hydrodynamic reflections upon its inversion and ensure the elements of $(\boldsymbol{\cal M}^\infty)^{-1}$ corresponding to the incompressible problem remain the same as the original SD.
However, the symmetry of $\boldsymbol{\cal R}$ is restored by copying the missing components in $\bR_{SU}$ and the lower triangular part of $\bR_{SE}$ from the transpose of $\bR_{FE}$ and the upper triangular part of of $\bR_{SE}$, respectively~\cite{swaroopthesis2010}.

Our simulations proceed as follows.  First, a random bidisperse hard-sphere packing at the desired composition is generated using the event-driven Lubachevsky-Stillinger algorithm~\cite{Lubachevsky1990,packing-gen-code_torquato_pre2006} with high compression rate. 
After the desired volume fraction $\phi$ is reached, the system is equilibrated for a short time (10 events per particle) without compression.  
This short equilibration stage is necessary as the compression pushes particles closer to each other, and prolonging this equilibration stage does not alter the resulting suspension structure significantly. 
After the grand resistance tensor $\boldsymbol{\cal R}$ is constructed based on the particle configuration $\br^N$, the short-time transport properties presented in section \ref{sec:bidisp-susp-poro} are extracted.

The simulations were performed for bidisperse systems of size ratio $\lambda = 2$ and $4$ as well as monodisperse systems.  To scan the parameter space, we first fix the mixture composition to $y_1 = 0.5$ and vary the total volume fraction $\phi$.  We then study the effects of $y_1$ with fixed $\phi$ at $\lambda = 2$.
Typically each configuration contains $800$ particles and at least $500$ independent configurations are studied for each composition. 
For systems with disparate size ratios, we ensure at least $10$ large particles are presented in the simulations.

\begin{figure}
  \begin{center}
  \subfloat[]{
    \centering
    \includegraphics[width=3in]{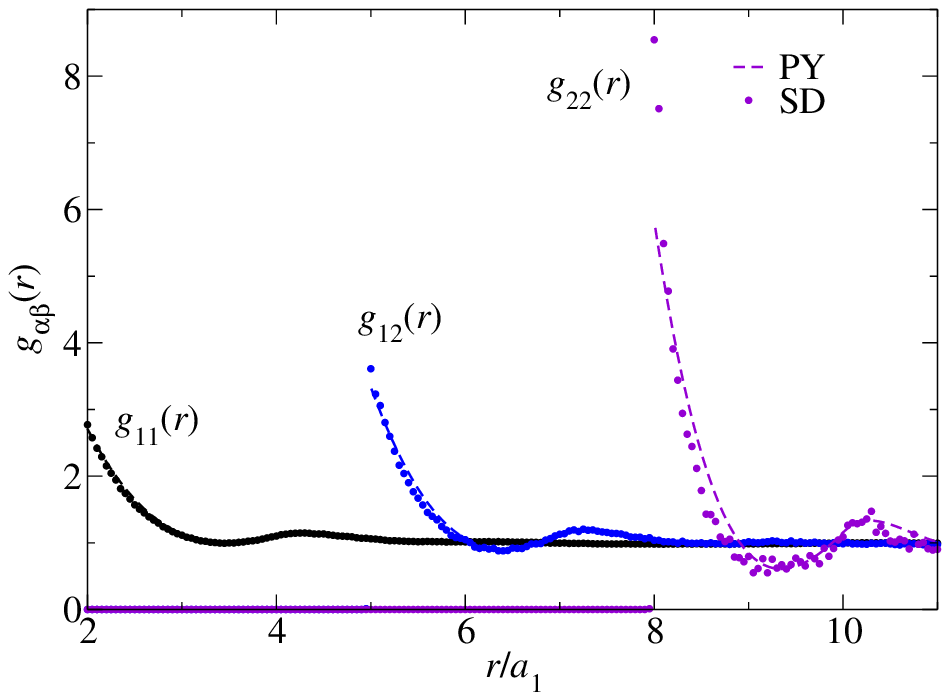}
    \label{fig:gr-py-sd}
  }\\
  \subfloat[]{
    \centering
    \includegraphics[width=3in]{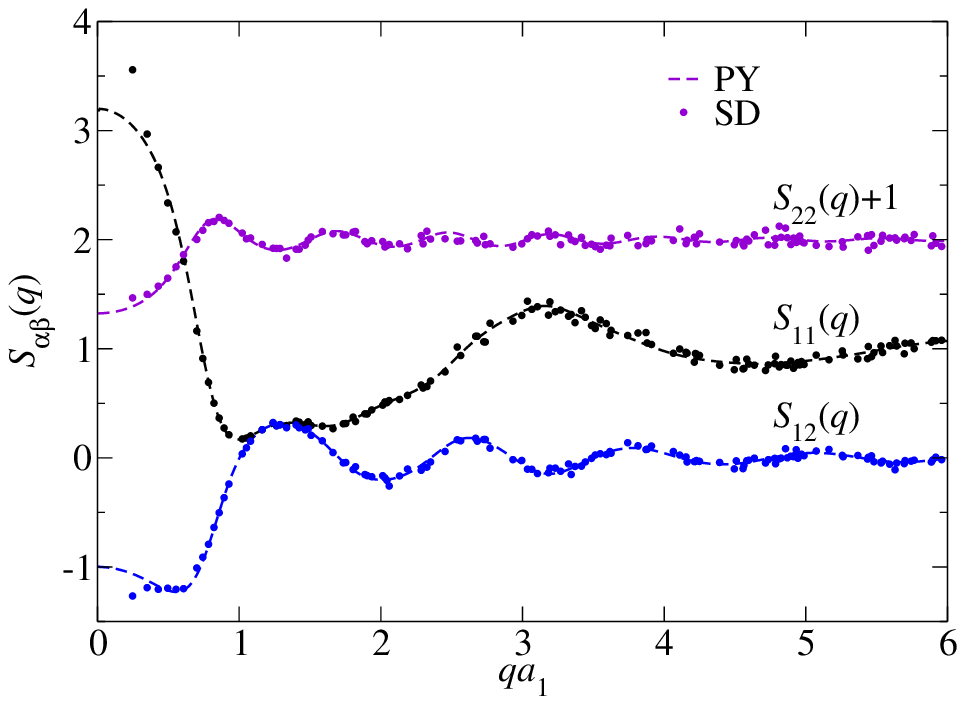}
    \label{fig:sk-py-sd}
  }
  \end{center}
  \caption{
(Color online) The structures of bidisperse suspensions with  $\lambda = 4$, $\phi = 0.4$, and $y_1 = 0.5$ directly measured from the SD simulations (dots) and computed from the Percus-Yevick (PY) integral equation (dashed lines): \protect\subref{fig:gr-py-sd} the partial radial distribution functions $g_{\alpha\beta}(r)$, and \protect\subref{fig:sk-py-sd} the partial static structure factors $S_{\alpha\beta}(q)$.
} \label{fig:gr-sk-py-sd}
\end{figure}

Fig.~\ref{fig:gr-sk-py-sd} shows the structural characterizations, $g_{\alpha\beta}(r)$ and $S_{\alpha\beta}(q)$, measured from the above simulation protocol for a bidisperse suspension of $\lambda = 4$, $y_1=0.5$, and $\phi = 0.4$.
The measurements from the SD simulations are compared with the Percus-Yevick (PY)\cite{Percus1958, py-hs-mixture_lebowitz_physrev1964} integral equation solutions. 
Note that at $y_1 = 0.5$, the mixture number composition is highly asymmetric, \ie, $x_1 = 0.985$.
For $g_{\alpha\beta}(r)$ in Fig.~\ref{fig:gr-py-sd}, the SD measurements can be accurately described by the PY solutions~\cite{code-gr-func_grundke_molphys2008, hs-mixture-distr-func_grundke_molphys1972} despite the small underestimation of the contact values for $g_{12}(r)$ and $g_{22}(r)$.
Although semi-empirical corrections~\cite{py-correction_verlet_pra1972, hs-mixture-distr-func_grundke_molphys1972} exist for this well-known symptom of the PY solutions~\cite{simpleliquids_hansen}, they are not applicable for dense mixtures with large size ratios.
In Fig.~\ref{fig:sk-py-sd}, $S_{\alpha\beta}(q)$ directly measured from SD agree well with the analytical PY solutions~\cite{sk-binary-mix_ashcroft_pr1967,  sk-binary-mix-errata_ashcroft_pr1968} except at small wavenumbers.
Note that the PY $\Sab(q)$ was shown to be valid for polydisperse mixtures at $\phi$ even beyond the monodisperse close packing~\cite{dense-colloids-scattering_mason_jpcm2009}. 
Fig.~\ref{fig:gr-sk-py-sd} validates the PY solution as a satisfactory description of the suspension structures in both the real and the wave spaces.

The transport properties extracted from $\bmu^{tt}$, \ie, $d^t_{s,\alpha}$, $U_{s,\alpha}$, and $H_{\alpha\beta}(q)$, exhibit a strong $\sqrt[3]{N}$ size dependence in the simulations due to the imposed periodic boundary conditions~\cite{hydro-trans-coeff_ladd_jcp1990,sd-ew-transport-coeff-pt1_phillips_pof1988,suspension-dws-simulation-correlation_ladd_pre1995,trans-prop-asd_banchio_jcp2008}
The finite size effect can be eliminated by considering $H_{\alpha\beta}(q)$ as a generalized sedimentation velocity with contributions from random suspensions and cubic lattices~\cite{sd-ew-transport-coeff-pt1_phillips_pof1988, suspension-dws-simulation-correlation_ladd_pre1995}.
For bidisperse suspensions, the finite size correction $\Delta_N H_{\alpha\beta}(q)$ for partial hydrodynamic functions from an $N$-particle system, $H_{\alpha\beta,N}(q)$, is 
\begin{equation}
  \label{eq:size-corr}
\Delta_N H_{\alpha\beta}(q) =  \frac{1.76 \mu_{0,1}
  S_{\alpha\beta}(q) }{(x_1+x_2 \lambda^3)^{\frac{1}{3}} }
\frac{\eta_0}{\eta_s} \left( \frac{\phi}{N} \right)^{\frac{1}{3}},
\end{equation}
so that in the thermodynamic limit the partial hydrodynamic function $H_{\alpha\beta}(q) = \Delta_N H_{\alpha\beta}(q)+H_{\alpha\beta,N}(q)$. In Eq.~(\ref{eq:size-corr}), $\eta_s/\eta_0$ is the suspension high-frequency dynamic shear viscosity obtained from the same simulations.
Note the scaling for $\Delta_N H_{\alpha\beta}(q)$ is $\mu_{0,1}$ regardless of the choice of $\alpha$ and $\beta$. 
The static structure factors $\Sab(q)$ are taken from the analytical PY solution~\cite{sk-binary-mix-errata_ashcroft_pr1968,sk-binary-mix_ashcroft_pr1967} in this work.
The correction for $d^t_{s,\alpha}$ and $U_{s,\alpha}$ corresponds to the large and small $q$ limit of Eq.~(\ref{eq:size-corr}), respectively.
We checked that other transport properties, including the shear viscosity $\eta_s/\eta_0$, change little with the system size. 

\begin{figure*}
  \begin{center}
    \includegraphics[width=7in]{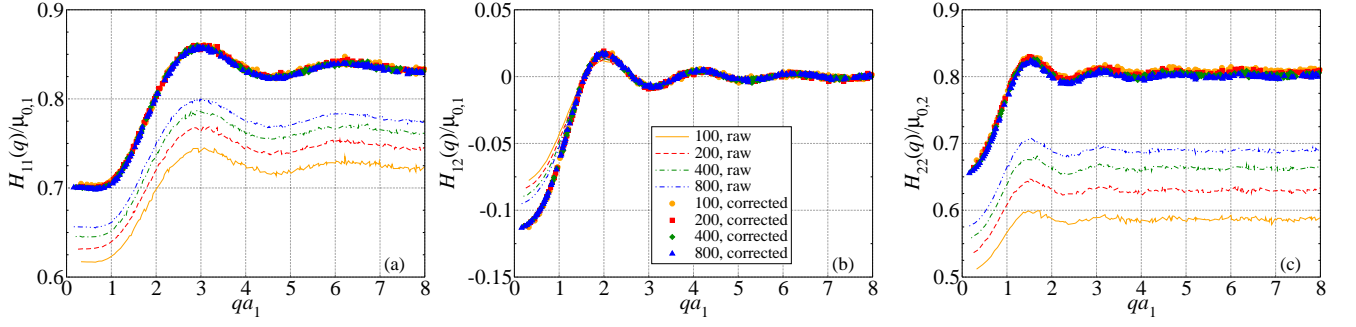}
  \end{center}
  \caption{(Color online) The raw and corrected partial hydrodynamic functions with different simulation system sizes, (a): $H_{11}(q)$, (b): $H_{12}(q)$, and (c): $H_{22}(q)$, for a bidisperse suspension at $\lambda = 2$, $\phi=0.1$, and $y_1=0.5$.}
  \label{fig:hq-finite-size}
\end{figure*}

The effectiveness of Eq.~(\ref{eq:size-corr}) is demonstrated in Fig.~\ref{fig:hq-finite-size} for all three partial hydrodynamic functions.  
Without the correction, simulations at different $N$ produce distinct $H_{\alpha\beta}(q)$ and the finite size effect is significant.
After applying Eq.~(\ref{eq:size-corr}), the data at different $N$ collapse for all $q$.
Note that the finite size collapse of $H_{22}(q)$ in Fig.~\ref{fig:hq-finite-size}c for small $N$ is slightly scattered due to the limited number of large particles, \eg, at $N=100$, there are only $11$ large particles in the mixture. 
The corrected results for $N=400$ and $800$ do agree with each other satisfactorily.
Eq.~(\ref{eq:size-corr}) spares us from extrapolating multiple simulations to eliminate the finite size effect, and we apply it for all the  presented results.

\section{The pairwise additive approximation}
\label{sec:pairw-addit-appr}

The pairwise additive (PA) approximation is convenient for estimating suspension transport properties at low volume fractions~\cite{dynamic-charge-susp_nagele_physrep_1996}.
It explicitly takes the mixture structures into account by incorporating the radial distribution functions (RDF) $g_{\alpha\beta}(r)$ into its formulation.
As is evident from Fig.~\ref{fig:gr-sk-py-sd}, the PY solution satisfactorily captures the suspension structures, and is therefore used in this work.

The PA approximations of the short-time translational and rotational self-diffusivities, $ d_{s,\alpha}^t$ and $d_{s,\alpha}^r$ respectively, for species $\alpha$ are~\cite{diff-dilute-poly_batchelor_jfm1983, tracer-diffusivity-bimodal_nagele_jcp2002}
\begin{align}
  \frac{d_{s,\alpha}^t }{ d_{0,\alpha}^t} & = 1 + \sum_{\beta} I^t_{\alpha\beta} \phi_\beta, \label{eq:pa-diff-t} \\
  \frac{d_{s,\alpha}^r}{ d_{0,\alpha}^r} & = 1 + \sum_{\beta}  I^r_{\alpha\beta} \phi_\beta,\label{eq:pa-diff-r}
\end{align}
where $d^t_{0,\alpha} = \kT\mu_{0,\alpha}$ is the single particle translational diffusivity, $d^r_{0,\alpha} = \kT/(8\pi\eta_0 a_\alpha^3)$ is the single particle rotational self-diffusivity. The integrals $I^t_{\alpha\beta}$ and  $I^r_{\alpha\beta}$ are
\begin{align}
I_{\alpha\beta}^t = & \frac{(1+\lba)^3}{8\lba^3} \int_2^\infty s^2 g_{\alpha\beta}(s) (x_{11}^a + 2y_{11}^a -3 ) \dd s, \label{eq:int-It}\\
I_{\alpha\beta}^r = & \frac{(1+\lba)^3}{8\lba^3} \int_2^\infty s^2 g_{\alpha\beta}(s) (x_{11}^c + 2y_{11}^c -3 ) \dd s, \label{eq:int-Ir}
\end{align}
where $s = 2r/(a_\alpha + a_\beta)$ and $\lba = a_\beta/a_\alpha$.  
Note that the RDF $g_{\alpha\beta}(s) = g_{\alpha\beta}(s,\lambda,\phi)$ depends on the mixture composition.
The mobility couplings of the dimensionless hydrodynamic functions $x^a$, $y^a$, $x^c$, \etc, are described in Kim \& Karrila~\cite{Kim2005}, and we adopt the scaling of Jeffrey \& Onish~\cite{resist-func_jeffrey_jfm1984}.

The PA approximation of the sedimentation velocity is a natural extension of Batchelor~\cite{sed-general_batchelor_jfm1982}:
\begin{equation}
\frac{U_{s,\alpha}}{U_{0,\alpha}} = 1+\sum_\beta S_{\alpha\beta}\phi_\beta,  
\end{equation}
and the integral~\cite{sed-disorder-suspension_brady_pof1988}
\begin{align}
S_{\alpha\beta} = & \left( \frac{1+\lba}{2\lba}\right)^3 \int_2^\infty s^2 g_{\alpha\beta}(s) (x_{11}^a + 2y_{11}^a - 3) \dd s \nonumber\\
&-\gamma(\lba^2+3\lba +1)  + \tfrac{3}{4}\gamma(1+\lba)^2\int_2^\infty s h_{\alpha\beta}(s) \dd s \nonumber  \\
& + \gamma \left( \frac{1+\lba}{2}\right)^2 \int_2^\infty s^2 g_{\alpha\beta}(s) (\hat{x}_{12}^a+ 2\hat{y}_{12}^a) \dd s \label{eq:sed-coeff},
\end{align}
where $h_{\alpha\beta}(s) = g_{\alpha\beta}(s)-1$.  The far field hydrodynamic functions take the form
\begin{align}
  \hat{x}_{12}^a(\lba,s) = & x_{12}^a(\lba,s) - \tfrac{3}{2}s^{-1} + \frac{2(1+\lba^2)}{(1+\lba)^2}s^{-3}, \\
\hat{y}_{12}^a(\lba,s) = &y_{12}^a(\lba,s) - \tfrac{3}{4}s^{-1} - \frac{1+\lba^2}{(1+\lba)^2}s^{-3}.
\end{align}

The PA approximation of the distinct part of the partial hydrodynamic function, $H_{\alpha\beta}^d$, is~\cite{hydro-effect-short-time_nagele_pre1993}
\begin{align}
  H_{\alpha\beta}^d(q) = & \mu_{0,\alpha}  \lba^{-\frac{3}{2}} \sqrt{\phi_\alpha\phi_\beta}
\left[ \tfrac{9}{8}(1+\lba)^2   H_{\alpha\beta}^{d,1}\right.\nonumber\\  
&\left. + \tfrac{3}{2}(1+\lba^2) H_{\alpha\beta}^{d,2}
 + \tfrac{3}{4}(1+\lba)^2 H_{\alpha\beta}^{d,3} \right], 
\end{align}
with 
\begin{align*}
H_{\alpha\beta}^{d,1} = &-2\frac{j_1(2\bar{q})}{\bar{q}} + \int_2^\infty s h_{\alpha\beta}(s) \left( j_0(\bar{q}s)-\frac{j_1(\bar{q}s)}{\bar{q}s} \right)  \dd s, \\
H_{\alpha\beta}^{d,2} = & \frac{j_1(2\bar{q})}{2\bar{q}} + 
\int_2^\infty h_{\alpha\beta}(s) \frac{j_2(\bar{q}s)}{\bar{q}s} \dd s, \\
H_{\alpha\beta}^{d,3} = & \int_2^\infty s^2 g_{\alpha\beta}(s) \times \\
& \left[
\hat{y}_{12}^aj_0(\bar{q}s) + 
(\hat{x}_{12}^a- \hat{y}_{12}^a ) \left(  j_0(\bar{q}s)-2\frac{j_1(\bar{q}s)}{\bar{q}s} \right)   \right]\dd s,
\end{align*}
where $\bar{q} = \tfrac{1}{2}(a_\alpha + a_\beta )q$ is the rescaled wavenumber, and $j_0(x)$, $j_1(x)$, and $j_2(x)$ are spherical Bessel functions of the first kind.

The shear viscosity for polydisperse suspensions is computed as \cite{bimodal-viscosity_wagner_jfm1994,batchelor_green_c2_jfm1972}
\begin{equation}
\frac{\eta_s}{\eta_0} = 1+\tfrac{5}{2}\phi+ \tfrac{5}{2}\phi^2 + \sum_{\alpha, \beta} I^\eta_{\alpha\beta}\phi_\alpha\phi_\beta, 
\end{equation}
where $\tfrac{5}{2}\phi$ is the Einstein viscosity correction and $\tfrac{5}{2}\phi^2$ is the sum of force dipoles in the suspension.
The integral $I^\eta_{\alpha\beta}$ is
\begin{equation}
\label{eq:ieta-def}
I^\eta_{\alpha\beta} = \tfrac{15}{32}(1+\lba)^3(1+\lba^{-3})\int_2^\infty s^2 g_{\alpha\beta}(s) \hat{J}(s,\lba)  \dd s,
\end{equation}
and the expression for $\hat{J}$ is presented Ref.~\onlinecite{bimodal-viscosity_wagner_jfm1994}.

The PA approximation of the suspension bulk viscosity is~\cite{bulk-viscosity_brady_jfm2006}
\begin{equation}
  \label{eq:kappa-pa}
\frac{\kappa_s}{\eta_0} = \frac{\kappa_0}{\eta_0} + \frac{4\phi}{3(1-\phi)} + \sum_{\alpha, \beta} I^\kappa_{\alpha\beta}\phi_\alpha\phi_\beta,
\end{equation}
where the integral
\begin{equation}
\label{eq:ikappa-def}
I^\kappa_{\alpha\beta} = \frac{(1+\lba)^6}{32\lba^3} \int_2^\infty s^2 g_{\alpha\beta} (s) \hat{J}_Q(s, \lba) \dd s.
\end{equation}
The definition of $\hat{J}_Q$  and its asymptotic forms are presented in Appendix~\ref{sec:addit-expr-pa}.

The integrals for the PA approximations are evaluated numerically using Gauss-Kronrod quadrature over the entire integration domain.  The integrands are calculated using twin-multipole expansions up to $s^{-300}$ for $2 \leq s \leq 30$ and far-field asymptotes, presented in Appendix \ref{sec:addit-expr-pa}, for $s>30$.
At $s = 30$, the difference between the exact and the asymptotic solutions is sufficiently small.

\begin{table}[tp]
  \caption{The PA approximation coefficients computed with $\gab = 1$.  For $S_{\alpha\beta}$, the density ratio is $\gamma = 1$. }
  \label{tab:pa-coeff}
  \centering
  \begin{tabularx}{0.9\columnwidth}{*{6}{Y}}
    \hline\hline
    $\lba$ & $-I_{\alpha\beta}^t$ & $-I_{\alpha\beta}^r$ & $-S_{\alpha\beta}$ & $I_{\alpha\beta}^\eta$ & $I_{\alpha\beta}^\kappa$ \\
    \hline
    \sfrac{1}{16} & 2.4152   & 2.2345   & 3.6033 & 1.8237 & 1.6412 \\
    \sfrac{1}{8}  & 2.3464   & 1.9952   & 3.7252 & 1.8661 & 1.6059 \\
    \sfrac{1}{4}  & 2.2424   & 1.6101   & 4.0146 & 2.0388 & 1.5716 \\
    \sfrac{1}{2}  & 2.0876   & 1.1083   & 4.7167 & 2.3312 & 1.5683 \\
    1             & 1.8315   & 0.63102  & 6.5464 & 2.5023 & 1.5835 \\
    2             & 1.4491   & 0.30980  & 11.966 & 2.3312 & 1.5683 \\
    4             & 1.0365   & 0.14186  & 29.392 & 2.0388 & 1.5716 \\
    8             & 0.68904  & 0.064479 & 88.930 & 1.8661 & 1.6059 \\
    16            & 0.43484  & 0.030028 & 304.60 & 1.8237 & 1.6412 \\
    \hline\hline
  \end{tabularx}
\end{table}

Table \ref{tab:pa-coeff} presents the PA approximation coefficients for suspension properties with  $g_{\alpha\beta} = 1$ and for $S_{\alpha\beta}$ the density ratio $\gamma = 1$. Note that $ I^\eta_{\alpha\beta}$ and $I^\kappa_{\alpha\beta}$ are symmetric with respect to $\lba$ and $\lba^{-1}$ in the table.
The PA computations agree well with the published results for monodisperse and polydisperse systems~\cite{diff-dilute-poly_batchelor_jfm1983,sed-polydisperse-numerical_batchelor_jfm1982, tracer-diffusivity-bimodal_nagele_jcp2002,validity-se-rot-diff_koenderink_faradiss2003,bimodal-viscosity_wagner_jfm1994,bulk-viscosity_brady_jfm2006}.  
As far as we are aware, the values of $I_{\alpha\beta}^r$ and $I^\kappa_{\alpha\beta}$ are presented for the first time using the exact two-body problem solutions.

\section{Analytical results beyond the PA level}
\label{sec:summ-exist-results}

\subsection{Suspension properties}

The short-time diffusive behaviors of monodisperse hard-sphere colloidal suspensions have been extensively studied in the past.  The short-time translational self-diffusivity,  $d_s^t$, can be accurately  estimated by the following semi-empirical expression for  $\phi\leq 0.5$~\cite{yukawa-short-time-transport_heinen_jcp2011, Abade2011}
\begin{equation}
\label{eq:dt-approx}
  \frac{d_s^t}{d_0^t} \approx 1 -  1.8315\phi\times(1+0.1195\phi - 0.70 \phi^2),
\end{equation}
where $d_0^t = \kT/(6\pi\eta_0 a)$ is the SES translational diffusivity for particles of radius $a$.  The quadratic term in Eq. (\ref{eq:dt-approx}) recovers the three-body coefficients with lubrication~\cite{self-diffusion-lubrication-3body_cichocki_jcp1999}, and the cubic term is fitted from the computation results of ASD~\cite{trans-prop-asd_banchio_jcp2008} and hydromultipole calculations~\cite{Abade2011}.
The short-time rotational self-diffusivity, $d_s^r$, has been calculated up to $\phi^2$ by including the three-body HIs with lubrication effects\cite{self-diffusion-lubrication-3body_cichocki_jcp1999},
\begin{equation}
  \label{eq:dr-approx}
  \frac{d_s^r}{d_0^r} \approx 1 -  0.631\phi - 0.726 \phi^2,
\end{equation}
where $d_0^r = \kT / (8\pi\eta_0 a^3)$ is the SES rotational diffusivity.

Extending the monodisperse results above to polydisperse colloidal suspensions is a non-trivial undertaking and the results beyond the PA level are limited to the case of bidisperse suspensions with one species presented in trace amount~\cite{tracer-diffusivity-bimodal_nagele_jcp2002}.
Alternatively, inspired by the form of Eq.~(\ref{eq:dt-approx}), we propose the following hybrid scheme for the polydisperse self-diffusivities: 
\begin{align}
\label{eq:dt-poly-approx}
\frac{d_{s,\alpha}^t}{d_{0,\alpha}^t} \approx & 1 + \big( \sum_{\beta} I^t_{\alpha\beta} \phi_\beta \big) \times (1+0.1195\phi - 0.70 \phi^2), \\  
\label{eq:dr-poly-approx}
\frac{d_{s,\alpha}^r}{d_{0,\alpha}^r} \approx & 1 + \big( \sum_{\beta} I^r_{\alpha\beta} \phi_\beta \big)\times (1+1.1505\phi),
\end{align}
with the coefficients $I^t_{\alpha\beta}$ and $I^r_{\alpha\beta}$ from Table~\ref{tab:pa-coeff}.  
Eq.~(\ref{eq:dt-poly-approx}) and (\ref{eq:dr-poly-approx}) are designed in such a way that, for monodisperse suspensions, we recover Eq.~(\ref{eq:dt-approx}) and (\ref{eq:dr-approx}), and for dilute polydisperse suspensions, we recover the PA approximation results with $g_{\alpha\beta} = 1$.
Moreover, it assumes that the particle size only affects the HIs on the pair level, and the many-body HIs are of a mean-field nature, depending only on the total volume fraction.
A similar decoupling idea was used for studying the translational and rotational diffusivities of permeable particle suspensions~\cite{Abade2011}.
In the companion paper~\cite{dg-sd-comp_wang_jcp_2014}, we have successfully applied Eq. (\ref{eq:dt-poly-approx}) to approximate the bidisperse partial hydrodynamic functions $H_{\alpha\beta}(q)$ with the monodisperse $\delta\gamma$ scheme \cite{self-diff-high-dens_beenakker_physa1983,hydro-func-approx_beenakker_physica1984} up to $\phi = 0.4$.

The analytical expression of the monodisperse sedimentation velocity including the three-body HIs is~\cite{sed-coll-diffusion-3body_cichocki_jcp2002}
\begin{equation}
\label{eq:cichocki-used}
  \frac{U_s}{U_0} \approx 1-6.546\phi + 21.918\phi^2,
\end{equation}
where $U_0 = F/(6\pi\eta_0 a)$ is the single particle sedimentation velocity.  A semi-empirical approximation of the polydisperse sedimentation velocities was proposed by Davis \& Gecol~\cite{sed-func-empirical_davis_jaiche1994}, and for bidisperse suspensions it is
\begin{equation}
  \label{eq:davis-used}
 \frac{U_{s,\alpha}}{U_{0,\alpha}} = (1-\phi)^{-S_{\alpha\alpha}} [1+(S_{\alpha\beta} - S_{\alpha\alpha})\phi_\beta],
\end{equation}
with the coefficients from Table~\ref{tab:pa-coeff}.  Eq.~(\ref{eq:davis-used}) recovers the PA approximation results with $g_{\alpha\beta} = 1$ in the dilute limit.

For the monodisperse hydrodynamic function $H(q)$, the principal peak occurs very near the wavenumber $q_m$ corresponding to the static structure factor peak.  The value $H(q_m)$ is well represented by a linear fit~\cite{visc-ses_banchio_jcp_1999, trans-prop-asd_banchio_jcp2008}
\begin{equation}
\label{eq:banchio-hm}
  H(q_m)/\mu_0 = 1 - 1.35 \phi.
\end{equation}

The analytical approximation for monodisperse suspension shear viscosity including the three-body HIs is~\cite{effect-viscosity-phi-cube_cichocki_jcp2003}
\begin{equation}
\label{eq:cichocki-eta}
    \frac{\eta_s}{\eta_0} \approx 1+2.5\phi + 5.0023 \phi^2 + 9.09\phi^3.
\end{equation}
Presently, we are not aware of any approximations of the suspension bulk viscosity beyond the PA approximation level.  Note that in Ref.~\onlinecite{bulk-viscosity_brady_jfm2006}, the quadratic term in the suspension bulk viscosity is $1.57$, and agrees with $1.58$ in Table \ref{tab:pa-coeff} for $\lba = 1$.

\subsection{Porous medium properties}

For monodisperse porous media, the following expression agreed with the Lattice Boltzmann simulation results within a $3\%$ error up to $\phi = 0.6$~\cite{bidisperse-drag-permeability_kuipers_jfm2005}:
\begin{equation}
  \label{eq:kuipers-f}
  F(\phi) = 10\frac{\phi}{(1-\phi^2)} + (1-\phi)^2(1+1.5\sqrt{\phi}).
\end{equation}
For polydisperse porous media, the species drag coefficient is well represented by the following equation~\cite{bidisperse-drag-permeability_kuipers_jfm2005},
\begin{equation}
\label{eq:kuipers-poly}
  F_\alpha=[(1-\phi)z_\alpha + \phi z_\alpha^2 + 0.064(1-\phi)z_\alpha^3] F(\phi), 
\end{equation}
where $z_\alpha$ is the species diameter fraction defined in Sec.~\ref{sec:transp-prop-poro}, and $F(\phi)$ is from Eq.~(\ref{eq:kuipers-f}).  

Few studies have been performed on the hindered diffusion in porous media.
As far as we are aware, only the translational hindered diffusivity for monodisperse porous media has been investigated, and it can be obtained by solving the following self-consistent equation~\cite{hindered-diffusion_muthukumar_jcp1978}:
\begin{equation}
  \label{eq:dhd-fm}
  (d_{\mathrm{HD}}^t)^{-1}  =  1+\sqrt{\tfrac{9}{2}\phi }(d_{\mathrm{HD}}^t)^{-\frac{1}{2}} + \tfrac{3}{2}\phi (d_{\mathrm{HD}}^t)^{-1} + \ldots
\end{equation}

\section{Results for suspensions}
\label{sec:results-suspensions}

\subsection{Short-time translational self-diffusivity}
\label{sec:transl-self-diff}




\begin{figure}
  \begin{center}
  \subfloat[]{
    \centering
    \includegraphics[width=3in]{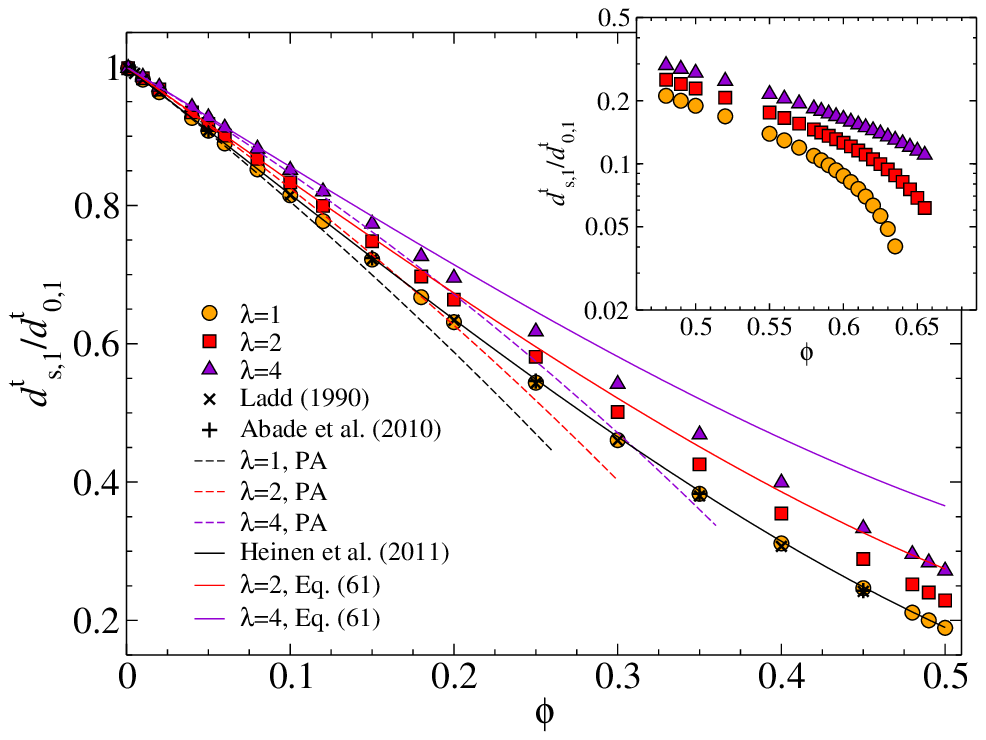}
    \label{fig:dinf-t1}
  }\\
  \subfloat[]{
    \centering
    \includegraphics[width=3in]{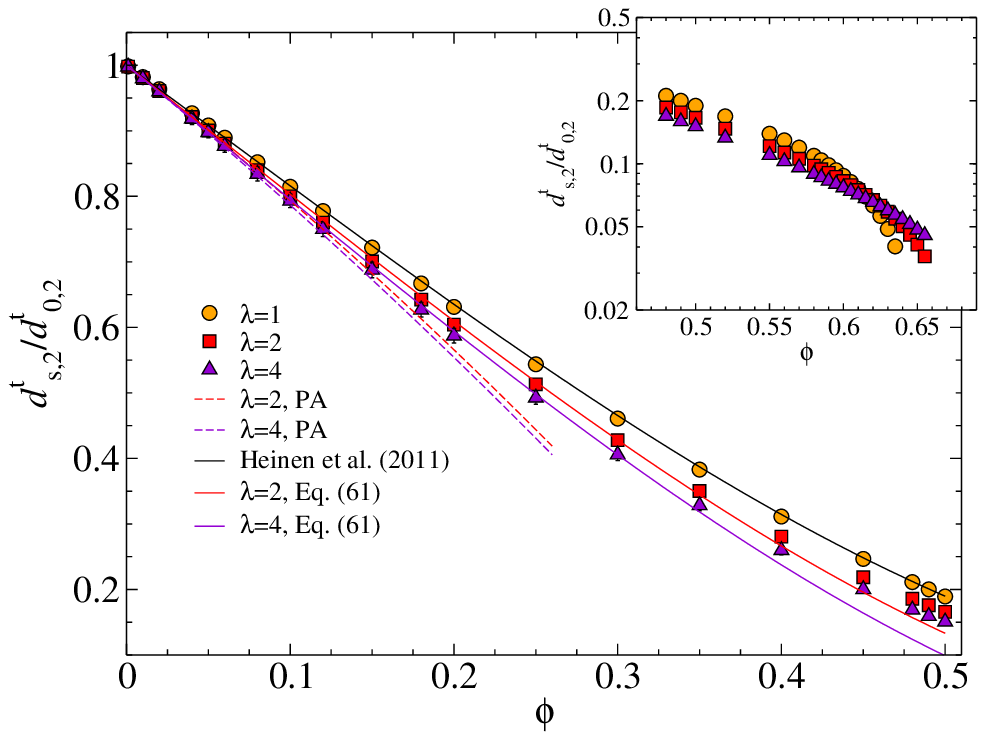}
    \label{fig:dinf-t2}
  }
  \end{center}
  \caption{(Color online) The short-time translational self-diffusivity \protect\subref{fig:dinf-t1}: $d_{s,1}^t$ and \protect\subref{fig:dinf-t2}: $d_{s,2}^t$ as a function of $\phi$ for bidisperse suspensions with $y_1 = 0.5$ and $\lambda = 1$, $2$, and $4$ [bottom to top in \protect\subref{fig:dinf-t1} and top to bottom in \protect\subref{fig:dinf-t2}, respectively].
The monodisperse simulation results from Ladd \cite{hydro-trans-coeff_ladd_jcp1990} and Abade \etal \cite{Abade2010} are also presented in \protect\subref{fig:dinf-t1}.
The PA approximations are shown in dashed lines and Eq. (\ref{eq:dt-poly-approx}),  which reduces to the expression of Heinen \etal \cite{yukawa-short-time-transport_heinen_jcp2011} at $\lambda=1$, is shown in solid lines.
The insets show the results at higher $\phi$.
}
  \label{fig:dinft-1}
\end{figure}

Fig.~\ref{fig:dinft-1} presents the short-time translational self-diffusivity $d^t_{s,\alpha}$ for both species as a function of the total volume fraction $\phi$ for bidisperse suspensions with $y_1 = 0.5$ and $\lambda=2$ and $4$, as well as for monodisperse suspensions.
In the same figure we also present the monodisperse computations of Ladd~\cite{hydro-trans-coeff_ladd_jcp1990} and Abade \etal~\cite{Abade2010}, which agree excellently with the SD results.
The semi-empirical expression of Heinen \etal~\cite{yukawa-short-time-transport_heinen_jcp2011}, Eq.~(\ref{eq:dt-approx}), accurately captures the monodisperse data up to $\phi = 0.5$.
The PA approximations, however, are valid only for $\phi< 0.1$, and begin to deviate from the simulation data afterwards.
At very high $\phi$, as is shown in the inset of Fig.~\ref{fig:dinf-t1}, the monodisperse $d^t_s$ decreases drastically when $\phi>0.60$, and vanishes as the volume fraction approaches $\phi\sim 0.64$~\cite{asd_sierou_jfm01}.

In a bidisperse suspension ($\lambda > 1$),  the relative short-time translational self-diffusivities of the smaller species is always higher than that of the larger species, \ie, $d_{s,1}^t/d_{0,1}^t > d_{s,2}^t/d_{0,2}^t$.  The diffusivity difference between the small and the large particles increases with the increasing suspension size ratio $\lambda$.
At a fixed $\phi$, for the smaller particles $d_{s,1}^t/d_{0,1}^t$ can be much higher than the monodisperse value $d_s^t/d_0^t$, particularly at high $\lambda$, as is shown in Fig.~\ref{fig:dinf-t1}, but $d_{s,2}^t/d_{0,2}^t$ does not differ significantly from $d_s^t/d_0^t$ even with a large size ratio, as is shown in Fig.~\ref{fig:dinf-t2}.
This suggests that the HIs for the two species are distinct: intuitively, the larger particles, which can be surrounded by multiple smaller particles, experience mean-field-like HIs, as if they were suspended in an effective medium formed by the solvent and the smaller particles.
The HIs for the smaller particles, on the other hand, are expected to be strongly affected by the presence of the large particles.

The PA approximations of $d^t_{s,\alpha}$, shown in dashed lines in Fig.~\ref{fig:dinft-1}, agree with the SD computations up to $\phi \approx 0.1$.  At higher volume fractions, the HIs beyond the two-body level begin to dominate and the PA approximations underestimate the diffusivities for both species.
The decoupling approximations of Eq.~(\ref{eq:dt-poly-approx})), shown in solid lines, exhibit superior agreement.  For the small particles in Fig.~\ref{fig:dinf-t1}, Eq.~(\ref{eq:dt-poly-approx}) is accurate up to $\phi \approx 0.25$ and $0.15$ for $\lambda = 2$ and $4$, respectively.  
The decoupling approximation works much better for the large particles, and remains valid for $\phi=0.4$ and $0.35$ for $\lambda = 2$ and $4$, respectively, as is shown in Fig.~\ref{fig:dinf-t2}.
Beyond their range of validity, the decoupling approximation overestimates the small particle diffusivity and underestimates the large particle diffusivity.

The SD calculations for very dense suspensions up to and beyond the monodisperse close packing volume fraction ($\phi \sim 0.64$) are shown in the insets of Fig.~\ref{fig:dinft-1}.
For the smaller particles in Fig.~\ref{fig:dinf-t1}, the reduction of $d_{s,1}^t$ with increasing $\phi$ is slower for $\lambda > 1$ compared to the monodisperse case.  In particular, at $\phi = 0.655$, the highest volume fractions we studied in this work, the diffusivity $d_{s,1}^t/d_{0,1}^t$ remains higher than $0.1$ at $\lambda = 4$.
More interestingly, for the larger particles shown in Fig.~\ref{fig:dinf-t2}, $d^t_{s,2}/d^t_{0,2}$ for $\lambda > 1$ crosses the monodisperse values near $\phi\approx 0.61$.  At higher $\phi$, the diffusivities $d^t_{s,2}/d^t_{0,2}$ for $\lambda=4$ is higher than those for $\lambda = 2$.
This is simply because the size polydispersity improves the particle packing and increases the suspension maximum packing density~\cite{bidisperse-packing_torquato_pre2013}, where the diffusivity $d_{s,\alpha}^t$ reduces to zero due to particle contact.
At a fixed $y_1$, increasing $\lambda$ increases the maximum packing density.
As a result, at sufficiently high $\phi$, the diffusivities of both species can exceed the monodisperse value, and the apparent diffusivity enhancement increases with $\lambda$.

\begin{figure}
  \begin{center}
  \subfloat[]{
    \centering
    \includegraphics[width=3in]{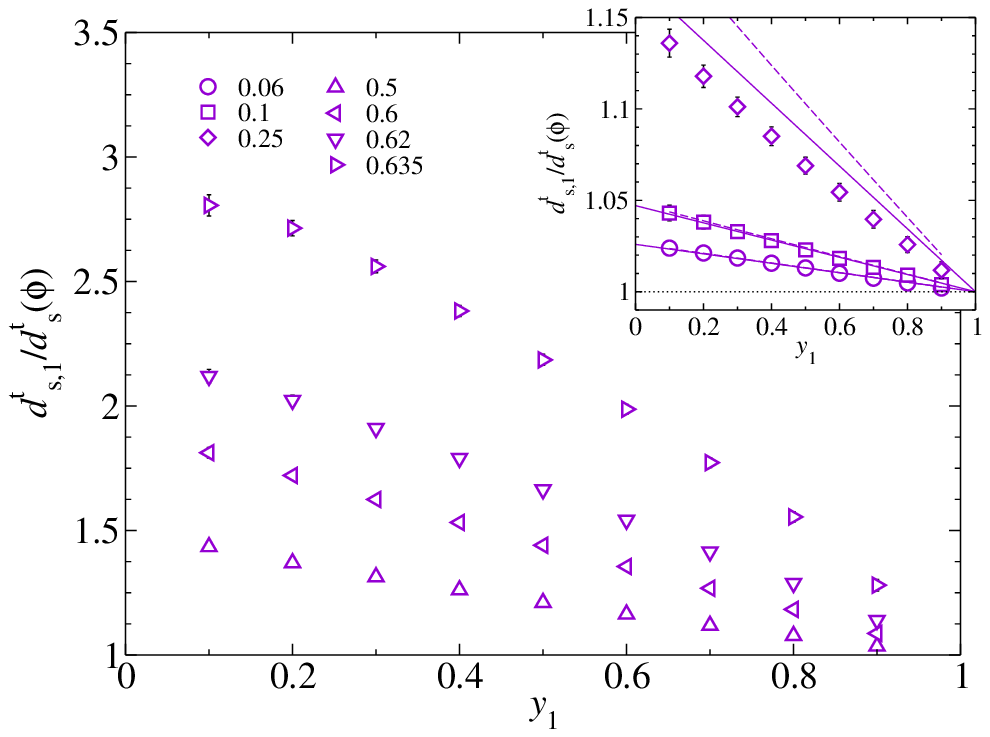}
    \label{fig:dst1-t2}
  }\\
  \subfloat[]{
    \centering
    \includegraphics[width=3in]{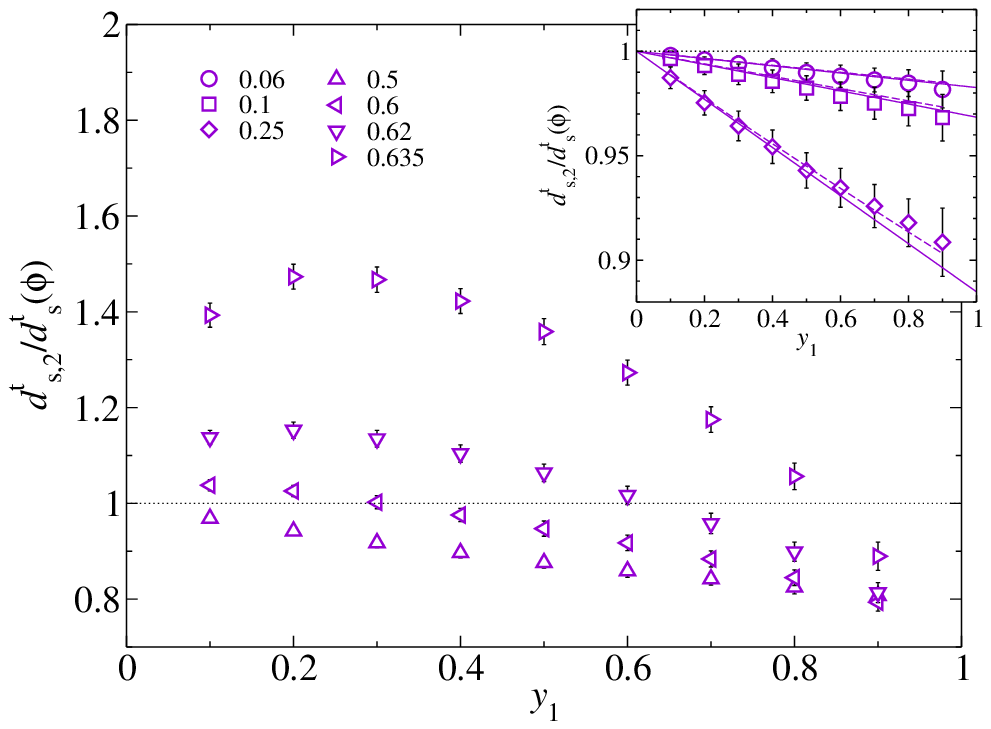}
    \label{fig:dst2-t2}
  }
  \end{center}
  \caption{The normalized translational diffusivities \protect\subref{fig:dst1-t2}: $d_{s,1}^t/d_s^t$ and \protect\subref{fig:dst2-t2}: $d_{s,2}^t/d_s^t$ as a function of $y_1$ at different $\phi$ for bidisperse suspensions of $\lambda = 2$. The monodisperse short-time translational self-diffusivity at the corresponding $\phi$ is $d_s^t$.  The insets also show the PA approximations (dashed lines) and Eq.~(\ref{eq:dt-poly-approx}) (solid lines).
}
  \label{fig:dinft-2}
\end{figure}

Fig.~\ref{fig:dinft-2} exams the ratio of the species diffusivity to the monodisperse value at the same volume fraction $\phi$, $d^t_{s,\alpha}/d^t_s$, as a function of the suspension composition $y_1$ at several $\phi$ for bidisperse suspensions of $\lambda = 2$.
The ratio $d^t_{s,\alpha}/d^t_s$ highlights the influence of suspension composition on the diffusivities, such that the ratio recovers $1$ when $y_1\rightarrow 0$ for the large species and $y_1\rightarrow 1$ for the small species.
For low to moderate $\phi$, as is shown in the insets of Fig.~\ref{fig:dst1-t2} and \ref{fig:dst2-t2}, the PA approximation and the decoupling approximation of Eq.~(\ref{eq:dt-poly-approx}) are also presented in dashed and solid lines, respectively.
Both approximation schemes capture the SD calculations up to $\phi = 0.25$ at all $y_1$ except overestimating $d_{s,1}^t/d_{s}^t$ at $\phi = 0.25$.
Within this volume fraction range, $d_{s,\alpha}^t/d_{s}^t$ for both species decreases almost linearly with increasing $y_{1}$, with $d_{s,1}^t/d_{s}^t$ towards and $d_{s,2}^t/d_{s}^t$ away from unity, respectively.
Physically, replacing smaller particles with larger particles at a fixed $\phi$ (decreasing $y_1$) increases the diffusivities of both species.  
Moreover, at a given $\phi$, the tracer diffusivity is the maximum diffusivity for the smaller particles and the minimum diffusivity for the larger particles.
At $\phi = 0.25$, the maximum diffusivity enhancement for the smaller particles is $15\%$ as $y_1 \rightarrow 0$, while the maximum reduction for the larger particles is $10\%$ as $y_1 \rightarrow 1$.

The ratio $d_{s,\alpha}^t/d_s^t$ exhibits more intriguing behaviors for dense suspensions.  The ratio $d_{s,1}^t/d_s^t$ for the smaller particles, as is shown in Fig.~\ref{fig:dst1-t2}, increases significantly with decreasing $y_1$.  In particular, at $\phi = 0.635$, the ratio $d_{s,1}^t/d_s^t \rightarrow 2.9$ as $y_1 \rightarrow 0$.  Moreover, the ratio $d_{s,1}^t/d_s^t$ is no longer linear with $y_1$ when $\phi$ is close to $0.635$, particularly when $y_1$ is small.
For the larger particles in Fig.~\ref{fig:dst2-t2}, the ratio $d_{s,2}^t/d_s^t$ is more surprising.
Contrary to the dilute behaviors shown in the inset, the ratio $d_{s,2}^t/d_s^t$ increases with increasing $\phi$ when $\phi>0.5$.
Moreover, with $\phi>0.6$, $d_{s,2}^t/d_s^t$ exceeds unity, and a maximum $d_{s,2}^t/d_s^t$ emerges at a non-trivial $y_1$.  For $\phi = 0.635$, the maximum occurs between $y_1=0.2$ and $0.3$, and corresponds to a $150\%$ diffusivity enhancement relative to the monodisperse value.
These peculiar behaviors correspond to the approaching and crossing of the monodisperse diffusivities in the inset of Fig.~\ref{fig:dinf-t2}, and are due to changes in both the HIs and the bidisperse particle packing.
A particularly interesting aspect of Fig.~\ref{fig:dinft-2} is that for a dense monodisperse suspensions near closing packing, replacing a small amount of large particles with small particles promotes diffusivities $d^t_{s,\alpha}$ of \emph{both} species.

\subsection{Short-time rotational self-diffusivity}
\label{sec:rotat-self-diff}


\begin{figure}
  \begin{center}
  \subfloat[]{
    \centering
    \includegraphics[width=3in]{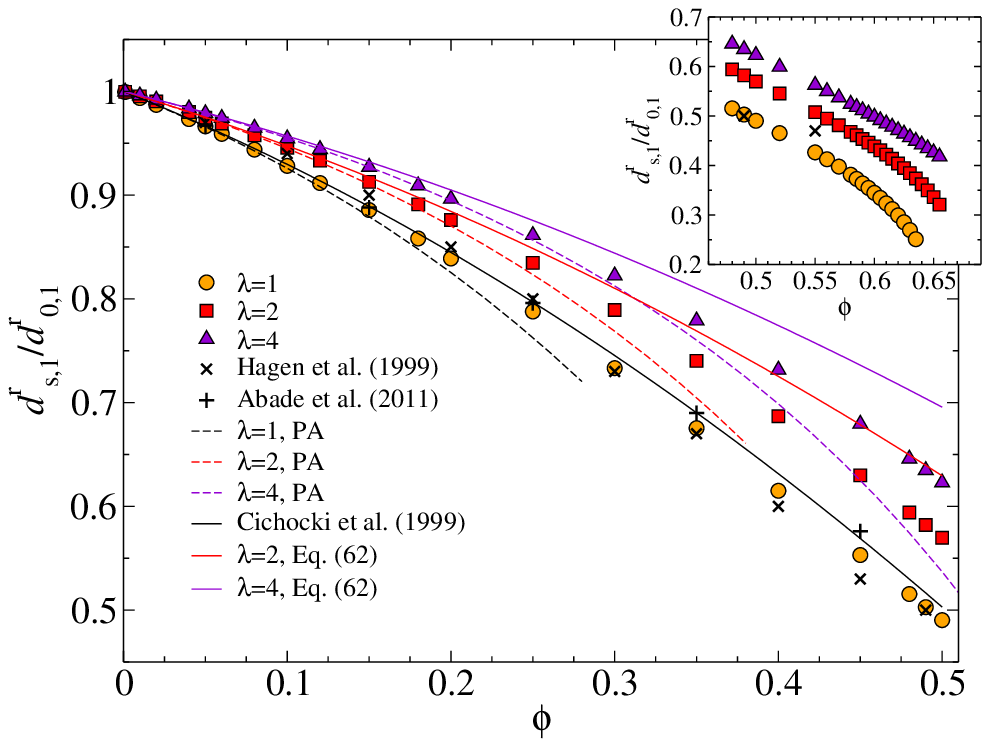}
    \label{fig:dinf-r1}
  }\\
  \subfloat[]{
    \centering
    \includegraphics[width=3in]{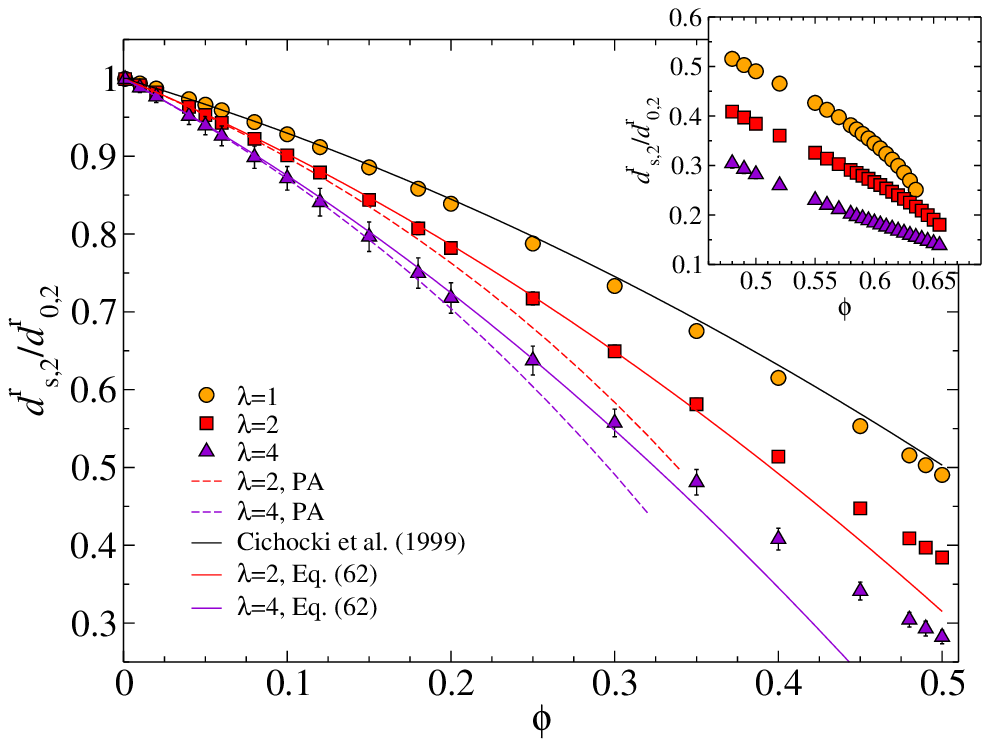}
    \label{fig:dinf-r2}
  }
  \end{center}
  \caption{(Color online) The short-time rotational self-diffusivity \protect\subref{fig:dinf-r1}: $d^r_{s,1}$ and \protect\subref{fig:dinf-r2}: $d^r_{s,2}$ as a function of $\phi$ for bidisperse suspensions with $y_1=0.5$ and $\lambda = 1$, $2$, and $4$ [bottom to top in \protect\subref{fig:dinf-r1} and top to bottom in \protect\subref{fig:dinf-r2}, respectively].
The monodisperse simulation results from Hagen \etal \cite{rot-diff-lb-sim_hagen_physa1999} and Abade \etal \cite{Abade2011} are also presented in \protect\subref{fig:dinf-r1}. 
The PA approximations are shown in dashed lines and Eq.~(\ref{eq:dr-poly-approx}), which reduces the results of Cichocki \etal~\cite{self-diffusion-lubrication-3body_cichocki_jcp1999} at $\lambda=1$, is shown in solid lines.  The insets show the results at higher $\phi$.
}
  \label{fig:dinfr-1}
\end{figure}

Fig.~\ref{fig:dinfr-1} shows the short-time rotational self-diffusivity for both species, $d_{s,\alpha}^r$, as a function of $\phi$ for bidisperse suspensions with $\lambda = 2$ and $4$ at $y_1 = 0.5$, as well as for monodisperse suspensions.
The Lattice-Boltzmann (LB) computations of Hagen \etal\cite{rot-diff-lb-sim_hagen_physa1999} and the hydromultipole calculations of Abade \etal\cite{Abade2011} for monodisperse suspensions are also presented.
The monodisperse $d_s^r/d_0^r$ shows a much weaker $\phi$ dependence compared to its translational counterpart $d_s^t/d_0^t$.
Up to $\phi = 0.2$, the monodisperse SD results agree well with the hydromultipole results, and at higher volume fractions, SD predicts lower $d_s^r/d_0^r$. 
The SD results lie between the LB and the hydromultipole results.
The PA approximation agrees with the SD results only up to $\phi = 0.1$, and underestimates the diffusivity at higher $\phi$.
The analytical expression of Cichocki~\etal\cite{self-diffusion-lubrication-3body_cichocki_jcp1999}, Eq.~(\ref{eq:dr-approx}), exhibits remarkable agreement with the simulations up to $\phi=0.5$.
Moreover, for very dense suspensions, as is shown in the inset of Fig.~\ref{fig:dinf-r1}, the diffusivity $d_s^r/d_0^r$ does not drop as rapidly as $d_s^t/d_0^t$, and retains a large value ($\sim 0.25$) even close to the maximum packing, undoubtedly owning to the weak logarithm lubrication singular behavior of rotation.

For bidisperse suspensions, the small and the large particle rotational diffusivities $d_{s,\alpha}^r/d_{0,\alpha}^r$ are shown in Fig.~\ref{fig:dinf-r1} and~\ref{fig:dinf-r2}, respectively.
Compared to the monodisperse results, $d_{s,1}^r/d_{0,1}^r$ are higher and $d_{s,2}^r/d_{0,2}^r$ are lower.
Unlike their translational counterparts, the rotational diffusivities of both species are noticeably different from the monodisperse values, and are sensitive to the size ratio $\lambda$, particularly at moderate to dense $\phi$.
On the other hand, they display less sensitivity to $\phi$ compared to $d_{s,\alpha}^t/d_{0,\alpha}^t$, as rotation is always easier than translation in a crowded environment. 
At very high $\phi$, as shown in the inset of Fig.~\ref{fig:dinf-r1} and \ref{fig:dinf-r2}, the diffusivities $d_{s,2}^r/d_{0,2}^r$ at higher $\lambda$ do not cross the monodisperse values even at $\phi = 0.635$.
Therefore, the suspension packing plays a less significant role on the rotational diffusivities.
Note that the weak $\phi$ and the strong $\lambda$ dependence of $d_{s,\alpha}^r/d_{0,\alpha}^r$ exhibited in Fig.~\ref{fig:dinfr-1} can be exploited experimentally as a structural probe for dense suspensions~\cite{tracer-diffusivity-bimodal_nagele_jcp2002,validity-se-rot-diff_koenderink_faradiss2003}.

The PA approximations, shown in dashed lines in Fig.~\ref{fig:dinfr-1} in respective colors, agree reasonably with the polydisperse SD results up to $\phi=0.15$, and then significantly underestimate the diffusivities due to the HIs beyond the pairwise level.
The decoupling approximation of Eq.~(\ref{eq:dr-poly-approx}), plotted as solid lines in the respective colors in Fig.~\ref{fig:dinfr-1}, shows a better agreement, and, similarly to the translational case, works better for the larger particles.
In particular, the decoupling approximation is valid up to $\phi = 0.2$ for the smaller particles with $\lambda = 2$ and $4$; for the larger particles, it is valid up to $\phi = 0.4$ for $\lambda = 2$ and up to $\phi = 0.3$ for $\lambda = 4$.
The success of the decoupling approximation again demonstrates that the HIs for the larger particles are mean-field-like.  For the smaller particles, the size effect is more complex and is beyond the decoupling approximation.

\begin{figure}
  \begin{center}
  \subfloat[]{
    \centering
    \includegraphics[width=3in]{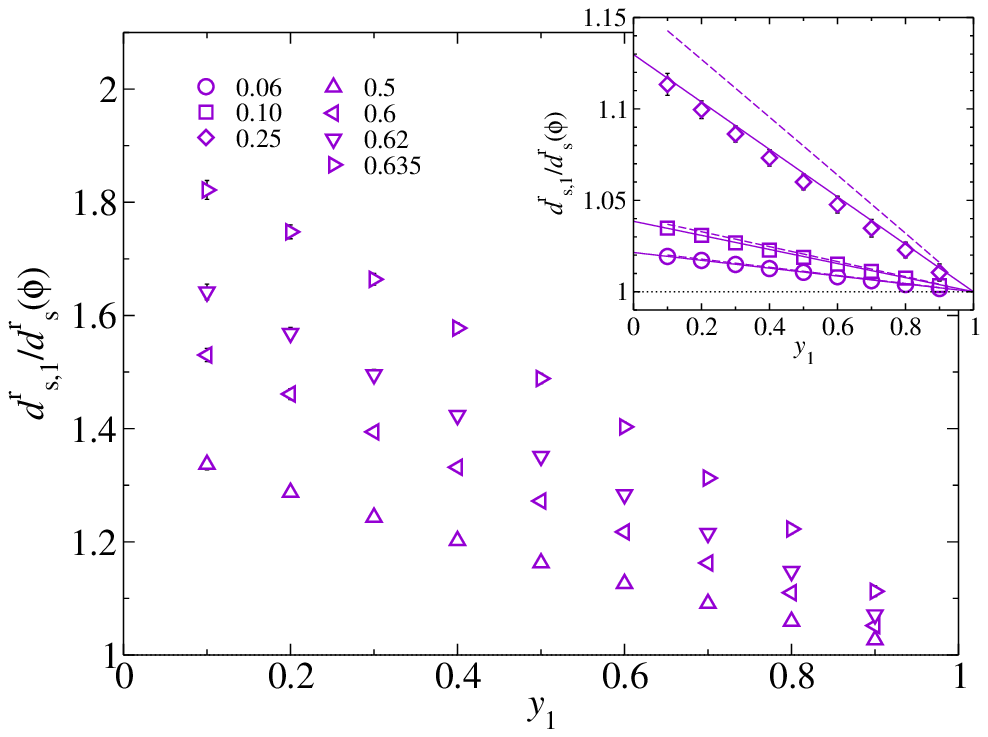}
    \label{fig:dsr1-t2}
  }\\
  \subfloat[]{
    \centering
    \includegraphics[width=3in]{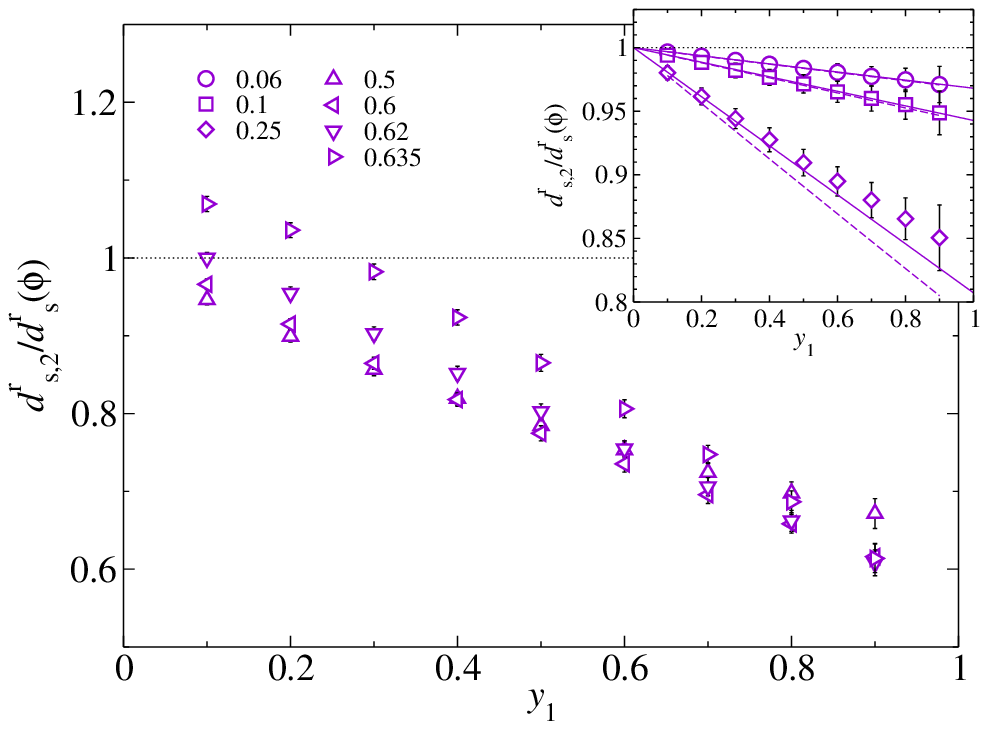}
    \label{fig:dsr2-t2}
  }
  \end{center}
  \caption{
The normalized rotational diffusivities \protect\subref{fig:dsr1-t2}: $d_{s,1}^r/d_s^r$ and \protect\subref{fig:dsr2-t2}: $d_{s,2}^r/d_s^r$ as a function of $y_1$ at different $\phi$ for bidisperse suspensions of $\lambda = 2$. The monodisperse short-time rotational self-diffusivity at the corresponding $\phi$ is $d_s^t$.  The insets also show the PA approximations (dashed lines) and Eq.~(\ref{eq:dr-poly-approx}) (solid lines).
} 
  \label{fig:dinfr-2}
\end{figure}

The influences of the composition $y_1$ on the ratio $d_{s,\alpha}^r/d_s^r$, with $d_s^r$ at the same $\phi$, are presented in Fig.~\ref{fig:dinfr-2} for bidisperse suspensions with $\lambda = 2$.
The effect of $y_1$ at dilute and moderate $\phi$ are shown in the insets of Fig.~\ref{fig:dsr1-t2} and \ref{fig:dsr2-t2} for the smaller and the larger particles, respectively.
Increasing the small particle composition $y_1$ with a fixed $\phi$ decreases  $d_{s,\alpha}^r$ of both species almost linearly, with the smaller particles towards the monodisperse value and the larger particles away from it.
The ratio $d^r_{s,1}/d^r_{0,1}$ exhibits a maximum for trace amount of smaller particles and $d^r_{s,2}/d^r_{0,2}$ exhibits a minimum for trace amount of larger particles.
As is shown in the insets of Fig.~\ref{fig:dinfr-2}, increasing $\phi$ increases the maximum of $d^r_{s,1}/d^r_{0,1}$ for the smaller particles and reduces the minimum of $d^r_{s,2}/d^r_{0,2}$ for the larger particles.
The PA approximation and the decoupling expression, Eq.~(\ref{eq:dr-poly-approx}), are presented as dashed and solid lines, respectively, in the insets of Fig.~\ref{fig:dinfr-2}.
Both approximation schemes capture the composition $y_1$ dependence of  $d_{s,\alpha}^r/d_s^r$ up to $\phi = 0.10$ for both species.  At $\phi = 0.25$, Eq.~(\ref{eq:dr-poly-approx}) also captures the $y_1$ dependence for both species, but the PA approximations overestimate the effect of composition change.

The ratio $d_{s,\alpha}^r/d_s^r$ at higher $\phi$ differ significantly from its translational counterpart.
For the smaller particles in Fig.~\ref{fig:dsr1-t2},  $d_{s,1}^r/d_s^r$ increases with increasing $\phi$ and remains linear with $y_1$ with fixed $\phi$.  At $\phi = 0.635$, the tracer diffusivity of the small particles is almost $190\%$ of the monodisperse values.
For the larger particles in Fig.~\ref{fig:dsr2-t2}, with $\phi\geq 0.5$, increasing $\phi$  also \emph{increases} $d_{s,2}^r/d_s^r$ altogether, and this is qualitatively different from the dilute behaviors in the inset.
At $\phi \geq 0.62$, the ratio $d_{s,2}^r/d_s^r$ can exceed unity, suggesting the rotational diffusivities of both species are enhanced due to the change in the particle packing.
Moreover, the ratio $d_{s,2}^r/d_s^r$ is very sensitive to $y_1$, and with the presented data, it appears almost linear with $y_1$.
This means $d_{s,2}^r/d_s^r$ must exhibit a maximum at $y_1\ll 0.1$. 
Therefore, for a dense monodisperse suspension near close packing, replacing trace amount of large particles with small particles can increase the rotational diffusivities of the both species.
Together with Fig.~\ref{fig:dinft-2}, Fig.~\ref{fig:dinfr-2} illustrates the distinctive behaviors of the HIs for the translational and rotational motions.

\subsection{Instantaneous sedimentation velocity}
\label{sec:short-time-sedim}


\begin{figure}
  \subfloat[]{
    \includegraphics[width=3in]{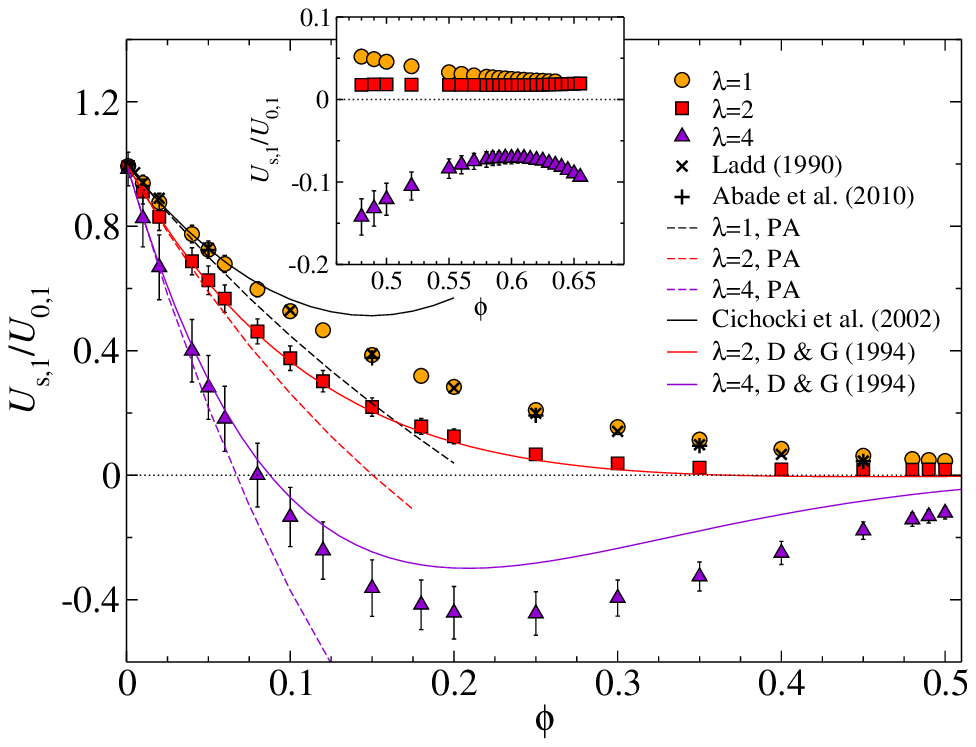}
    \label{fig:sed-t1}
  }\\
  \subfloat[]{
    \includegraphics[width=3in]{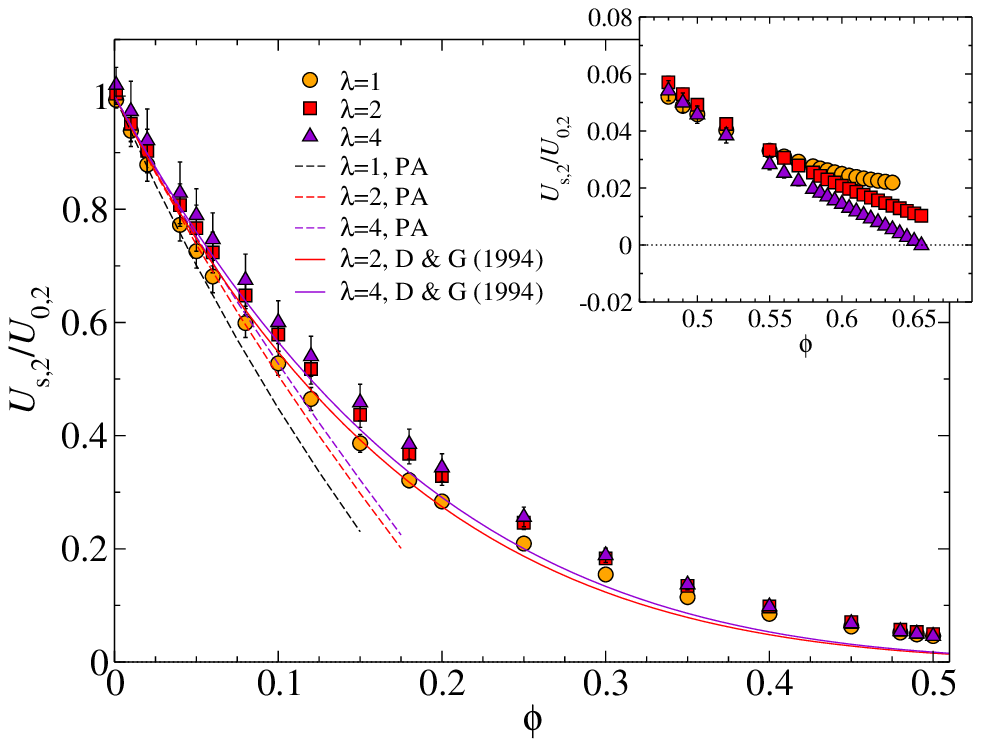}
    \label{fig:sed-t2}
  }
  \caption{
(Color online) The instantaneous sedimentation velocities \protect\subref{fig:sed-t1}: $U_{s,1}$ and \protect\subref{fig:sed-t2}: $U_{s,2}$ as a function of $\phi$ for bidisperse suspensions with $y_1 = 0.5$ and $\lambda = 1$, $2$, and $4$ [top to bottom in \protect\subref{fig:sed-t1} and bottom to top in \protect\subref{fig:sed-t2}, respectively].
The monodisperse simulation results from Ladd \cite{hydro-trans-coeff_ladd_jcp1990} and Abade \etal \cite{Abade2010} are also presented in \protect\subref{fig:dinf-t1}.
The PA approximations are shown in dashed lines. 
The theoretical results of Cichocki \etal \cite{sed-coll-diffusion-3body_cichocki_jcp2002} for $\lambda = 1$, and the semi-empirical expression of Davis \& Gecol~\cite{sed-func-empirical_davis_jaiche1994}, Eq.~(\ref{eq:davis-used}), for $\lambda = 2$ and $4$ are presented in solid lines.
The insets show the results at higher $\phi$.
}
  \label{fig:usedt-1}
\end{figure}

The instantaneous sedimentation velocities $U_{s,\alpha}/U_{0,\alpha}$ of bidisperse suspensions with equal density materials at $\lambda = 2$ and $4$ and $y_1 = 0.5$, as well as monodisperse suspensions, are presented in Fig.~\ref{fig:usedt-1}.
For monodisperse suspensions, $U_s/U_0$ from Ladd~\cite{hydro-trans-coeff_ladd_jcp1990} and Abade \etal~\cite{Abade2010} are also shown in Fig.~\ref{fig:sed-t1} for comparison.
The SD results agree with the earlier computational studies~\cite{hydro-trans-coeff_ladd_jcp1990,Abade2010} up to $\phi = 0.2$, and then yield higher values.
Although the absolute magnitude of the differences appears to be small, the relative difference is significant, up to $36\%$ at $\phi = 0.45$.
The origin of the discrepancy, as pointed out by Brady \& Durlofsky~\cite{sed-disorder-suspension_brady_pof1988}, is that the multipole expansions up to the mean-field quadrupole level used in SD is not sufficient to capture the collective HIs in sedimentation problems.
On the other hand, SD closely captures the qualitative aspects of $U_s/U_0$, and remains positive over the entire volume fraction range.
As mentioned earlier, the incorporation of the mean-field quadrupole in the mobility tensor construction improves the accuracy of conventional SD compared to ASD~\cite{asd_sierou_jfm01}.

The monodisperse PA approximations and the analytical results of Cichocki \etal~\cite{sed-coll-diffusion-3body_cichocki_jcp2002}, Eq.~(\ref{eq:cichocki-used}), are shown in dashed and solid lines in Fig.~\ref{fig:sed-t1}, respectively.
The agreement between the simulations and the analytical expressions is unsatisfactory.
The PA approximation is valid only up to $\phi = 0.05$, and Eq.~(\ref{eq:cichocki-used}), which incorporates three-body effect, shows a minor improvement and agrees with the simulations only up to $\phi = 0.08$.
Such lack of agreement at higher $\phi$ clearly illustrates the challenges in developing theories for sedimentation problems.

For bidisperse suspensions, the species sedimentation velocities are shown in Fig.~\ref{fig:sed-t1} and \ref{fig:sed-t2} for the small and the large particles, respectively.
With equal densities for both species, $U_{s,1}/U_{0,1}$ of the smaller particles is lower than the monodisperse values, and $U_{s,2}/U_{0,2}$ of the larger particles is higher.
Interestingly, at $\lambda = 4$, the small particle sedimentation velocity $U_{s,1}$ changes sign when $\phi\geq 0.08$.
In this case, the fall of large particles generates a strong upward backflow that  offsets the effects of the downward force on the small particles, making them move with the fluid in the opposite direction.  
The small particle $U_{s,1}$ first reaches a minimum, then increases with increasing $\phi$.  
At $\lambda = 2$, $U_{s,1}$ approaches zero for $\phi>0.35$, suggesting that the combination of the imposed force and the back flow makes the particles almost stationary.
Apparently, the HIs for the small particles are strongly affected by $\phi$ and $\lambda$.
On the other hand, for the larger particles, $U_{s,2}$ closely follows the monodisperse values, and shows little variation with different $\lambda$.

The SD results of the sedimentation velocity $U_{s,\alpha}/U_{0,\alpha}$ for very dense systems are shown in the insets of Fig.~\ref{fig:usedt-1}. For the smaller particles near close packing, $U_{s,1}$ is positive for $\lambda = 1$ and $2$, and remains negative for $\lambda = 4$.
For the larger particles, the sedimentation velocities $U_{s,2}$ cross each other. 
As a result, at $\phi>0.6$, the monodisperse sedimentation velocity is the highest, and the magnitude of $U_{s,2}/U_{0,2}$ decreases with increasing $\lambda$, an opposite trend compared to the dilute suspensions.

The polydisperse PA approximation and the semi-empirical expression of Davis \& Gecol\cite{sed-func-empirical_davis_jaiche1994}, Eq.~(\ref{eq:davis-used}), are presented in dashed and solid lines in respective colors in Fig.~\ref{fig:usedt-1}, respectively.
The PA approximations capture $U_{s,\alpha}$ of both species up to $\phi = 0.05$ for $\lambda = 2$ and $4$, and then underestimate the SD results.
In Fig.~\ref{fig:sed-t1}, the semi-empirical approximation of Eq.~(\ref{eq:davis-used}) shows a remarkable overall agreement with the SD results for $\lambda =2$ at all $\phi$, and for $\lambda = 4$, it captures the velocity direction change but overestimates the sedimentation velocity at higher $\phi$.
For the larger particles in Fig.~\ref{fig:sed-t2}, Eq.~(\ref{eq:davis-used}) captures the qualitative trend in the SD results of $U_{s,2}$.  However, at higher $\phi$, the quantitative difference becomes apparent.

\begin{figure}
  \begin{center}
  \subfloat[]{
    \centering
    \includegraphics[width=3in]{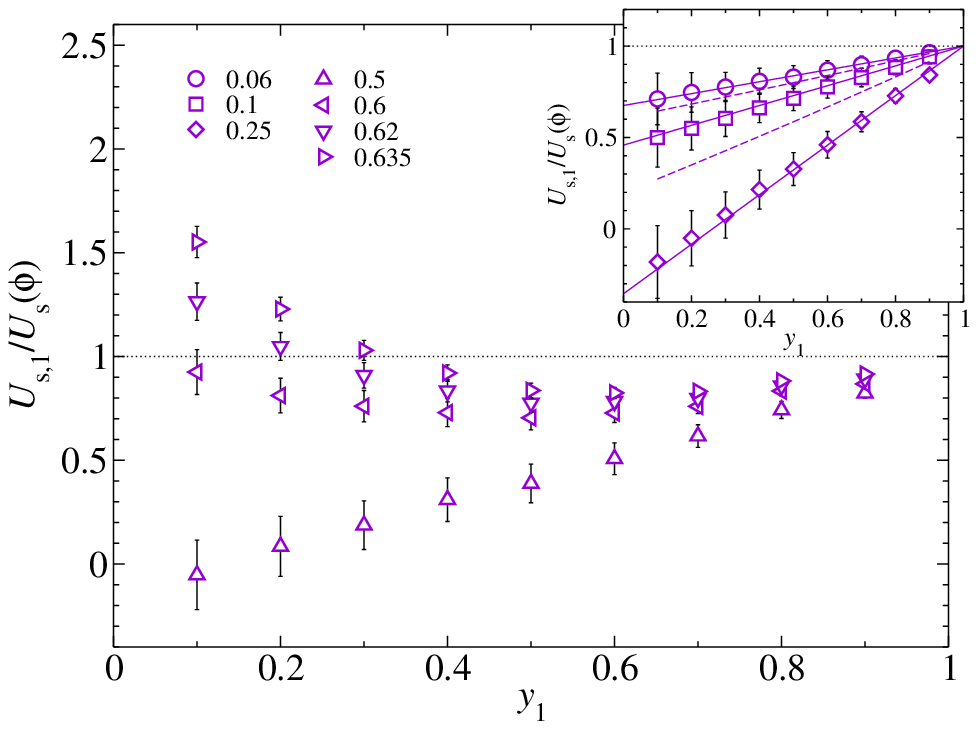}
    \label{fig:sed1-t2}
  }\\
  \subfloat[]{
    \centering
    \includegraphics[width=3in]{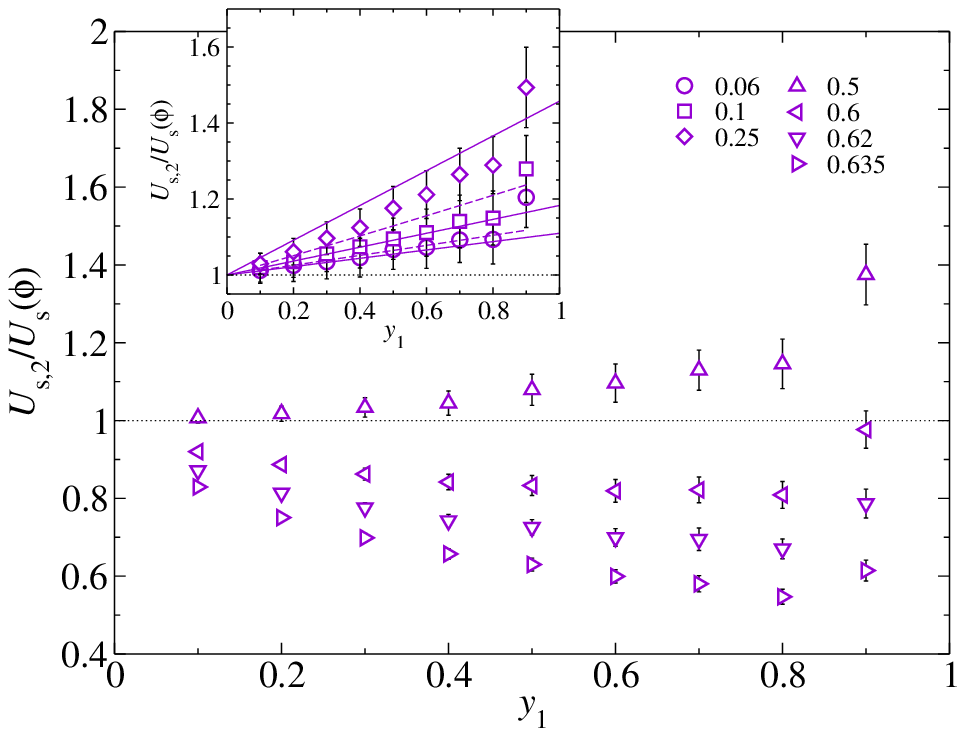}
    \label{fig:sed2-t2}
  }
  \end{center}
  \caption{
The normalized instantaneous sedimentation velocities \protect\subref{fig:sed1-t2}: $U_{s,1}/U_s$ and \protect\subref{fig:sed2-t2}: $U_{s,2}/U_s$ as a function of $y_1$ at different $\phi$ for bidisperse suspensions of $\lambda = 2$. The monodisperse instantaneous sedimentation velocity at the corresponding $\phi$ is $U_s$.  The insets also show the PA approximations (dashed lines) and the approximations of Davis \&  Gecol~\cite{sed-func-empirical_davis_jaiche1994}, Eq.~(\ref{eq:davis-used}) (solid lines).
} 
  \label{fig:usedt-2}
\end{figure}

Fig.~\ref{fig:usedt-2} presents the effect of composition $y_1$ on the ratio $U_{s,\alpha}/U_s$, where $U_s$ is the the monodisperse value at the same $\phi$, for bidisperse suspension with $\lambda = 2$ at various volume fractions.
For volume fractions up to $\phi=0.25$, the data are shown in the insets of Fig.~\ref{fig:sed1-t2} and \ref{fig:sed2-t2} for the small and the large species, respectively.
At dilute and moderate $\phi$, increasing $y_1$ increases the ratio $U_{s,\alpha}/U_s$ for both species almost linearly.
For the smaller particles the ratio moves towards unity and for the larger particles away from unity.
The ratio $U_{s,\alpha}/U_s$ exhibits a minimum as $y_1\rightarrow 0$ for the smaller particles and a maximum as $y_1\rightarrow 1$ for the larger particles.
Increasing the total volume fraction $\phi$ reduces the minimum in $U_{s,1}/U_s$ and increases the maximum in $U_{s,2}/U_s$ due to stronger HIs.
When $\phi$ is large enough, the small particle velocity ratio $U_{s,1}/U_s$ can change sign as the backflow from the other species becomes strong enough to reverse the particle motion.
On the other hand, the enhancement of $U_{s,2}/U_s$ for the larger particles as $y_1\rightarrow 1$ is more modest.
In this limit, a large particle sees the small particles and the solvent as an effective medium with a higher viscosity, leading to the sedimentation velocity enhancement relative to the monodisperse case.
The PA approximations, shown as the dashed lines in Fig.~\ref{fig:usedt-2}, capture the effect of $y_1$ on $U_{s,\alpha}/U_s$ only up to $\phi = 0.06$ and then overestimate the effect of suspension composition.
On the other hand, the semi-empirical expression of Davis \& Gecol~\cite{sed-func-empirical_davis_jaiche1994} works up to $\phi=0.25$ for the smaller particles and $\phi = 0.1$ for the larger particles.

At higher $\phi$, the sedimentation behaviors are different from the dilute limit.
For example, in Fig.~\ref{fig:sed1-t2}, the ratio $U_{s,1}/U_s$ in the dilute limit $y_1\rightarrow 0$ \emph{increases} with increasing $\phi$ when $\phi \geq 0.5$.
When $\phi \geq 0.6$, $U_{s,1}/U_s$ is no longer monotonic in $y_1$, and exceeds unity for small $y_1$. 
In Fig.~\ref{fig:sed2-t2}, the $y_1\rightarrow 1$ limit of $U_{s,2}/U_s$ exhibits a trend opposite to dilute suspensions, and decreases with increasing $\phi$. 
At $\phi\geq 0.6$, the ratio $U_{s,2}/U_s$ becomes less than $1$ and also exhibits non-linear behaviors with respect to $y_1$, most likely due to changes in the suspension packing.

\subsection{Hydrodynamic functions}
\label{sec:hydr-funct}

\begin{figure*} 
  \begin{center}    
    \includegraphics[width=7in]{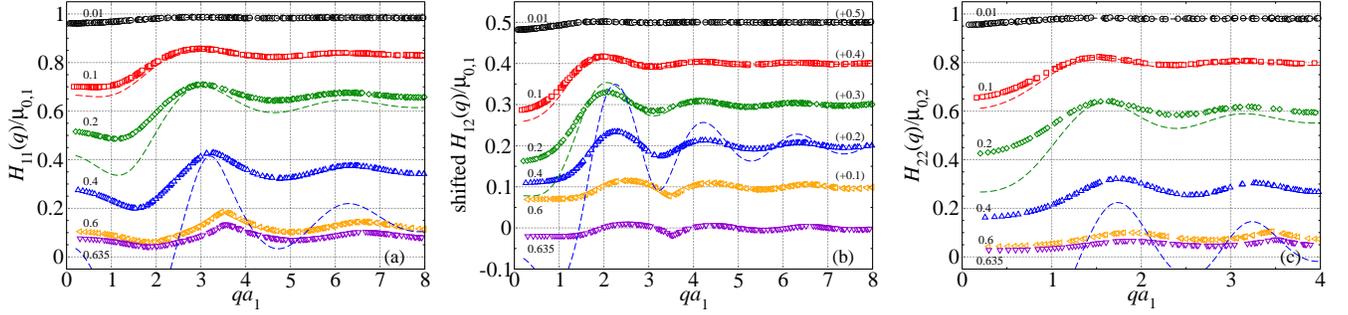}
  \end{center}
  \caption{(Color online)
The partial hydrodynamic functions (a): $H_{11}(q)$, (b): $H_{12}(q)$, and (c): $H_{22}(q)$ for bidisperse suspensions with $\lambda = 2$ and $y_1=0.5$ at $\phi=0.01$ ($\bigcirc$), $0.1$ ($\Box$), $0.2$ ($\diamond$), $0.4$ ($\triangle$), $0.6$ ($\triangleleft$), and $0.635$ ($\triangledown$).  In (b) the interspecies partial hydrodynamic functions $H_{12}(q)$ are shifted by the amount indicated in the parentheses for clarity.
The PA approximations up to $\phi=0.4$ are shown in dashed lines with the same color as the simulation results.
}
  \label{fig:hq-0}
\end{figure*}

The $q$-dependent partial hydrodynamic functions $\Hab(q)$ for bidisperse suspensions with $\lambda = 2$ at various $\phi$ are presented in Fig.~\ref{fig:hq-0}.
The interspecies partial hydrodynamic function $H_{12}(q)$ in Fig.~\ref{fig:hq-0}b are shifted for clarity.
Physically, the partial hydrodynamic function $\Hab(q)$ corresponds to the wave space component of a generalized sedimentation velocity of species $\alpha$ in response to a spatially periodic external force on species $\beta$.
Therefore, at small $q$, the species hydrodynamic functions $H_{11}(q)$ and $H_{22}(q)$ are always positive since the other species remains force-free.
This interpretation also explains the negative interspecies $H_{12}(q)$ at small $q$: the external force on species $2$ generates a backflow that moves the force-free species $1$ in an opposite direction.

The species hydrodynamic functions $H_{11}(q)$ and $H_{22}(q)$, as shown in Fig.~\ref{fig:hq-0}a and \ref{fig:hq-0}c respectively, are always less than unity for all $q$ and decrease with increasing $\phi$.
At $\phi = 0.01$ and $0.1$, $H_{11}(q)$ and $H_{22}(q)$ are similar to each other for the scaled wavenumber $qa_\alpha$.
At higher $\phi$, $H_{11}(q)$ exhibits a minimum ahead of the dominant peak at the wavenumber corresponding to the principal peak of $H_{22}(q)$.
The modulations of $H_{11}(q)$ and $H_{22}(q)$ are the strongest at moderate $\phi$, where the HIs are the most sensitive to the suspension structures.
At $\phi\geq 0.6$, the magnitude and the $q$-modulations of $H_{\alpha\alpha}(q)$ become small.
Therefore, for very dense suspensions, the HIs are mean-field-like and are insensitive to different length scales.
Note that the peak of $H_{11}(q)$ develops to a cusp-shape at $qa_1 \approx 3.5$, most likely due to the packing of particles.

The interspecies hydrodynamic functions $H_{12}(q)$, shown in Fig.~\ref{fig:hq-0}b, exhibit the most significant modulation at moderate volume fractions between $\phi = 0.1$ and $0.4$.  
Comparing to that of $H_{11}(q)$ and $H_{22}(q)$, however, the modulation is relatively weak.
When $\phi\geq 0.6$, $H_{12}(q)$ becomes almost constant in $q$.

The PA approximations of $\Hab(q)$, shown as dashed lines in respective colors in Fig.~\ref{fig:hq-0}, capture the SD results satisfactorily up to $\phi = 0.1$.  
The largest difference between the PA approximation and the SD results is in the low $q$ limit.
At $\phi = 0.2$, the PA approximations capture the shape of $\Hab(q)$, but are quantitatively inaccurate.
The method completely fails at $\phi = 0.4$, where the estimated $\Hab(q)$ becomes negative and exhibits too much modulations.
Note that, for $H_{11}(q)$ at $\phi = 0.4$, the peak values from the PA approximation coincide the SD results.

\begin{figure} 
  \begin{center}
    \includegraphics[width=3in]{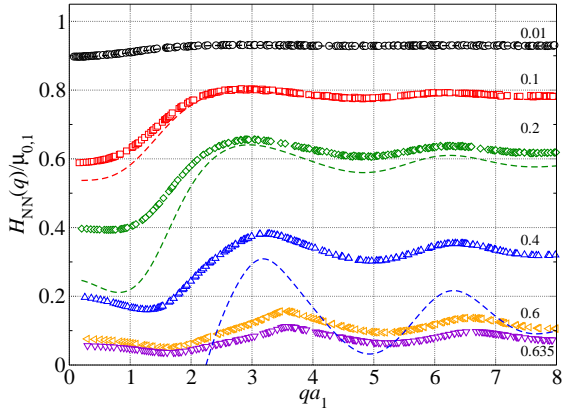}
  \end{center}
  \caption{
(Color online) The number-number mixture hydrodynamic functions $H_{NN}(q)$ for bidisperse suspensions with $\lambda = 2$ and $y_1=0.5$ at $\phi=0.01$ ($\bigcirc$), $0.1$ ($\Box$), $0.2$ ($\diamond$), $0.4$ ($\triangle$), $0.6$ ($\triangleleft$), and $0.635$ ($\triangledown$).  The PA approximations up to $\phi=0.4$ are shown in dashed lines with the same color as the simulation results.
}
  \label{fig:hqnn-l2}
\end{figure}

Fig.~\ref{fig:hqnn-l2} presents the number-number mixture hydrodynamic function $H_{NN}(q)$ constructed from $\Hab(q)$ in Fig.~\ref{fig:hq-0}.
The corresponding PA approximations up to $\phi = 0.4$ are also shown in the respective colors, and exhibit a similar degree of agreement as in Fig.~\ref{fig:hq-0}.
Note that $H_{NN}(q)$ is the simplest form of the mixture hydrodynamic function $H_{M}(q)$, and treats the bidisperse suspension as a single entity with equal and constant scattering intensities for both species.
Evidently, $H_{NN}(q)$ is strongly affected by $H_{11}(q)$ since the number composition corresponding to $y_1 = 0.5$ is $x_1 = 0.889$.
The mobility of the most mobile structures in the suspension corresponds to the principal peak of $H_{NN}(q)$, and the respective wavenumber $q_m$ identifies the length scale of such structure.
In Fig.~\ref{fig:hqnn-l2}, the length scale corresponding to $q_m$,  $\ell_m\sim 2\pi/q_m$, approximately reflects the average spacing between neighboring small particles in the mixture.
It suggests that the collective particle motion on the length scale of the nearest neighbor cage experiences the least hydrodynamic resistance.
The wavenumber $q_m$ increases with $\phi$, suggesting the cage shrinks.
Moreover, a minimum appears ahead of $H_{NN}(q)$ principal peak when $\phi>0.2$.  Such a minimum is a unique feature of polydisperse mixture hydrodynamic functions.

\begin{figure*} 
  \begin{center}    
    \includegraphics[width=7in]{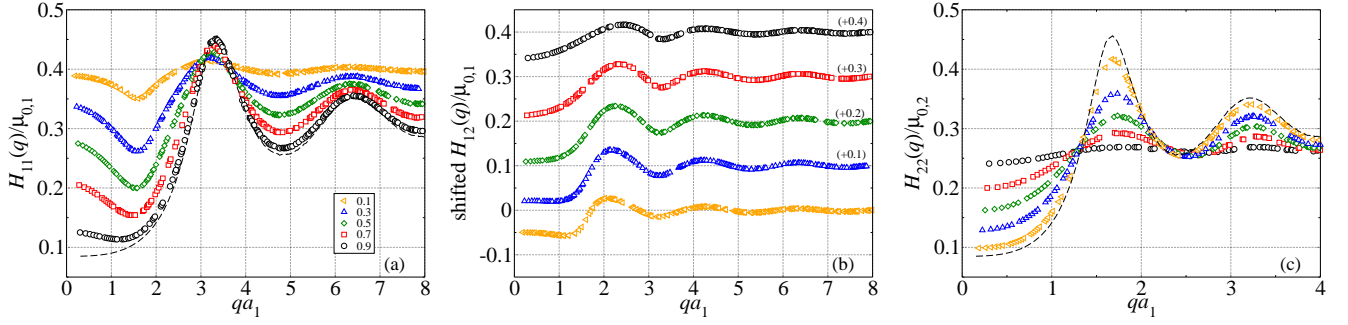}
  \end{center}
  \caption{
(Color online)
The partial hydrodynamic functions (a): $H_{11}(q)$, (b): $H_{12}(q)$, and (c): $H_{22}(q)$ for bidisperse suspensions with $\lambda = 2$ and $\phi=0.4$ at different $y_1$, with the legends shown in (a). 
The monodisperse results are shown in dashed lines.  
In (b) the interspecies partial hydrodynamic functions $H_{12}(q)$ are shifted by the amount indicate in the parentheses for clarity.
}
  \label{fig:hq-1}
\end{figure*}

Fig.~\ref{fig:hq-1} illustrates the influence of $y_1$ on $\Hab(q)$ for bidisperse suspensions at $\phi = 0.4$ and $\lambda = 2$. 
Note that the corresponding monodisperse hydrodynamic functions are presented as dashed lines in Fig.~\ref{fig:hq-1}a and \ref{fig:hq-1}c, and $H_{12}(q)$ is shifted for different $y_1$ in Fig.~\ref{fig:hq-1}b.
When present in small quantity, $H_{11}(q)$ at $y_1=0.1$ is distinct from $H_{22}(q)$ at $y_1 = 0.9$ in several aspects.
First, the average magnitude of $H_{11}(q)/\mu_{0,1}$ is almost $60\%$ higher than that of $H_{22}(q)/\mu_{0,2}$, suggesting a higher intrinsic mobility of the smaller particles.
Meanwhile, the modulation of $H_{11}(q)/\mu_{0,1}$ is stronger: $H_{11}(q)$ at $y_1 = 0.1$ exhibits distinct maximum and minimum with respect to $q$, but $H_{22}(q)$ at $y_2 = 0.9$ is almost flat.
Therefore, the smaller particles are sensitive to the local suspension environment, while the larger ones experience mean-field-like HIs.
The transition of $H_{\alpha\alpha}(q)$ towards the monodisperse $H(q)$ also illustrates the distinct HIs for the small and the large particles.
In essence, the large wavenumber limit of $H_{11}(q)$ reduces with increasing $y_1$, but the limiting value of $H_{22}(q)$ grows with decreasing $y_1$, \ie, increased presence of the larger particles.
For the interspecies partial hydrodynamic function $H_{12}(q)$, the modulation reaches a maximum at $y_1 = 0.5$, but the magnitude of the modulation remains small compared to $H_{\alpha\alpha}(q)$.

\begin{figure} 
  \begin{center}
    \includegraphics[width=3in]{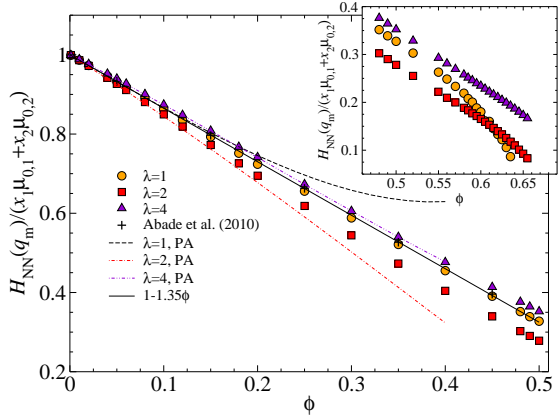}
  \end{center}
  \caption{(Color online)
The peak value of the rescaled number-number mixture hydrodynamic function $H_{NN}(q_m)/(x_1\mu_{0,1} + x_2 \mu_{0,2})$ as a function of $\phi$ for bidisperse suspensions with $\lambda = 1$, $2$, and $4$ at composition $y_1 = 0.5$.  
The monodisperse simulation results from Abade \etal \cite{Abade2010} and the analytical fitting of Banchio \& N\"{a}gele \cite{trans-prop-asd_banchio_jcp2008} are also presented.
The PA approximations are shown in dashed line for $\lambda=1$, dash-dotted line for $\lambda=2$, and dash-double-dotted line for $\lambda=4$.  
The insets show the results at higher $\phi$.
}
  \label{fig:hqm-1}
\end{figure}

The principal peaks of $H_{NN}(q_m)$ as a function of $\phi$ for bidisperse suspensions with $\lambda = 2$ and $4$ and $y_1 = 0.5$, as well as for monodisperse suspensions, are shown in Fig.~\ref{fig:hqm-1}.
The peak wavenumber $q_m$ is directly measured from the computed $H_{NN}(q)$, and using $q_m$ corresponding to the principal peak of the number-number static structure factor $S_{NN}(q)$ of the PY closure yields virtually the same results.
Note that we scale the results with $(x_1\mu_{0,1} + x_2\mu_{0,2})$ for proper dilute behaviors.
The monodisperse SD results agree well with the computations of Abade \etal~\cite{Abade2010}, also presented in Fig.~\ref{fig:hqm-1}.
For $\phi<0.5$, the monodisperse data are well described by the linear expression of Eq.~(\ref{eq:banchio-hm})~\cite{trans-prop-asd_banchio_jcp2008}.
For bidisperse suspensions, the $\phi$ evolution of the principal peak value $H_{NN}(q_m)$ follows closely the monodisperse results, with the data for $\lambda = 2$ below and the data for $\lambda = 4$ above, and is also almost linear.
The PA approximations exhibit varying degrees of agreement with the SD computations: they are valid up to $\phi = 0.15$ for $\lambda \leq 2$, and show exceptional agreement up to $\phi = 0.4$ for $\lambda = 4$.
This agreement, however, is incidental and similar to the peak value agreement observed in Fig.~\ref{fig:hq-0}a for $H_{11}(q)$.
For very dense suspensions ($\phi>0.45$) shown in the inset of Fig.~\ref{fig:hqm-1}, the peak value drops drastically near close packing, and the $\lambda = 2$ data cross the monodisperse results at $\phi \approx 0.61$ due to changes in the suspension packing structure.

\begin{figure} 
  \begin{center}
    \includegraphics[width=3in]{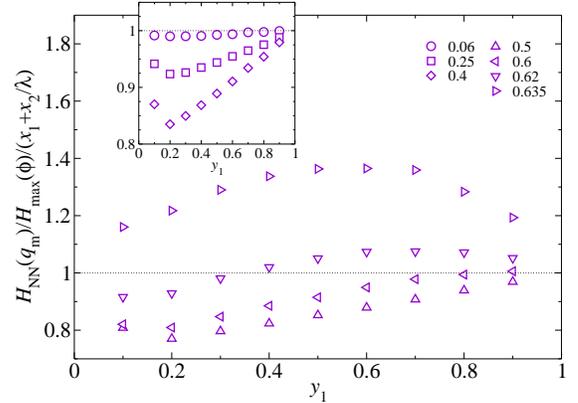}
  \end{center}
  \caption{
The normalized number-number mixture hydrodynamic function peaks $[H_{NN}(q_m)/H_{\max}] / (x_1 + x_2 \lambda^{-1})$ as a function of $y_1$ at different $\phi$ for bidisperse suspensions with $\lambda = 2$.  The $q_m$ is directly measured from the simulations.  The monodisperse peak values at the corresponding $\phi$ is $H_{\max}$.
}
  \label{fig:hqm-2}
\end{figure}

\begin{figure} 
  \begin{center}
    \includegraphics[width=3in]{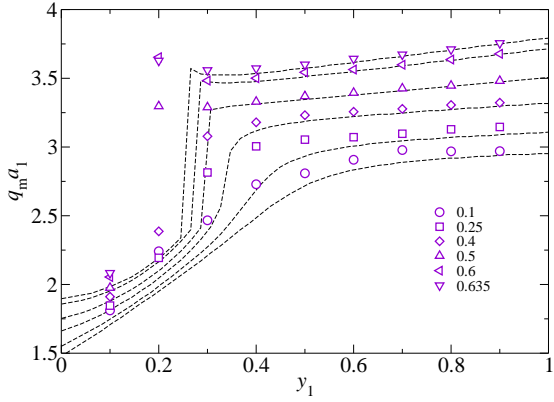}
  \end{center}
  \caption{
The scaled wavenumber $q_ma_1$ corresponding to the maximum of $H_{NN}(q)$ measured from the simulations (symbols) as a function of $y_1$ at different $\phi$ for bidisperse suspensions with $\lambda = 2$.  The $q_m$ from the PY number-number mixture static structure factor $S_{NN}(q)$ are shown in dashed lines, with ascending $\phi$ indicated in the legend from bottom to top.
}
  \label{fig:qm}
\end{figure}

The effects of the composition $y_1$ on the normalized peak of the hydrodynamic function are shown in Fig. \ref{fig:hqm-2} at various $\phi$ for bidisperse suspensions with $\lambda = 2$.
The peak values and the corresponding wavenumbers are directly measured from the computed $H_{NN}(q)$, and the scaling $H_{NN}(q_m)/H_{\max}/(x_1 +x_2\lambda^{-1})$ ensures the ratio goes to $1$ as $y_1 \rightarrow 0$ or $1$.
The inset of Fig.~\ref{fig:hqm-2} shows normalized peaks for $\phi \leq 0.4$.
In this range, the presence of the second species always reduces the peak value relative to the monodisperse suspensions, and the reduction increases with increasing $\phi$, \eg, at $\phi=0.4$ the maximum reduction is $20\%$.
The corresponding composition is $y_1 = 0.2$, suggesting an asymmetric influence on $H_{NN}(q)$ for different species.
For more dense suspensions, shown in the main figure of Fig.~\ref{fig:hqm-2}, increasing the volume fraction $\phi$ lessens the extent of the normalized peak reduction, and when $\phi \geq 0.6$, introducing a second species into the suspension can bring the normalized peak beyond unity.
Here, the particle packing at different $y_1$ clearly plays a vital role in the behavior of the mixture hydrodynamic function principal peak.

The wavenumbers $q_m$ corresponding to the principal peak of $H_{NN}(q)$ in Fig.~\ref{fig:hqm-2} are shown in Fig.~\ref{fig:qm}.
Also plotted as dashed lines are the wavenumbers of the principal peak of $S_{NN}(q)$ from the PY closure.
For very dense suspensions up to $\phi = 0.635$, we have verified that the bidisperse PY static structure factor $S_{NN}(q)$ adequately describes suspension structures at finite wavenumbers~\cite{dense-colloids-scattering_mason_jpcm2009}.
For monodisperse suspensions, the maximum of $H(q)$ is practically at the maximum of the static structure factor~\cite{trans-prop-asd_banchio_jcp2008}.
However, as shown in Fig.~\ref{fig:qm}, this is not the always the case for bidisperse suspensions.
For $y_1$ close to $1$, $q_m$ for the principal peak of $H_{NN}(q)$ and $S_{NN}(q)$ indeed coincide.
However, with decreasing $y_1$, the peak location $q_m$ for $H_{NN}(q)$ and $S_{NN}(q)$ begin to deviate from each other, and the most significant difference is found at $y_1 = 0.2$ at high $\phi$.
Here, the $S_{NN}(q)$ peak corresponds to the mean distance between large particles, while the $H_{NN}(q)$ peak corresponds to the mean distance between small particles.
The decoupling of the hydrodynamic and structural descriptions of dense suspensions illustrates the care needed when treating the HIs of dense mixtures.

\subsection{High-frequency dynamic shear viscosity}
\label{sec:shear-viscosity}

\begin{figure}
  \begin{center}
    \includegraphics[width=3in]{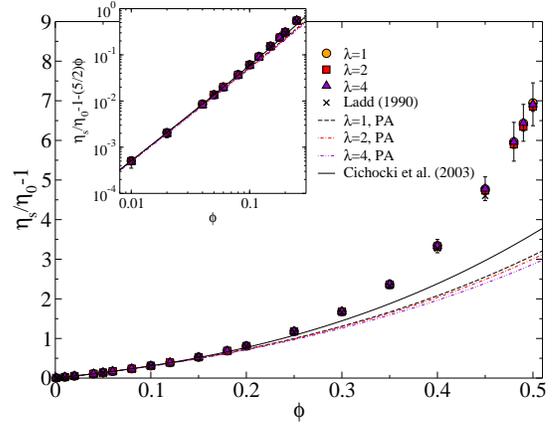}
\end{center}
  \caption{(Color online)
The particle shear viscosity $\eta_s/\eta_0 -1$ as a function of $\phi$, up to $\phi=0.5$, for bidisperse suspensions with $y_1 = 0.5$ and $\lambda = 1$, $2$, and $4$.
The monodisperse simulation results from Ladd~\cite{hydro-trans-coeff_ladd_jcp1990} and the analytical result of Cichocki \etal \cite{effect-viscosity-phi-cube_cichocki_jcp2003} are also shown.
The PA approximations for $\lambda=1$ (dashed), $2$ (dash-dotted), and $4$ (dash-double-dotted) are also presented.
The inset shows the interaction contribution to the suspension shear viscosity, $\eta_s/\eta_0 - 1 - \tfrac{5}{2}\phi$, in the dilute limit.
}
  \label{fig:eta-11}
\end{figure}

The high-frequency dynamic shear viscosities $\eta_s$ for volume fractions up to $\phi = 0.5$ of bidisperse suspensions with $\lambda =2$ and $4$ and $y_1 = 0.5$, as well as monodisperse suspensions, are shown in Fig.~\ref{fig:eta-11}.
The monodisperse SD results exhibit excellent agreement with the computations of Ladd \cite{hydro-trans-coeff_ladd_jcp1990}, also shown in the figure.
The analytical expression of Cichocki \etal~\cite{effect-viscosity-phi-cube_cichocki_jcp2003}, Eq.~(\ref{eq:cichocki-eta}), is valid up to $\phi = 0.25$.
The bidisperse $\eta_s$ closely follows the monodisperse results, and is almost indistinguishable from the monodisperse results until $\phi>0.45$.
The weak size dependence of $\eta_s$ is also evident from the weak $\lba$ dependence of $I^\eta_{\alpha\beta}$ in Table~\ref{tab:pa-coeff}.
The PA approximations with proper suspension structures, also shown in Fig.~\ref{fig:eta-11}, exhibit very weak $\lambda$ dependence for $\lambda<4$, and agree with the SD computations up to $\phi = 0.2$.
The inset of Fig.~\ref{fig:eta-11} examines the pairwise HI contributions to the high-frequency dynamic shear viscosity, $\eta_s/\eta_0 - 1 -\tfrac{5}{2}\phi$, in the dilute limit.
Here, the SD results closely follow the PA approximations, and grow $\sim \phi^2$ when $\phi\ll 1$.

\begin{figure}
  \begin{center}
    \includegraphics[width=3in]{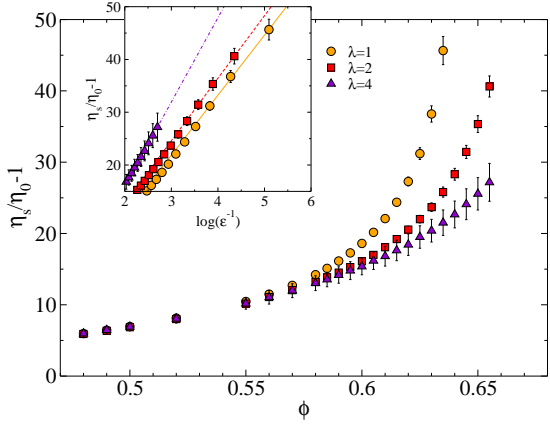}
  \end{center}
  \caption{(Color online)
The particle shear viscosity $\eta_s/\eta_0 -1$ as a function of $\phi$ for very dense bidisperse suspensions with $y_1 = 0.5$ and $\lambda = 1$, $2$, and $4$.
The inset shows the logarithmic shear viscosity divergence, with $\varepsilon=1-\phi/\phi_m$.  Also presented in lines are the asymptotic behaviors, Eq.~(\ref{eq:eta-div}), with fitted constants from Table~\ref{tab:phim} for $\lambda = 1$(solid), $2$(dashed), and $4$(dash-dotted).
}
  \label{fig:eta-12}
\end{figure}

\begin{table}[tp]
  \caption{The limiting volume fraction $\phi_m$ in $\varepsilon=1-\phi/\phi_m$, and the constants in Eq. (\ref{eq:eta-div}) and (\ref{eq:kappa-div}), characterizing the asymptotic $\eta_s$ and $\kappa_s$ divergences, respectively, fitted from the SD computations for bidisperse suspensions with $y_1= 0.5$.}
  \label{tab:phim}
  \centering
  \begin{tabularx}{0.9\columnwidth}{*{6}{Y}}
    \hline\hline
    $\lambda$ & $\phi_m$ & $B_\eta$ & $-C_\eta$ & $B_\kappa$ & $-C_\kappa$ \\
    \hline 
    $1$             & $0.639$   & $11.38$ & $12.37$ & $32.69$ & $71.86$ \\
    $2$             & $0.664$   & $11.84$ & $10.93$ & $31.29$ & $61.46$ \\
    $4$             & $0.702$   & $15.98$ & $16.00$ & $-$   & $-$ \\
    \hline\hline
  \end{tabularx}
\end{table}

The results for dense suspensions with $\phi>0.45$ are shown in Fig.~\ref{fig:eta-12}.
When $\phi>0.55$, the viscosity $\eta_s$ increases drastically, and increasing $\lambda$ reduces $\eta_s$ significantly.
As revealed by experiments \cite{exp-bimodal-visc_shikata_jor1998, exp-bimodal-latex-visc_wolfe_lang1992, bimodal-colloids-viscosity_chong_japs1971} and simulations \cite{sd-bimodal-mc_powell_pof1994}, the viscosity reduction is primarily due to the improved packing for polydisperse suspensions, \ie, the average particle spacing increases with $\lambda$, leading to a viscosity reduction.
The divergent behavior of $\eta_s$ is well represented by the asymptotic expression~\cite{asd_sierou_jfm01},
\begin{equation}
\label{eq:eta-div}
\frac{\eta_s}{\eta_0} \approx B_\eta \log(\varepsilon^{-1})  + C_\eta + \cdots,  (\varepsilon\ll 1),
\end{equation}
where $B_\eta$ and $C_\eta$ are constants, and $\varepsilon= 1 - \phi/\phi_m$, with $\phi_m$ the limiting volume fraction.
The parameter $\varepsilon$ characterizes the mean interparticle gap spacing relative to the particle size.
Note that $B_\eta$, $C_\eta$, and $\phi_m$ depend on the bidisperse suspension composition~\cite{colloidal-crystal_hofman_pre2000, asd_sierou_jfm01}, and the fitted values from the SD computations are shown in Table~\ref{tab:phim}.
The inset of Fig.~\ref{fig:eta-12} shows the $\eta_s$ asymptotic behaviors based on Eq.~(\ref{eq:eta-div}), and that the SD results and the fitted expression agree well.
However, the numerical values of $B_\eta$ and $C_\eta$ for monodisperse suspensions differ from earlier ASD results~\cite{asd_sierou_jfm01}.
This is likely because the asymptotic behaviors near close packing are very sensitive to the suspension structures, and any differences in the packing generation protocol, or even different parameters within the same protocol, can lead to quantitative differences.
However, the asymptotic form suggested by Eq.~(\ref{eq:eta-div}) remains valid.

\begin{figure}
  \begin{center}
    \includegraphics[width=3in]{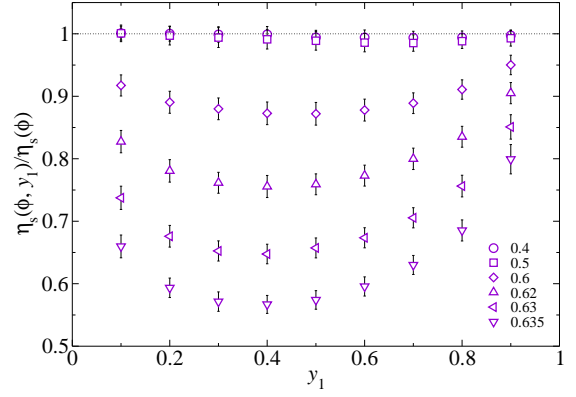}
  \end{center}
  \caption{
The normalized high-frequency dynamic shear viscosity $\eta_s(\phi, y_1)/\eta_s(\phi)$ as a function of $y_1$ at different $\phi$ for bidisperse suspensions with $\lambda = 2$.  The monodisperse high-frequency dynamic shear viscosity at the corresponding $\phi$ is $\eta_s(\phi)$.  
}
  \label{fig:eta-2}
\end{figure}

The effects of composition $y_1$ on the normalized shear viscosity $\eta_s(\phi, y_1)/\eta_s(\phi)$, where $\eta_s(\phi)$ is the monodisperse shear viscosity at the same $\phi$, are shown in Fig.~\ref{fig:eta-2} for bidisperse suspensions at $\lambda = 2$.
For the moderate $\lambda$ studied here, the effect of size ratio is not apparent until $\phi = 0.4$.
At higher $\phi$ near the monodisperse close packing, the presence of a second species with a different size leads to significant viscosity reduction.
Moreover, the normalized shear viscosity in Fig.~\ref{fig:eta-2} is not symmetric for $y_1$: the smaller particles are more effective at viscosity reduction.
For example, at $\phi = 0.635$, at $y_1 = 0.1$ and $0.9$ the viscosity is $66\%$ and $80\%$ of the monodisperse value, respectively.

\subsection{High-frequency dynamic bulk viscosity}
\label{sec:bulk-viscosity}

\begin{figure}
  \begin{center}
    \includegraphics[width=3in]{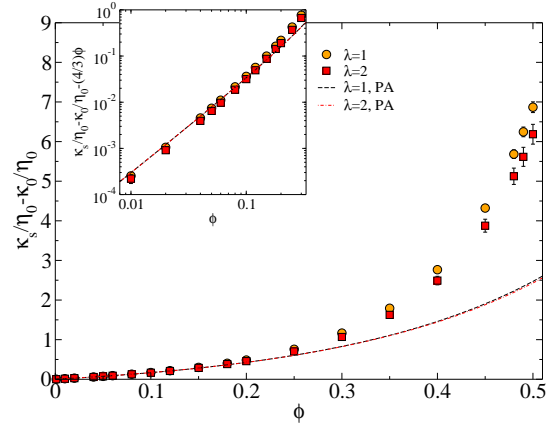}
\end{center}
  \caption{(Color online)
The particle bulk viscosity $(\kappa_s-\kappa_0)/\eta_0$ as a function of $\phi$, up to $\phi=0.5$, for bidisperse suspensions with $y_1 = 0.5$ and $\lambda = 1$ and $2$.
The PA approximations for $\lambda=1$ (dashed) and $2$ (dash-dotted) are also presented.
The inset shows the interaction contribution to the suspension bulk viscosity, $(\kappa_s-\kappa_0)/\eta_0 - \tfrac{4}{3}\phi$, in the dilute limit.
}
  \label{fig:kappa-11}
\end{figure}

Fig.~\ref{fig:kappa-11} presents the the high-frequency dynamic bulk viscosity $\kappa_s$ as a function of $\phi$ for monodisperse and bidisperse suspensions with $\lambda = 2$ and $y_1 = 0.5$.
Note that the level of approximation in SD is insufficient for the bulk viscosity at large size ratios, and therefore the results for $\lambda = 4$ are not shown.
For monodisperse suspensions, the SD results agree with earlier studies~\cite{swaroopthesis2010}.
The bulk viscosity of bidisperse suspensions are slightly smaller than the monodisperse values.
The particle size ratio $\lambda$ weakly affects $\kappa_s$, but the influence is stronger compared to $\eta_s$.
At $\phi = 0.3$, differences in $\lambda$ can be found between $\lambda = 2$ and $1$, while for $\eta_s$, this is not apparent until $\phi = 0.45$.
The PA approximations, also presented in Fig.~\ref{fig:kappa-11},
show little size dependence, as also indicated in Table~\ref{tab:pa-coeff}, and agree with the SD computations up to $\phi = 0.2$.
The inset of Fig.~\ref{fig:kappa-11} presents the dilute behaviors of pairwise HI contribution to the bulk viscosity, $(\kappa_s-\kappa_0)/\eta_0 - \tfrac{4}{3}\phi$.
The results show quadratic growth with $\phi$, and agree well with the PA approximations.

\begin{figure}
  \begin{center}
    \includegraphics[width=3in]{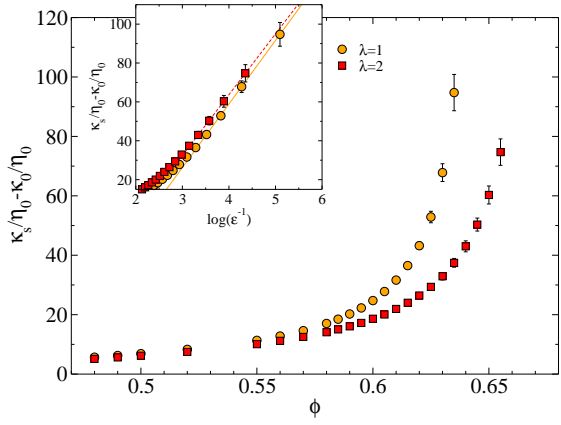}
  \end{center}
  \caption{(Color online)
The particle bulk viscosity $(\kappa_s-\kappa_0)/\eta_0$ as a function of $\phi$ for very dense bidisperse suspensions with $y_1 = 0.5$ and $\lambda = 1$ and $2$.
The inset shows the logarithmic bulk viscosity divergence, with $\varepsilon=1-\phi/\phi_m$.  Also presented in lines are the asymptotic behaviors, Eq.~(\ref{eq:kappa-div}), with fitted constants from Table~\ref{tab:phim} for $\lambda = 1$(solid) and $2$(dashed).
}
  \label{fig:kappa-12}
\end{figure}

The results of $\kappa_s$ for $\phi>0.45$ are presented in Fig.~\ref{fig:kappa-12}.
At $\phi>0.55$, significant differences emerge between the monodisperse and bidisperse results.
To identify the divergent behavior of $\kappa_s$, we fitted the SD results for dense suspensions ($\phi>0.6$) with asymptotic terms $\varepsilon^{-1}$ and $\log(\varepsilon^{-1})$ using the same $\phi_m$ in Table~\ref{tab:phim} for $\varepsilon$, since both $\eta_s$ and $\kappa_s$ are computed from the same configurations.
It was consistently found that the coefficient of the $\varepsilon^{-1}$ terms are orders of magnitudes smaller than those of the $\log(\varepsilon^{-1})$ terms.
Therefore, we conclude from the SD data that the $\kappa_s$ divergence is best described by
\begin{equation}
  \label{eq:kappa-div}
\frac{\kappa_s}{\eta_0}-\frac{\kappa_0}{\eta_0} \approx B_\kappa \log(\varepsilon^{-1})  + C_\kappa + \cdots,  (\varepsilon\ll 1),
\end{equation}
where the constants $B_\kappa$ and $C_\kappa$ are functions of the suspension compositions and packing generation protocol, and their fitted values are presented in Table~\ref{tab:phim}.   Eq.~(\ref{eq:kappa-div}) and the SD data agree well, as shown in the inset of Fig.~\ref{fig:kappa-12}.
The weak logarithmic divergence of $\kappa_s$ first appears odd given the inverse gap spacing ($\sim \xi^{-1}$) divergence of the hydrodynamic function $T^Q$~\cite{compres-res_khair_pof2006}.
In the $\varepsilon\ll 1$ limit, the HIs are dominated by the lubrication forces, and $\eta_s$ and $\kappa_s$ can be estimated from the HIs between nearest neighbors with appropriate geometric information~\cite{acrivos1967,sphere-visc_denn_rheoacta_1985, lub-model-susp_jongschaap_jstatphys_1991}.
This approach is particularly useful for estimating the divergence behavior of colloidal lattices~\cite{visc-periodic-suspension_nunan_jfm_1984,colloidal-crystal_hofman_pre2000}.
For random suspensions, however, such divergence behavior also depends on the geometric statistics such as the nearest neighbor gap spacing distribution $P(\xi)\dd \xi$~\cite{eff-visc-lub_patlazhan_physA_1993}.
If the probability density function $P(\xi)$ is somewhat uniformly distributed~\cite{pair-corr-jammed-packing_donev_pre_2005}, with a lower bound proportional to $\varepsilon$, properties dominated by $\xi^{-1}$ HIs can show logarithm asymptotic behavior since $\int_\varepsilon \xi^{-1} P(\xi)\dd \xi \sim \log(\varepsilon^{-1})$.
This simple argument provides an explanation of the logarithm divergence of $\kappa_s$ despite the $\xi^{-1}$ divergence of $T^Q$.
The same argument also explains the logarithm divergence of the high-frequency dynamic shear viscosity $\eta_s$, shown in Eq.~(\ref{eq:eta-div}) and in Fig.~\ref{fig:eta-12}, since for two nearly touching spheres, $\eta_s$ is dominated by the two-body resistance function $X^M\sim \xi^{-1}$.
We defer the formal study involving structural analysis of hard-sphere packings to a future work.

\begin{figure}
  \begin{center}
    \includegraphics[width=3in]{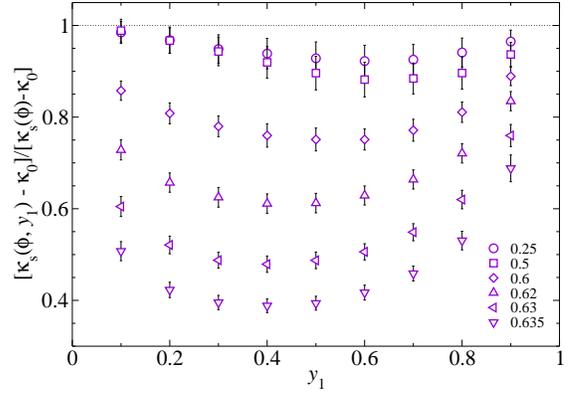}
  \end{center}
  \caption{
The normalized high-frequency dynamic bulk viscosity $[\kappa_s(\phi, y_1)-\kappa_0]/[\kappa_s(\phi) - \kappa_0]$ as a function of $y_1$ at different $\phi$ for bidisperse suspensions with $\lambda = 2$.  The monodisperse high-frequency dynamic bulk viscosity at the corresponding $\phi$ is $\kappa_s(\phi)$.  
}
  \label{fig:kappa-2}
\end{figure}

Fig.~\ref{fig:kappa-2} shows the influence of composition $y_1$ on the ratio $[\kappa_s(\phi, y_1)-\kappa_0]/[\kappa_s(\phi)-\kappa_0]$, where $\kappa_s(\phi)$ in the denominator is the monodisperse bulk viscosity at $\phi$, for bidisperse suspensions with $\lambda = 2$.
At moderate volume fraction $\phi = 0.25$, the effect of introducing a differently sized species on  $\kappa_s$ is slight.
At higher volume fraction, particularly near the monodisperse close packing, the bulk viscosity reduces significantly due to the introduction of a second species.
For example, at $\phi = 0.635$, the mixture $\kappa_s$ can be as low as $39\%$ of the monodisperse value at $y_1 = 0.4$.  
The shape of the curve is asymmetric to $y_1 = 0.5$, indicating that the larger and the smaller particles affect $\kappa_s$ differently.

\section{Results for porous media}
\label{sec:results-porous-media}

\subsection{Permeability (mean drag coefficient)}
\label{sec:permeability}




\begin{figure} 
  \begin{center}
    \includegraphics[width=3in]{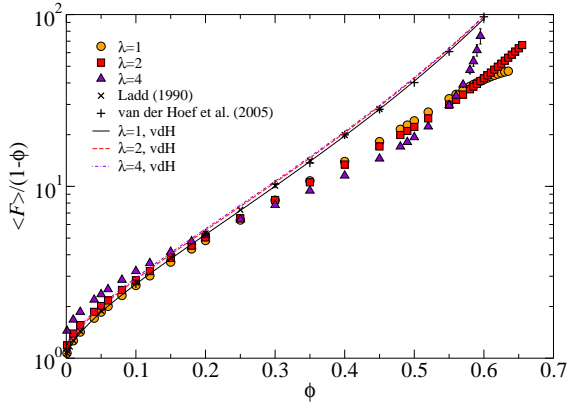}
  \end{center}
  \caption{(Color online)
The mean drag coefficient $\avg{F}/(1-\phi)$ as a function of $\phi$ for bidisperse porous media with $y_1 = 0.5$ and size ratios $\lambda=1$, $2$, and $4$.  The monodisperse simulation results from Ladd~\cite{hydro-trans-coeff_ladd_jcp1990} and van der Hoef \etal \cite{bidisperse-drag-permeability_kuipers_jfm2005} are also shown.  The semi-empirical correlations~\cite{bidisperse-drag-permeability_kuipers_jfm2005}, Eq.~(\ref{eq:kuipers-f}) and (\ref{eq:kuipers-poly}), are also presented for comparison.
}
  \label{fig:permt-1}
\end{figure}

The permeability, presented in terms of the mean particle drag coefficient $\avg{F}$ in Eq.~(\ref{eq:avgf-def}), is shown in Fig.~\ref{fig:permt-1} for bidisperse porous media of $\lambda = 2$ and $4$ and $y_1 = 0.5$, as well as for monodisperse media.
The monodisperse results of Ladd~\cite{hydro-trans-coeff_ladd_jcp1990} and van der Hoef~\etal~\cite{bidisperse-drag-permeability_kuipers_jfm2005} are also shown in the figure.
Note that near close packing, $\avg{F}$ does not diverge as the fluid can pass through the interstitial spaces between particles.
The SD results agree with earlier studies for $\phi<0.25$, and underestimate $\avg{F}$ at higher $\phi$.  
At $\phi = 0.6$, the drag coefficient from SD is only $40\%$ of the LB computations of van der Hoef \etal~\cite{bidisperse-drag-permeability_kuipers_jfm2005} in Fig.~\ref{fig:permt-1}.
This is because the drag coefficient in porous medium are strongly affected by the many-body HIs; the lubrication interactions only play a small role.
As a result, the computation of $\avg{F}$ relies on the accurate estimation of the grand mobility tensor.
The multipole expansion to the mean-field quadrupole level used in SD is insufficient to capture the HIs between stationary particles, similar to the errors associated with the sedimentation velocity $U_{s,\alpha}$ in Sec.~\ref{sec:short-time-sedim}.

For bidisperse suspensions, SD remains valid for $\phi<0.25$, and at higher $\phi$ it is expected to capture the qualitative aspect of the particle size effects.
Since each stationary particle in a porous medium acts as a force monopole, the particle size plays a relatively minor role.
This is confirmed in Fig.~\ref{fig:permt-1}, where the bidisperse $\avg{F}$ closely follows the monodisperse data.
At low $\phi$, the mean drag coefficient increases slightly with the size ratio $\lambda$.
The behavior for $\phi>0.25$ arises from the complex interplay between the HIs and the particles configurations.

The semi-empirical expressions for the drag coefficient, Eq.~(\ref{eq:kuipers-f}) and (\ref{eq:kuipers-poly}), are also plotted in Fig.~\ref{fig:permt-1}.
They accurately capture the earlier simulation results \cite{hydro-trans-coeff_ladd_jcp1990,bidisperse-drag-permeability_kuipers_jfm2005} up to dense suspensions for monodisperse suspensions.
For bidisperse porous media, comparing to the SD results at low $\phi$, the empirical expressions work well for $\lambda = 2$, but underestimate the size effects for $\lambda = 4$.
This may be because in constructing Eq.~(\ref{eq:kuipers-poly}), van der Hoef \etal~\cite{bidisperse-drag-permeability_kuipers_jfm2005} did not consider the case of $\lambda = 4$ at low to moderate $\phi$ in their simulations.

\begin{figure} 
  \begin{center}
    \includegraphics[width=3in]{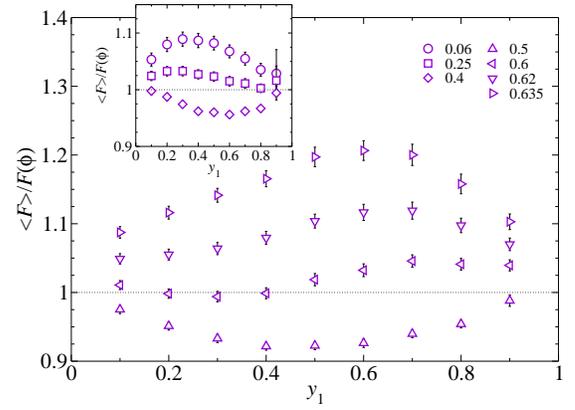}
   \end{center}
  \caption{The normalized mean drag coefficient $\avg{F}/F(\phi)$ as a function of $y_1$ at different  $\phi$ for bidisperse porous media with $\lambda = 2$.  The monodisperse drag coefficient at the corresponding $\phi$ is $F(\phi)$.}
  \label{fig:permt-favg}
\end{figure}

The effects of composition $y_1$ on the drag coefficient ratio $\avg{F}/F(\phi)$, where $F(\phi)$ is the monodisperse drag coefficient, for bidisperse mixtures at $\lambda = 2$, are presented in Fig.~\ref{fig:permt-favg}.
The empirical expressions Eq.~(\ref{eq:kuipers-f}) and (\ref{eq:kuipers-poly}) are not shown because they do not recover to the correct limit when $y_1\rightarrow 0$ or $1$.
Over the wide range of $\phi$ presented, except when $\phi> 0.62$, the mean drag coefficient $\avg{F}$ for the mixture differs from the monodisperse results by at most $10\%$.
Introducing a differently sized second species to a monodisperse porous medium first increases the mean drag coefficient for $\phi < 0.4$, while at higher volume fractions, the second species reduces $\avg{F}$ for $\phi<0.6$ and then increases the mean drag coefficient again near the monodisperse close packing.
At $\phi = 0.635$, $\avg{F}$ is merely $21\%$ higher than the monodisperse drag coefficient $F(\phi)$.
The relative insensitivity of $\avg{F}$ to $y_1$ suggests that the particle size plays a minor role in the permeability of porous media. 
Fig.~\ref{fig:permt-1} and \ref{fig:permt-favg} show that SD remains a useful tool~\cite{porous-permeability-bidisperse_otomo_tpm2013} to assess qualitative aspects of polydisperse porous media.

\subsection{Translational hindered diffusivity}
\label{sec:transl-hind-diff}

 
\begin{figure} 
  \begin{center}
    \includegraphics[width=3in]{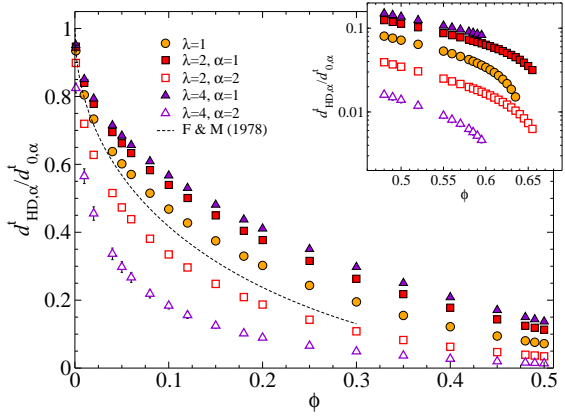}
  \end{center}
  \caption{(Color online)
The translational hindered diffusivities $\Dhdt{\alpha}$, $\alpha\in\{1,2\}$, of both species for bidisperse porous media with $y_1=0.5$ and $\lambda = 1$,  $2$, and $4$.  The result of Freed \& Muthukumar~\cite{hindered-diffusion_muthukumar_jcp1978}, Eq.~(\ref{eq:dhd-fm}), is shown in dashed line.  The inset shows the results at high $\phi$.
}
  \label{fig:dhdt-1}
\end{figure}

Fig.~\ref{fig:dhdt-1} presents the translational hindered diffusivities, $\Dhdta$, as a function of the volume fraction $\phi$ for bidisperse porous media with $y_1 = 0.5$ and $\lambda = 2$ and $4$, as well as for monodisperse porous media.
The self-consistent expression of Eq.~(\ref{eq:dhd-fm})~\cite{hindered-diffusion_muthukumar_jcp1978}, also presented in the figure, agrees with the SD computation for $\phi <0.05$ and underestimate the results at higher $\phi$.
Note that the hindered diffusive properties describe particle relative motions in a stationary matrix, and therefore the lubrication effects are important.

Compared to the suspension short-time translational self-diffusivity $d_{s,\alpha}^t$ in Sec.~\ref{sec:transl-self-diff}, the hindered diffusivity $\Dhdta$ exhibits a stronger $\phi$ and $\lambda$ dependence due to stronger HIs in porous media.
In particular, $\Dhdt{\alpha}$ decreases quickly with $\phi$ with an initial $\sim \sqrt{\phi}$ reduction.
The hindered diffusivity for small particles, $\Dhdt{1}$, exhibits moderate enhancement relative to the monodisperse systems similar to $d_{s,1}^t$.
Moreover, at a fixed $\phi$, the large particle hindered diffusivity $\Dhdt{2}$ reduces appreciably with increasing $\lambda$, in contrast to the $\lambda$-insensitive $d_{s,2}^t$ in suspensions.
The increased sensitivity is simply because the fixed particle matrix exerts much stronger HIs on a mobile particle inside.
For very dense systems shown in Fig.~\ref{fig:dhdt-1} inset, the hindered diffusivities for both species display dramatic reductions at $\phi>0.6$ as the nearby stationary particles get closer, and the reduction is most pronounced near the close packing volume fraction.
Moreover, the large particle $\Dhdt{2}$ approaches the monodisperse value at $\phi \approx 0.63$, suggesting an enhancement of $\Dhdt{2}$ due to more efficient particle packing in bidisperse systems.

\begin{figure} 
  \begin{center}
  \subfloat[]{
    \centering
    \includegraphics[width=3in]{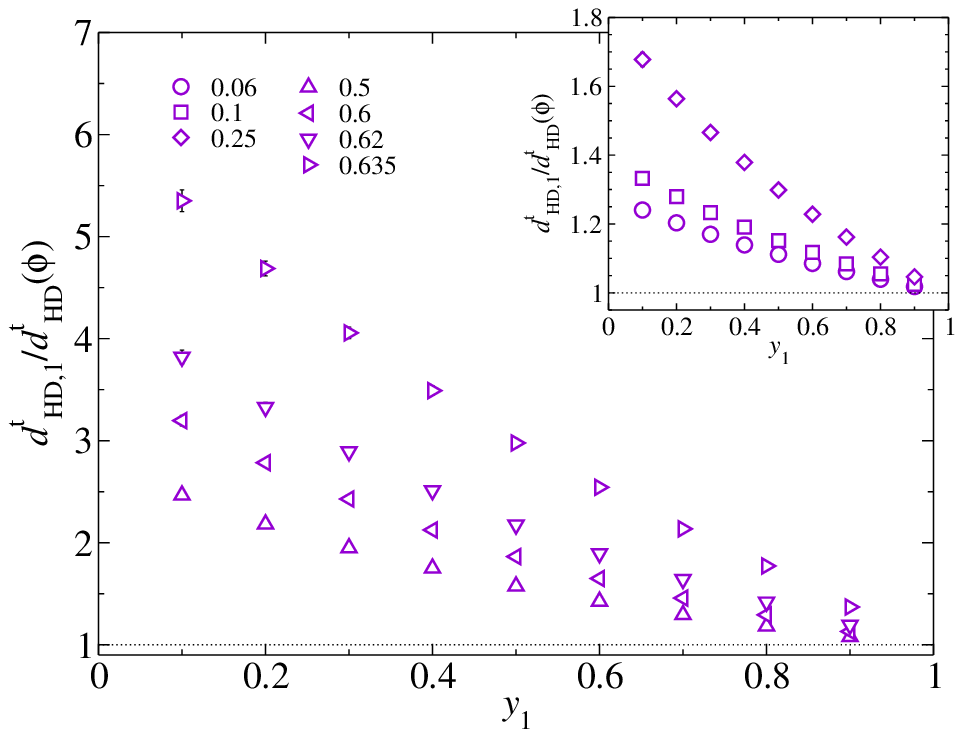}
    \label{fig:dhdt1-t2}
  }\\
  \subfloat[]{
    \centering
    \includegraphics[width=3in]{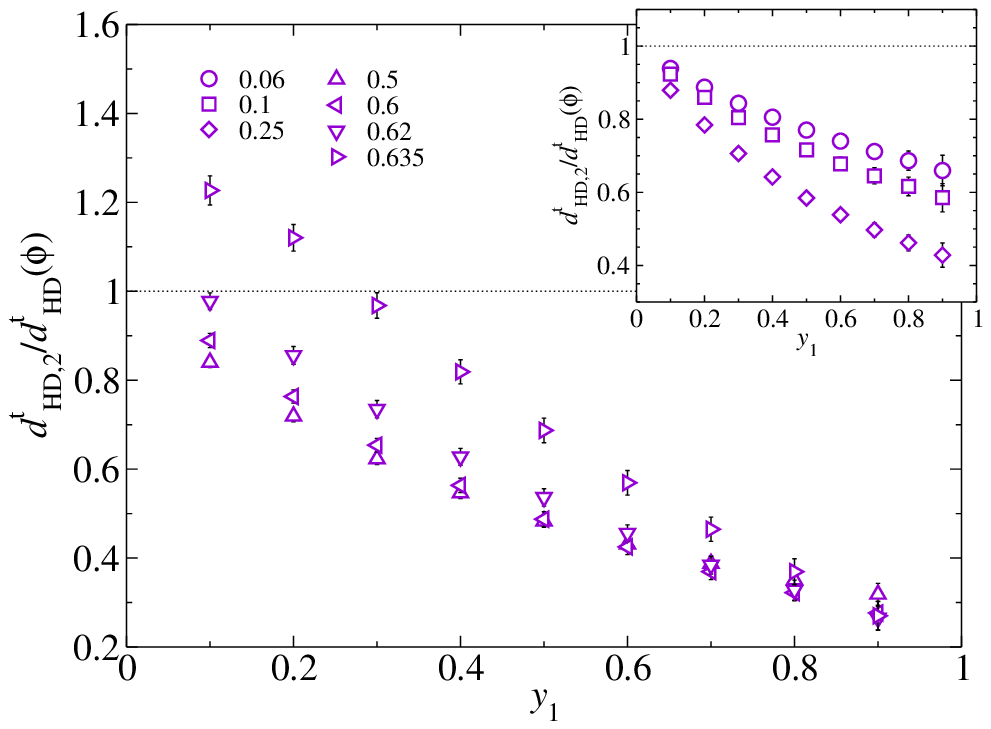}
    \label{fig:dhdt2-t2}
  }
  \end{center}
  \caption{
The normalized translational hindered diffusivities \protect\subref{fig:dhdt1-t2}: $\Dhdt{1}/d_{\mathrm{HD}}^t$ and \protect\subref{fig:dhdt2-t2}: $\Dhdt{2}/d_{\mathrm{HD}}^t$ as a function of $y_1$ at different $\phi$ for bidisperse porous media of $\lambda=2$.  The monodisperse translational hindered diffusivity at the corresponding $\phi$ is $d_{\mathrm{HD}}^t$.
}
  \label{fig:dhdt-2}
\end{figure}

The effects of porous media composition $y_1$ on the diffusivity ratio $\Dhdt{\alpha}/d^t_{\mathrm{HD}}$ are shown in Fig.~\ref{fig:dhdt-2}.
The translational hindered diffusivity for monodisperse porous media at the same $\phi$ is $d^t_{\mathrm{HD}}$.
At any $\phi$, the diffusivities $\Dhdta$ for both species decreases monotonically with increasing $y_1$, towards the monodisperse results for the smaller particles and away from it for the larger particles.
When presented in trace amount at a fixed $\phi$, the ratio $\Dhdt{1}/d^t_{\mathrm{HD}}$ of the smaller particles reaches a maximum while $\Dhdt{2}/d^t_{\mathrm{HD}}$ for the larger particles reaches a minimum.
Compared to the suspension $d^t_{s,1}/d^t_s$, the maximum of $\Dhdt{1}/d^t_{\mathrm{HD}}$ is significantly higher due to stronger HIs.
Moreover, the increase of $\Dhdt{1}/d^t_{\mathrm{HD}}$ with decreasing $y_1$ is clearly stronger than linear when $y_1\rightarrow 0$.
For the larger particles, at low to moderate $\phi$, as shown in the inset of Fig.~\ref{fig:dhdt2-t2}, introducing the smaller particles to the system reduces its hindered diffusivity, and the reduction enhances with increasing $\phi$.
However, for dense porous medium, particularly when $\phi>0.5$, increasing $\phi$ at fixed $y_1$ increases $\Dhdt{2}$.
For $\phi>0.6$, the hindered diffusivity for the large particles $\Dhdt{2}$ becomes extremely sensitive to the small particles.  In Fig.~\ref{fig:dhdt2-t2} at $\phi = 0.635$, the maximum for the ratio  $\Dhdt{1}/d^t_{\mathrm{HD}}$ is present at $y_1\ll 0.1$.
In contrast, the suspension ratio $d^t_{s,2}/d^t_s$ exhibits less sensitivity.
Note that only at $\phi = 0.635$, the presence of the smaller particles enhances the hindered diffusivities of both species in the porous medium.

\subsection{Rotational hindered diffusivity}
\label{sec:rotat-hind-diff}


\begin{figure} 
  \begin{center}
    \includegraphics[width=3in]{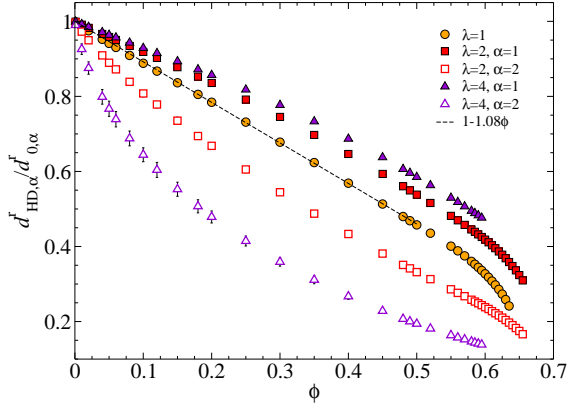}
   \end{center}
  \caption{(Color online)
The rotational hindered diffusivities $\Dhdr{\alpha}$, $\alpha\in\{1,2\}$, of both species for bidisperse porous media with $y_1=0.5$ and $\lambda = 1$,  $2$, and $4$. The linear fit in Eq.~(\ref{eq:dhdr-fit}) is also presented in dashed line.
}
  \label{fig:dhdr-1}
\end{figure}

Finally, the $\phi$ dependence of the rotational hindered diffusivities $\Dhdra$ for bidisperse porous media with $y_1 = 0.5$ at $\lambda = 2$ and $4$ and for monodisperse porous media is shown in Fig.~\ref{fig:dhdr-1}.
The monodisperse rotational hindered diffusivity $d^r_{\mathrm{HD}}$ agrees with the earlier study~\cite{sd-ew-transport-coeff-pt2_phillips_pof1988} and decreases much slower with $\phi$ comparing to its translational counterpart $d^t_{\mathrm{HD}}$.
The SD results up to $\phi = 0.5$ can be satisfactorily described by a linear fit,
\begin{equation}
  \label{eq:dhdr-fit}
  \frac{d^r_{\mathrm{HD}}}{d^r_0} = 1-1.08\phi,
\end{equation}
also shown in Fig.~\ref{fig:dhdr-1}.
This is a stronger dependence on $\phi$ compared to the suspension short-time rotational self-diffusivity $d_s^r$ in Sec.~\ref{sec:rotat-self-diff}.
Approaching the close packing volume fraction, the diffusivity $d^r_{\mathrm{HD}}$ decreases but largely remains finite, as nearby stationary particles can only weakly affect the rotation of the mobile particle.

In bidisperse porous media, $\Dhdr{\alpha}$ for both species is highly sensitive to the size ratio $\lambda$.
The bidisperse $\Dhdr{\alpha}$ differs significantly from the monodisperse results, and no longer displays the almost linear relation with $\phi$.
For the smaller particles, the diffusivity $\Dhdr{1}$ is higher than the monodisperse results, while the for the larger particles $\Dhdr{2}$ is always lower.
The deviation from the monodisperse results grows with increasing particle size ratio $\lambda$, and is more significant for the larger particles.
This is because the average number of neighboring particles, which produces the most significant HI to the mobile particle, scales as $\lambda^3$ for the larger particles.

\begin{figure} 
  \begin{center}
  \subfloat[]{
    \centering
    \includegraphics[width=3in]{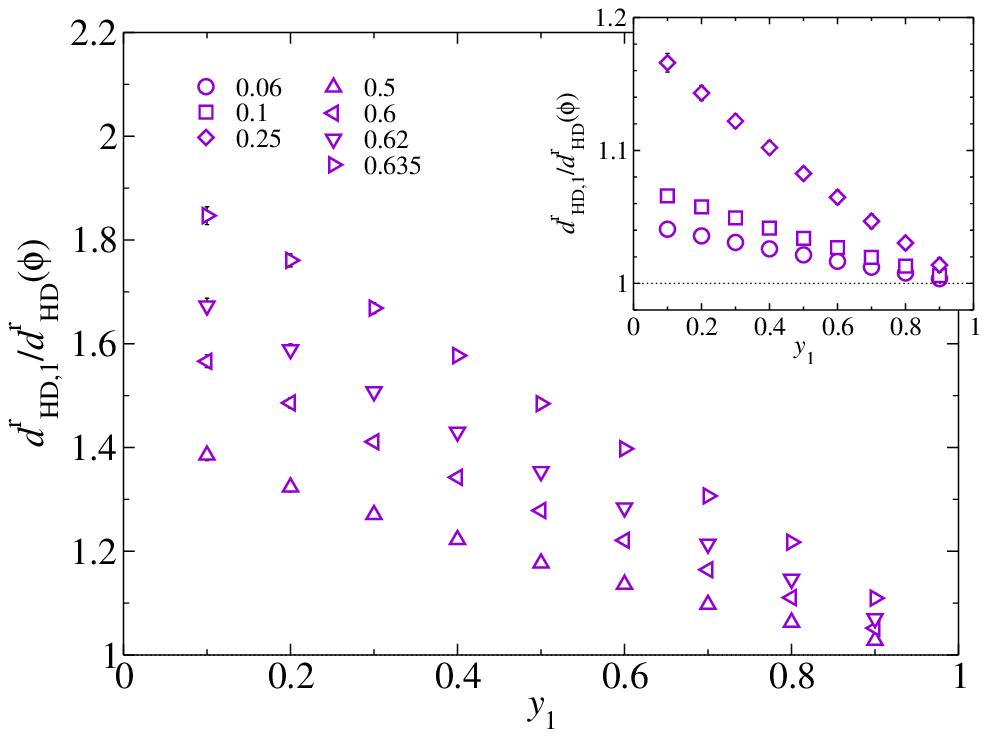}
    \label{fig:dhdr1-t2}
  }\\
  \subfloat[]{
    \centering
    \includegraphics[width=3in]{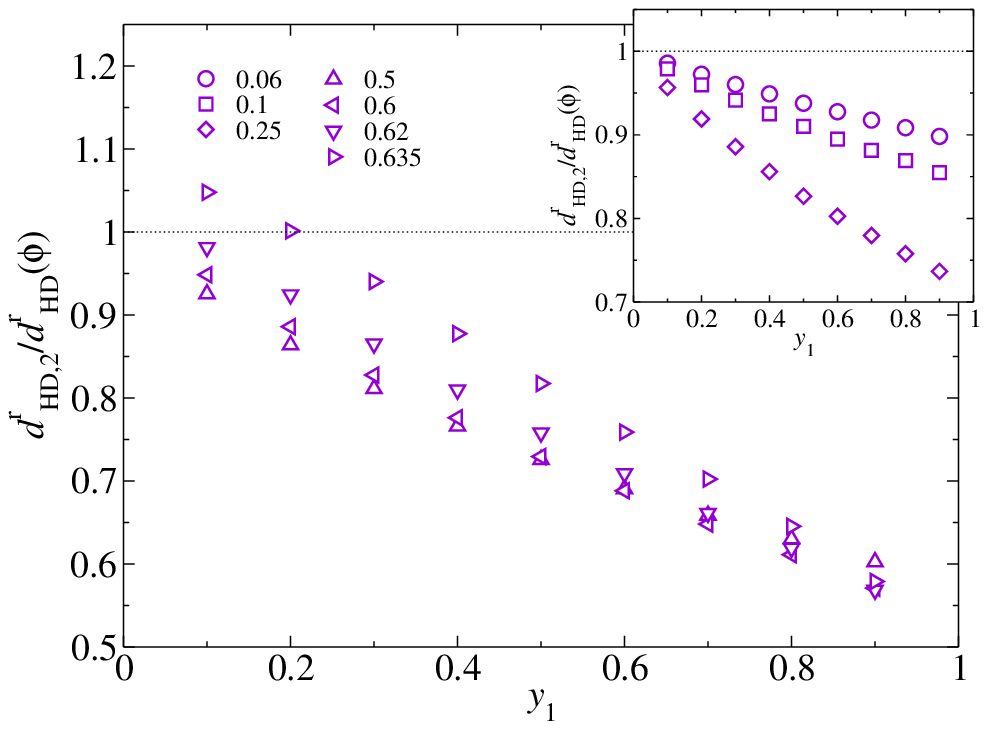}
    \label{fig:dhdr2-t2}
  }
  \end{center}
  \caption{
The normalized rotational hindered diffusivities \protect\subref{fig:dhdr1-t2}: $\Dhdr{1}/d_{\mathrm{HD}}^r$ and \protect\subref{fig:dhdr2-t2}: $\Dhdr{2}/d_{\mathrm{HD}}^r$ as a function of $y_1$ at different $\phi$ for bidisperse porous media of $\lambda=2$.  The monodisperse rotational hindered diffusivity at the corresponding $\phi$ is $d_{\mathrm{HD}}^r$.
}\label{fig:dhdr-2}
\end{figure}

The effects of the medium composition $y_1$ on the ratio $\Dhdr{\alpha}/d^r_{\mathrm{HD}}$ for $\lambda = 2$, where $d^r_{\mathrm{HD}}$ is the monodisperse data at the same $\phi$, are shown in Fig.~\ref{fig:dhdr-2}.
The results are qualitatively similar to $d_{s,\alpha}^r/d_s^r$ in Fig.~\ref{fig:dinfr-2}.
Quantitatively, the effect of $y_1$ at fixed $\phi$ on $\Dhdr{\alpha}$ is slightly stronger.
At low to moderate $\phi$, $\Dhdr{\alpha}/d^r_{\mathrm{HD}}$ for both species decreases monotonically with increasing $y_1$.
At a fixed $\phi$, a trace amount of the small particle yields the maximum of $\Dhdr{1}/d^r_{\mathrm{HD}}$, while a trace amount of the large particles leads to the minimum of $\Dhdr{2}/d^r_{\mathrm{HD}}$.
At very  high $\phi$, the most notable feature is the mutual enhancement of $\Dhdr{1}$ and $\Dhdr{2}$ with a small amount of small particles, \eg, at $y_1 = 0.1$ and $\phi = 0.635$.  The extent of the enhancement, however, is much weaker than the translational counterpart $\Dhdt{\alpha}$, but is similar to the suspension counterpart $d_{s,\alpha}^r$.
The similarity between $\Dhdr{\alpha}$ and $d_{s,\alpha}^r$ suggests that the HIs of rotational motions are weak but sensitive to the environment through $\phi$ and $\lambda$.

\section{Concluding remarks}
\label{sec:conclusions}

In this work we presented a detailed study of the short-time transport properties of bidisperse suspensions and porous media over a wide range of parameter space using conventional Stokesian Dynamics.
For suspensions, our study includes the short-time translational and rotational self-diffusivities, the instantaneous sedimentation velocity, the partial hydrodynamic functions, and the high-frequency dynamic shear and bulk viscosities, and for porous media, our study includes the mean drag coefficient (permeability) and the translational and rotational hindered diffusivities.

Our computational survey shows that introducing a second species of different size to a monodisperse suspension or porous medium leads to significant changes in the hydrodynamic interactions, and different transport properties respond differently. 
For dense suspensions, the changes in particle structures can significantly affect the HIs, leading to surprising mutual enhancement of diffusivities and reduction of viscosities.
The peak locations of the mixture hydrodynamic function $H_{NN}(q)$ differ from those of the mixture static structure factor $S_{NN}(q)$, suggesting great care is needed when studying the HIs of dense systems.
The $\log(\varepsilon^{-1})$ divergences of both the shear and bulk viscosities, where $\varepsilon=1-\phi/\phi_m$ with $\phi_m$ the limiting volume fraction, show the subtle and complex interplay between the lubrication interactions and the suspension structures.

To estimate suspension properties, the PA approximations can reliably predict various transport properties up to $\phi=0.15$.  The method breaks down at higher volume fractions, even with proper suspension structural input. 
For diffusivities, we found that the decoupling approximations in Eq.~(\ref{eq:dt-poly-approx}) and (\ref{eq:dr-poly-approx}) work better than the PA approximations.  They are particularly effective in estimating the diffusivities of larger particles up to $\phi = 0.4$, but the range of validity for the smaller particles is more restricted, indicating the HIs for the two species are different.  
For polydisperse sedimentation velocities, the approximation of Davis \& Gecol~\cite{sed-func-empirical_davis_jaiche1994} is quantitatively accurate at low to moderate $\phi$.

The limitation of the Stokesian Dynamics algorithm is also assessed in this work.  The low moment multipole expansions in SD cannot accurately capture the HIs corresponding to collective particle motions and with very large size ratios.  As a result, the SD computations of the suspension sedimentation velocity and porous media permeability are significantly different from other methods for $\phi>0.25$.  However, even in this range, SD is expected to capture the qualitative aspects of the size ratio effects.

The present work can serve as a concrete starting point for future experimental and computational investigations of polydisperse systems.  
Extension of this work includes improved approximation scheme for various transport properties~\cite{dg-sd-comp_wang_jcp_2014}, 
investigations of systems with different interaction potentials, \eg, screened electrostatic interactions, and long-time dynamic studies.

\begin{acknowledgments}
We thank Marco Heinen for helpful discussions.
M.W. gratefully acknowledges supports from the Natural Sciences and Engineering Research Council of Canada (NSERC) by a Postgraduate Scholarship (PGS), and the National Science Foundation (NSF) grant CBET-1337097.
\end{acknowledgments}

\appendix

\section{Additional expressions for the PA approximations}
\label{sec:addit-expr-pa}

The PA approximation of the polydisperse bulk viscosity requires first defining the functions $x_{\alpha\beta}^p$, which are the mobility counterpart of the resistance functions $X_{\alpha\beta}^P$ in Ref.~\onlinecite{pres-moment_jeffrey_pof1993}:
\begin{widetext}
  \begin{equation}
\begin{pmatrix}
  x_{11}^p &  \tfrac{1}{2}(1+\lba)x_{12}^p\\
   \tfrac{1}{2}(1+\lba)x_{21}^p &  \lba x_{22}^p
\end{pmatrix}
=\tfrac{1}{3}
\begin{pmatrix}
  x_{11}^a &  \frac{2}{1+\lba}x_{12}^a\\
   \frac{2}{1+\lba}x_{12}^a &  \frac{1}{\lba} x_{22}^a
\end{pmatrix}
\begin{pmatrix}
  X_{11}^P &  \tfrac{1}{4}(1+\lba)^2 X_{21}^P\\
  \tfrac{1}{4}(1+\lba)^2   X_{12}^P &  \lba^2 X_{22}^P
\end{pmatrix}
.   
  \end{equation}
\end{widetext}
With the definition of particle stresslet in Eq.~(\ref{eq:part-se-def}), we have the function $J_Q(s,\lba)$, which is essential for the suspension bulk viscosity,
\begin{align}
J_Q = & \frac{8}{(1+\lba)^3} \left[ \left(T_{11}^Q + \tfrac{1}{8}(1+\lba)^3T_{12}^Q  \right) 
- \left( X_{11}^Px_{11}^p \right. \right.\nonumber \\
&  + \tfrac{1}{8}(1+\lba)^3X_{12}^P x_{21}^p
+\tfrac{1}{2}(1+\lba) X_{11}^P x_{12}^p  \nonumber \\
& \left.\left.+ \tfrac{1}{4}\lba(1+\lba)^2 X_{12}^Px_{22}^p
\right)
\right],
\end{align} 
where $T^Q_{\alpha\beta}$ are computed in Ref.~\onlinecite{compres-res_khair_pof2006}. Finally, we have
\begin{equation}
\hat{J}_Q(s, \lba) = \tfrac{1}{2}[J_Q(s,\lba) + J_Q(s, \lba^{-1})]
\end{equation}
for computing the integral in Eq.~(\ref{eq:ikappa-def}).

We use the following asymptotic expressions for $s\rightarrow \infty$ in the PA approximations:
\begin{align}
 x_{11}^a + 2y_{11}^a -3 \approx &  -\frac{60\lba^3}{(1+\lba)^4} s^{-4} + \frac{480\lba^3-264\lba^5}{(1+\lba)^6} s^{-6}, \\
 x_{11}^c + 2y_{11}^c -3 \approx & -\frac{480\lba^3}{(1+\lba)^6} s^{-6} -  \frac{5760\lba^3}{(1+\lba)^8} s^{-8}, \\
\hat{x}_{12}^a+ 2\hat{y}_{12}^a \approx&  \frac{1200\lambda^3}{(1+\lambda)^6}s^{-5},\\
\hat{J} \approx &\frac{480 \lba^3}{(1+\lba)^6} s^{-6}, \\
\hat{J}_Q \approx&  \frac{1280\lba^3(1-\lba+\lba^2)}{(1+\lba)^8} s^{-6}.
\end{align}

\bibliography{trans-bidisperse}

\end{document}